\def\aap{\mbox{\bf{Astron.\ Astrophys.}}}

\def\apjl{\mbox{\bf{Astrophys.\ J.\ L.}}}

\def\apjs{\mbox{\bf{Astrophys.\ J.\ Suppl.}}}

\documentclass{aa}

\usepackage{longtable, lscape}
\usepackage{graphicx}
\usepackage{txfonts}
\usepackage{natbib}
\usepackage{color}

\bibpunct{(}{)}{;}{a}{}{,} 

\newcommand{\Msun}{\mbox{$\mathrm{M}_{\odot}$}}
\newcommand{\chisq}{\mbox{$\chi^2$}}
\newcommand{\chisqmin}{\mbox{$\chi^2_{\rm{min}}$}}
\newcommand{\chir}{\mbox{$\chi^2$}}
\newcommand{\chirmin}{\mbox{$\chi^2_{\rm{min}}$}}

\newcommand{\Rdon}{\ensuremath{R_\mathrm{D}}}
\newcommand{\Racc}{\ensuremath{R_\mathrm{A}}}

\defcitealias{2003MNRAS.339..157H}{H03}
\defcitealias{2005MNRAS.tmp...20H}{H05}

\begin{document}

   \title{Efficiency of mass transfer in massive close binaries }
   \subtitle{Tests from double-lined eclipsing binaries in the SMC}

   \titlerunning{Efficiency of mass transfer in massive close binaries}
   \author{S.E. de Mink \inst{1}, O.R. Pols \inst{1} \and R. W. Hilditch \inst{2}}
   \institute{Astronomical Institute, Utrecht University,
              PO Box 80000, NL-3508 TA Utrecht, The Netherlands,\\
              \email{S.E.deMink@astro.uu.nl}
   \and
   School of Physics and Astronomy. University of St Andrews,
              North Haugh, St Andrews, Fife KY16 9SS, Scotland, U.K.
             }
   \date{Received : 22/12/2006; accepted : 05/03/2007 }

  \abstract
  {}
  {One of the major uncertainties in close binary evolution is the
  efficiency of mass transfer $\beta$: the fraction of transferred
  mass that is accreted by a secondary star. We attempt to constrain
  the mass-transfer efficiency for short-period massive binaries
  undergoing case A mass transfer.}
  {We present a grid of about 20,000 detailed binary evolution tracks
  with primary masses 3.5--35~\Msun, orbital periods 1--5~days at a
  metallicity $Z = 0.004$, assuming both conservative and
  non-conservative mass transfer. We perform a systematic comparison,
  using least-squares fitting, of the computed models with a sample of
  50 double-lined eclipsing binaries in the Small Magellanic Cloud,
  for which fundamental stellar parameters have been determined. About
  $60\%$ of the systems are currently undergoing
  slow mass transfer. }
  { In general we find good agreement between our models and the
  observed detached systems. However, for many of the semi-detached
  systems the observed temperature ratio is more extreme than our
  models predict. For the 17 semi-detached systems that we are able to
  match, we find a large spread in the best fitting mass-transfer
  efficiency; no single value of $\beta$ can explain all systems. We
  find a hint that initially wider systems tend to fit better to less
  conservative models. We show the need for more accurate temperature
  determinations and we find that determinations of surface abundances
  of nitrogen and carbon can potentially constrain the mass-transfer
  efficiency further.  }
  {}
   \maketitle

\section{Introduction }\label{sec:intro}

    Evolutionary calculations of massive close binaries have been
    conducted with various stellar evolution codes since the 1960s,
    e.g. \cite{1966AcA....16..231P}, \cite{1969A&A.....3...83K},
    \cite{1992A&AS...96..653D}, \cite{1994A&A...290..119P},
    \cite{1994A&A...292..463D}, \cite{2001A&A...369..939W} and
    \cite{2001ApJ...552..664N}.
    One of the major uncertainties in these calculations is the
    efficiency of mass transfer: what fraction $\beta$ of the
    transferred mass is actually accreted by the secondary star?
    Conservative evolution, i.e. the case where mass and angular 
    momentum of the
    binary system are conserved, corresponds to $\beta=1$.
    Non-conservative evolution has been considered in most
    calculations by assuming a constant $\beta < 1$ while the amount
    of angular momentum loss is described using a second parameter.


    This question is relevant for many current astrophysical problems
    as it affects, for example, the formation rate of progenitor
    models of long-soft gamma-ray burst \citep{2005A&A...435..247P},
    of double neutron star binaries \citep{2006MNRAS.368.1742D,
    2007Dewisubmitted}, which are thought
    to be the progenitors of short-duration gamma-ray bursts, and Type
    Ia supernovae \citep{2005A&A...435..967Y}, which are used as
    standard candles to measure universal expansion.


    Two effects of mass transfer on the accreting star
    are likely to result in mass loss from the system: expansion and
    spin up.  \citet{1970PhDT........14B},
    \cite{1976ApJ...206..509U} and others have shown that when mass
    transfer occurs on a timescale on the order of the thermal
    timescale of the donor star, the less massive accreting star, with
    a longer thermal timescale, is driven out of thermal equilibrium
    and expands. In systems with initial mass ratios very different
    from unity, this leads to a contact or common-envelope
    configuration.  Significant mass and angular momentum loss from
    the system is then expected \citep{1977ApJ...212..533F}.
     The second effect, spin up, was pointed out by
     \cite{1981A&A...102...17P}: after gaining only a few percent of
     its original mass through disk accretion, enough angular momentum
     is transferred to spin up an isolated accreting star to critical
     rotation. In principle this can lead to significant mass loss
     from the system \citep[e.g.][]{2003IAUS..212..275L, 2004IAUS..215..535L},
     depending on how efficient tidal effects can keep the accreting
     star rotating in synchrony with the orbit.


     No consensus has been reached on this topic. Many relevant
     processes are still not yet well understood. Hydrodynamical
     simulations are the most promising approach, but are still far
     too time-consuming to study the dependence of the mass transfer
     efficiency on the binary parameters.
     At the moment the most fruitful way to address the efficiency
     problem is to parametrize the process of mass transfer and to use
     observations to calibrate the parameters.  The most stringent tests
     come from double-lined eclipsing binaries for which the stellar
     parameters can be determined with accuracies of a few
     percent. Semi-detached systems undergoing their first phase of
     mass transfer provide the best test objects, to avoid
     uncertainties introduced by an unknown mass loss history.


      Various studies in which theory is compared to observations
      indeed indicate evidence for non-conservative mass transfer. 
      \citet{1974A&A....36..113R} showed that
      it is very likely that the semi-detached binary AS Eri is the
      result of non-conservative evolution.
      \citet{1993MNRAS.262..534S} studied the semidetached system
      $\beta$~Per (Algol) and claimed that it
      has lost about 15 percent of its initial total mass and 30
      percent of its initial total angular momentum.
      Also for $\beta$~Lyr moderate mass loss was inferred
      \citep{1994A&A...291..786D}.
      \cite{1994A&A...283..144F} compared Galactic OB binaries, of
      which 8 are semi-detached, to stellar evolution models. They
      estimate that these systems have lost between 30\% and 60\% of
      the transferred mass.
      Many Wolf-Rayet binaries with O-type companions require a highly
      non-conservative first mass-transfer phase to explain their
      orbital periods and masses \citep{2005A&A...435..247P}.
%
%
      \citet{2006A&A...446.1071V} compared a grid of calculated binary
      models statistically to observed orbital periods and mass ratios
      of Algols. They need to assume a significant amount of mass loss
      to obtain agreement between models and observations.


     Other studies, however, show that mass transfer is fairly
     conservative, at least for some systems.
     \citet{2001ApJ...552..664N} compared observations of semi-detached
     binaries with intermediate mass, in which both components have
     spectra in the range G0 to B1, to models assuming conservative
     evolution and they found an acceptable agreement
     overall. However, they did not compare the observed systems to
     non-conservative models.
     Almost conservative evolution during the first phase of mass
     transfer is also needed to explain the formation of several types
     of evolved binaries, including
     some massive X-ray binaries, e.g. Wray 977
     \citep{1995A&A...300..446K},
     and eccentric binary systems consisting of a white dwarf and a
     neutron star \citep{1999MNRAS.309...26P, 2000A&A...355..236T}.
%
%

     In view of the observational evidence for non-conservative mass
     transfer, in many calculations of binary evolution a single
     constant mass transfer efficiency has been assumed for all
     systems, often $\beta = 0.5$
     (e.g. \citealp{1992A&AS...96..653D,2002MNRAS.335..948C}). However,
     neither the above-mentioned observational comparisons nor the
     theoretical considerations outlined earlier provide a basis for
     such an assumption -- in fact, they clearly indicate that $\beta$
     is not a constant but probably depends on the masses and orbital
     properties of a binary. No clear picture of this dependence has
     emerged yet, partly hampered by the fact that much of the
     evidence is based on incidental studies or heterogeneous
     observational samples.


     In this paper we attempt to shed more light on this question by
     means of a systematic comparison between binary evolution models
     and a homogeneous sample of binaries undergoing mass transfer.
     Recently \citet[hereafter
     \citetalias{2003MNRAS.339..157H}]{2003MNRAS.339..157H} and
     \citet[hereafter
     \citetalias{2005MNRAS.tmp...20H}]{2005MNRAS.tmp...20H} presented
     the fundamental stellar parameters for a sample of 50
     double-lined eclipsing binaries in the Small Magellanic Cloud
     (SMC).  It is the largest single set of fundamental parameters
     determined for high mass stars in any galaxy. More than $50\%$ of
     the systems are semi-detached and believed to be currently
     undergoing the slow phase of case A mass transfer (mass transfer
     in a system consisting of two main sequence stars). As case A
     mass transfer is the first phase of mass transfer after the stars
     left the zero-age main sequence, there are no uncertainties
     induced by previous mass transfer phases. The detached systems in
     the sample enable us to test our models in the pre-mass transfer
     phase.


     No suitable set of case A binary models are available at this
     moment. Large grids have been calculated before, for example by
     \cite{2001ApJ...552..664N} and \cite{2001A&A...369..939W}, but
     not at the metallicity of the SMC, i.e. $Z=0.004$.
     With this work we present a large grid of detailed case A binary
     evolution models at the metallicity of the SMC
     with different assumptions for the mass transfer efficiency.
     We address the problem of mass transfer efficiency by fitting
     binary evolution tracks to each individual system in the observed
     sample presented by \citetalias{2003MNRAS.339..157H} and
     \citetalias{2005MNRAS.tmp...20H}. We then investigate if
     correlations can be found between the best fitting mass transfer
     efficiency parameter $\beta$ and the initial binary parameters.


     \section{Stellar evolution code}\label{sec:code}

     To calculate detailed binary evolution tracks, we used the STARS
     stellar evolution code, a variant of the code originally
     developed by \citet{1971MNRAS.151..351E,1972MNRAS.156..361E}.
     An important update \citep[ and references
     therein]{1995MNRAS.274..964P} was the improvement of the original
     equation of state \citep{1973A&A....23..325E} by inclusion of
     pressure ionization and Coulomb interactions, OPAL opacity tables
     and recent nuclear reaction and neutrino loss rates.
     A recent addition to the code, which we have used in the
     calculations presented here, is the so-called TWIN mode in which
     the structure and composition equations for both stars in a
     binary are solved simultaneously with equations for the spin and
     orbital angular momentum, the orbital eccentricity and the mass
     flux between the stars \citep{2006epbm.book.....E}\footnote{A
     description of the current version of the code can be obtained on
     request from {\tt ppe@igpp.ucllnl.org}.}.


     Convective mixing is modeled by a diffusion equation for each of
     the composition variables. 
     A mixing length ratio $l/H_\mathrm{p}=2.0$ is assumed. 
     Convective overshooting is taken into account as in
     \citet{1997MNRAS.285..696S} with a free parameter
     $\delta_\mathrm{ov}=0.12$ calibrated against accurate stellar
     data from non-interacting binaries \citep{1997MNRAS.285..696S,
     1997MNRAS.289..869P}.


    In close binary systems tidal interaction tends to circularize the
    orbit and to synchronize the orbital period with the spin periods
    of the stars. Spin-orbit interaction by tides is treated according
    to the equilibrium tide theory
    \citep{1981A&A....99..126H, 1998ApJ...499..853E}.
    We enforce synchronized rotation in all our models by decreasing
    the tidal friction timescale by a factor of $10^{-4}$ with respect to
    the default value given by
    \citet{2002ApJ...575..461E}.
    The effect of rotation on stellar structure is taken into account
    as a reduction of the effective gravity assuming rigid rotation.


    The prescription for mass transfer implemented in the TWIN version
    of the code allows us to treat both semi-detached and contact
    binaries. Mass transfer by Roche-lobe overflow is modelled as a
    function of the potential difference $\phi_\mathrm{s}$ between the
    stellar surface and the Roche-lobe surface, for each star. As long
    as neither star exceeds its Roche lobe ($\phi_s < 0$) the
    mass-transfer rate is zero. If one of the stars overfills its
    Roche lobe, the mass transfer rate is calculated by solving an
    additional differential equation for the mass flux at each mesh
    point outside the Roche surface potential, simultaneously with the
    other structure equations,
    \begin{equation} \label{eq:Mdot}
      \frac{d \dot{M}}{d m} = - C \times \frac{\sqrt{2\phi_\mathrm{s}}}{r}, 
    \end{equation}
    where $m$ is the mass coordinate and $r$ the radius.  The mass
    transfer rate is then given by the integral of eq.~(\ref{eq:Mdot})
    over all mesh points outside the Roche surface potential.  By
    choosing a sufficiently large value for the free parameter $C$,
    i.e. $10^4$, we make sure that the radius stays close to the Roche
    lobe radius in semi-detached binaries and the mass transfer rate
    is self-regulating. In the case of stable mass transfer the mass
    loss rate does not depend on this function, it is set by the
    evolutionary expansion rate and the change in orbital
    separation. The exact form of eq.~(\ref{eq:Mdot}) is therefore not
    important.
    The main rationale for using it is that we can treat the case
    where both stars overfill their Roche lobes in a similar way. The
    direction of mass flow now depends on the difference in surface
    potentials $\phi_{s1}-\phi_{s2}$ between the stars,
    \begin{equation}
      \frac{d \dot{M_1}}{d m} = - C \times \mathrm{sign}(\phi_{s1}-\phi_{s2})
            \frac{\sqrt{2|\phi_{s1}-\phi_{s2}|}}{r},
    \end{equation}
    with the equivalent equation for star 2 having the reverse sign.
    Hence mass flows from star 1 to star 2 if $\phi_{\mathrm s1} >
    \phi_{\mathrm s2}$ and vice versa. 
%

    We note that the physics and evolution of contact binaries is a
    complex and currently unsolved problem in stellar evolution.
    Massive contact binaries are quite common, and although they are
    less well studied than their low-mass equivalents, the W\,UMa
    systems, both classes appear to share the property of having
    components with nearly equal effective temperatures. This requires
    a mechanism for heat transfer between the stars. A simple physical
    model for heat transfer in low-mass contact binaries with
    convective envelopes was developed and included in the TWIN
    version of the code by Yakut \& Eggleton (2005). However, the
    physics of heat transfer in binaries with radiative envelopes is
    probably quite different. For want of a quantitative model, we
    ignored the possibility of heat transfer between the stars in the
    calculations presented in this paper.
%

    Mass and angular momentum loss during mass transfer is modeled by
    treating the mass transfer efficiency as a free parameter,
    \begin{equation}
        \beta= -\dot{M_A}/\dot{M_D},
    \end{equation}
    where $A$ stands for accreting star and $D$ for the mass
    donor. $\beta$ is assumed to be constant throughout the
    evolution. We assume that the mass lost from the system takes away
    the specific angular momentum of the orbit of the accreting star,
    \begin{equation} 
       h = \left( \frac{M_D}{M_{\rm tot}}\right)^2 a^2\omega, 
    \end{equation} 
    where $M_{\rm tot}$ represents the total mass of the system, $a$
    the separation between the two stars and $\omega$ the orbital
    angular velocity.  Reverse mass transfer, i.e., mass transfer in
    an evolved system from the initially less massive component to the
    initially more massive component, is assumed to be
    conservative. This assumption does not affect the comparison with
    the observed sample as all stars are on the main sequence and
    reverse mass transfer occurs only in evolved systems.

    The rate of stellar wind mass loss decreases with metallicity
    \citep{2001A&A...369..574V,2006A&A...456.1131M}. For comparison
    with the observed systems our main interest is the core hydrogen
    burning phase during which the amount of wind mass loss is
    small compared to later stages of evolution and
    compared to the amount of mass loss during Roche lobe
    overflow. Therefore we assume no mass loss in the form of a
    stellar wind.


    We calculated our models at a metallicity of $Z = 0.004$. For the
    hydrogen and helium abundance we assume
    \begin{equation}
    \begin{array}{ccl}
    X &=& 0.76 -3.0 Z, \\
    Y &=& 0.24 + 2.0 Z. \\
    \end{array}
    \end{equation}
    The abundances of the heavier
    elements are assumed to scale to solar and meteoric
    abundance ratios as determined by
    \citet{1989GeCoA..53..197A}
    so that they are consistent with the opacity tables
    \citep{1992ApJS...79..507R} 
    even though the actual abundances in the SMC
    deviate to some extent from solar proportions
    \citep{1992ApJ...384..508R}. 
%


    \section{Binary evolution tracks} \label{sec:models}

        \begin{figure*}
           \centering 
           \includegraphics[ bb=80 0 540 620,clip,angle=-90,width=0.5\textwidth]
	   {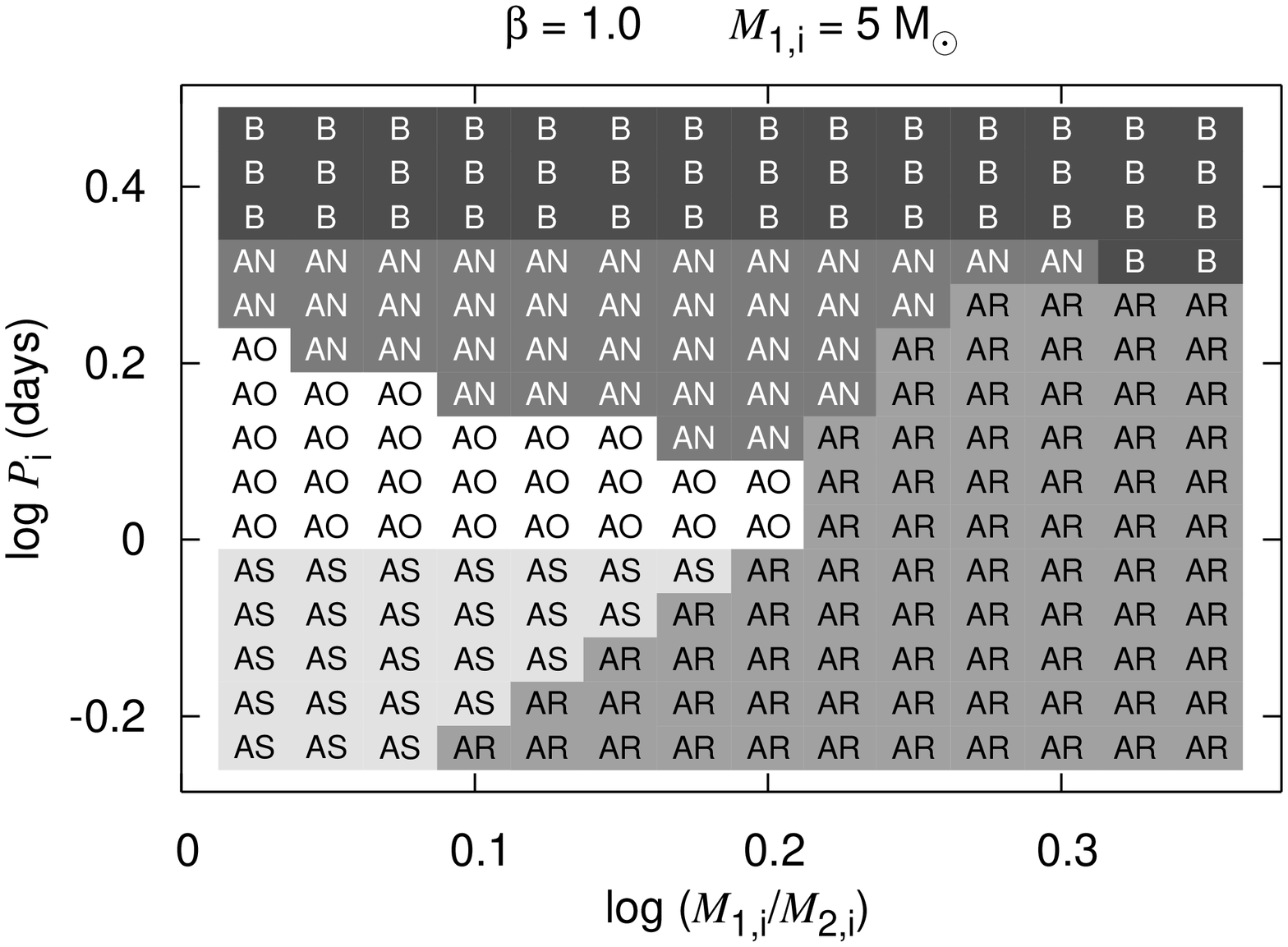}%
	   \includegraphics[ bb=80 0 540 620,clip,angle=-90,width=0.5\textwidth]
           {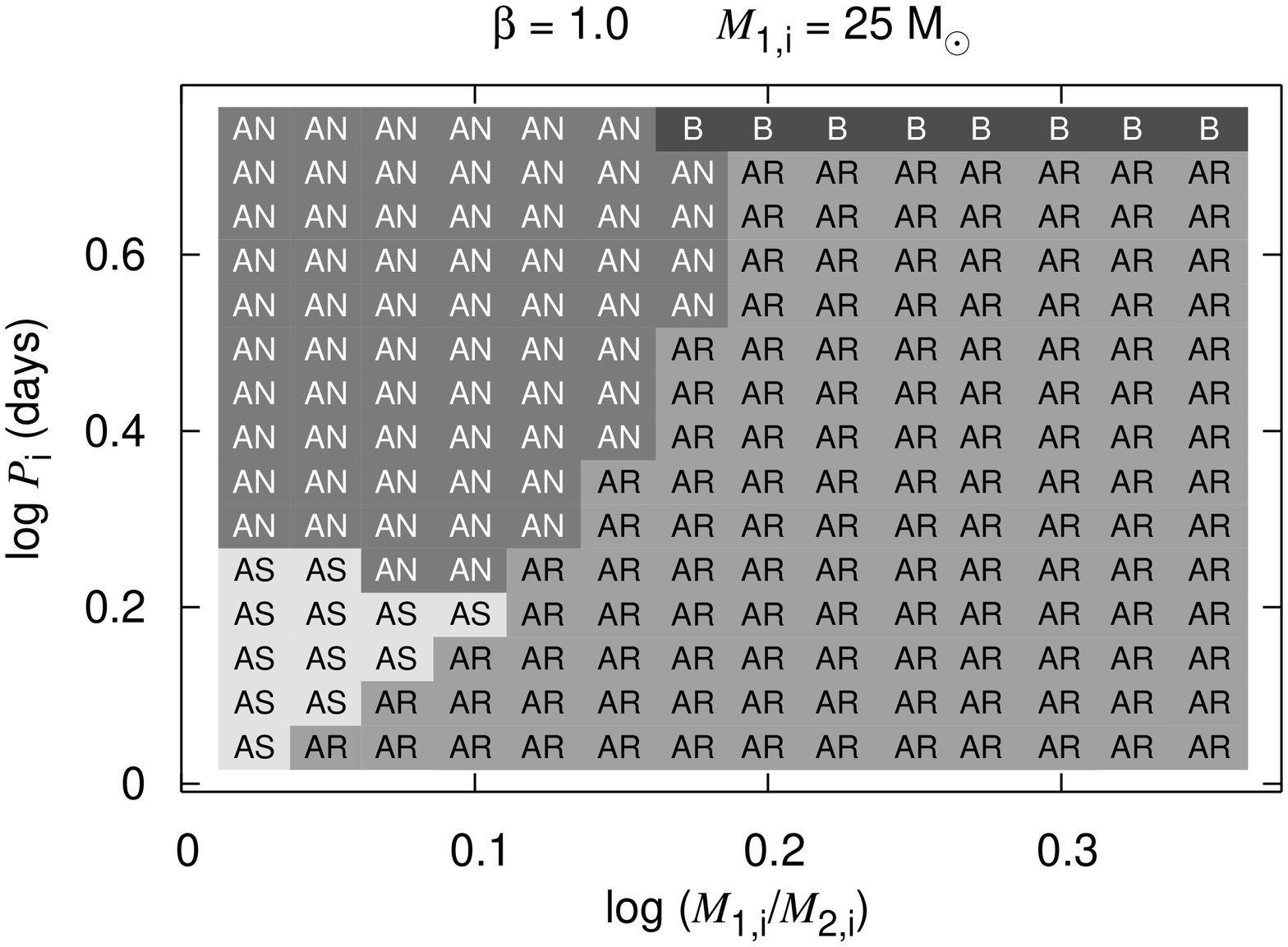} 
           \includegraphics[ bb=80 0 540 620,clip,angle=-90,width=0.5\textwidth]
	   {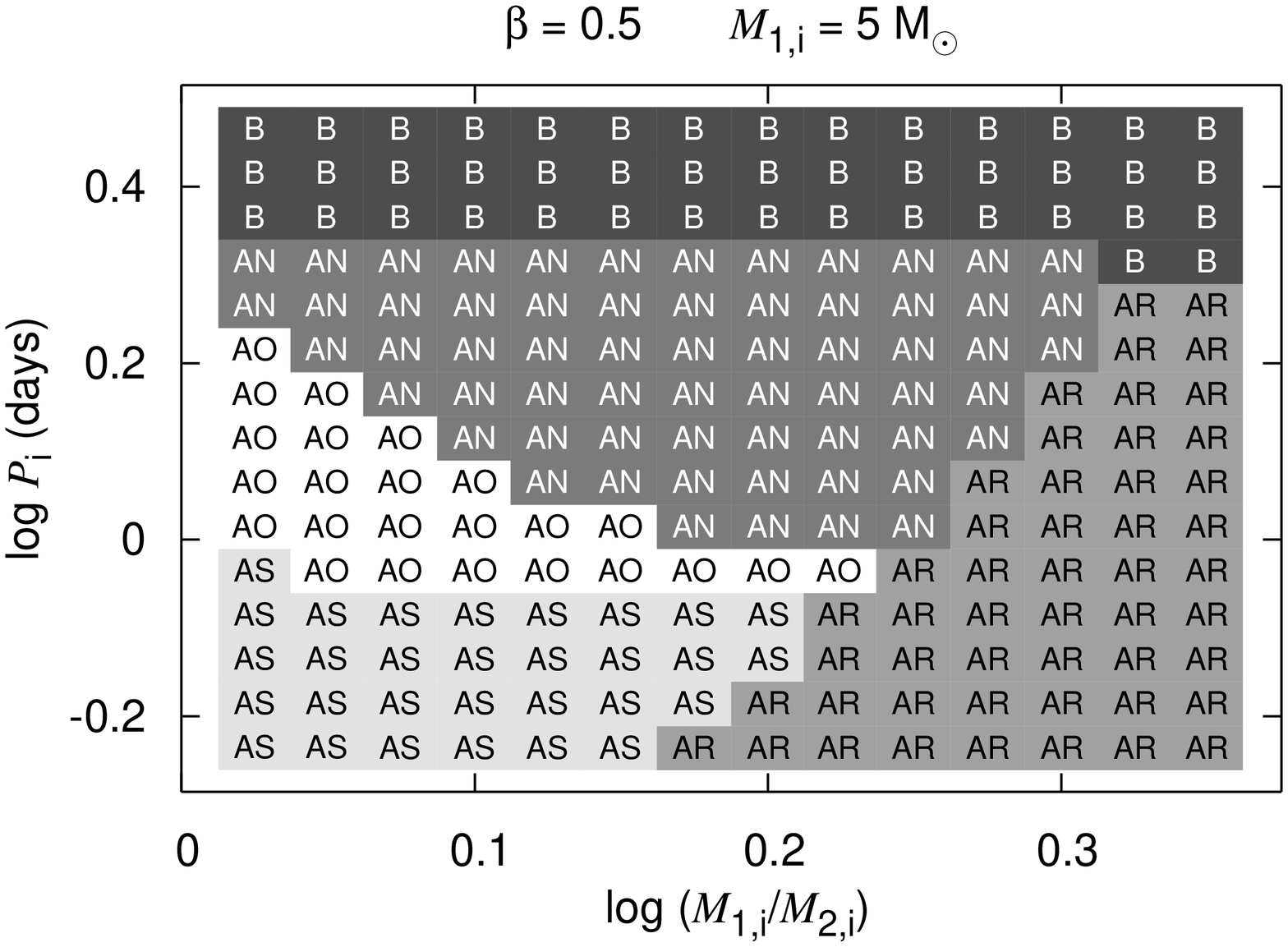}%
           \includegraphics[ bb=80 0 540 620,clip,angle=-90,width=0.5\textwidth]
	   {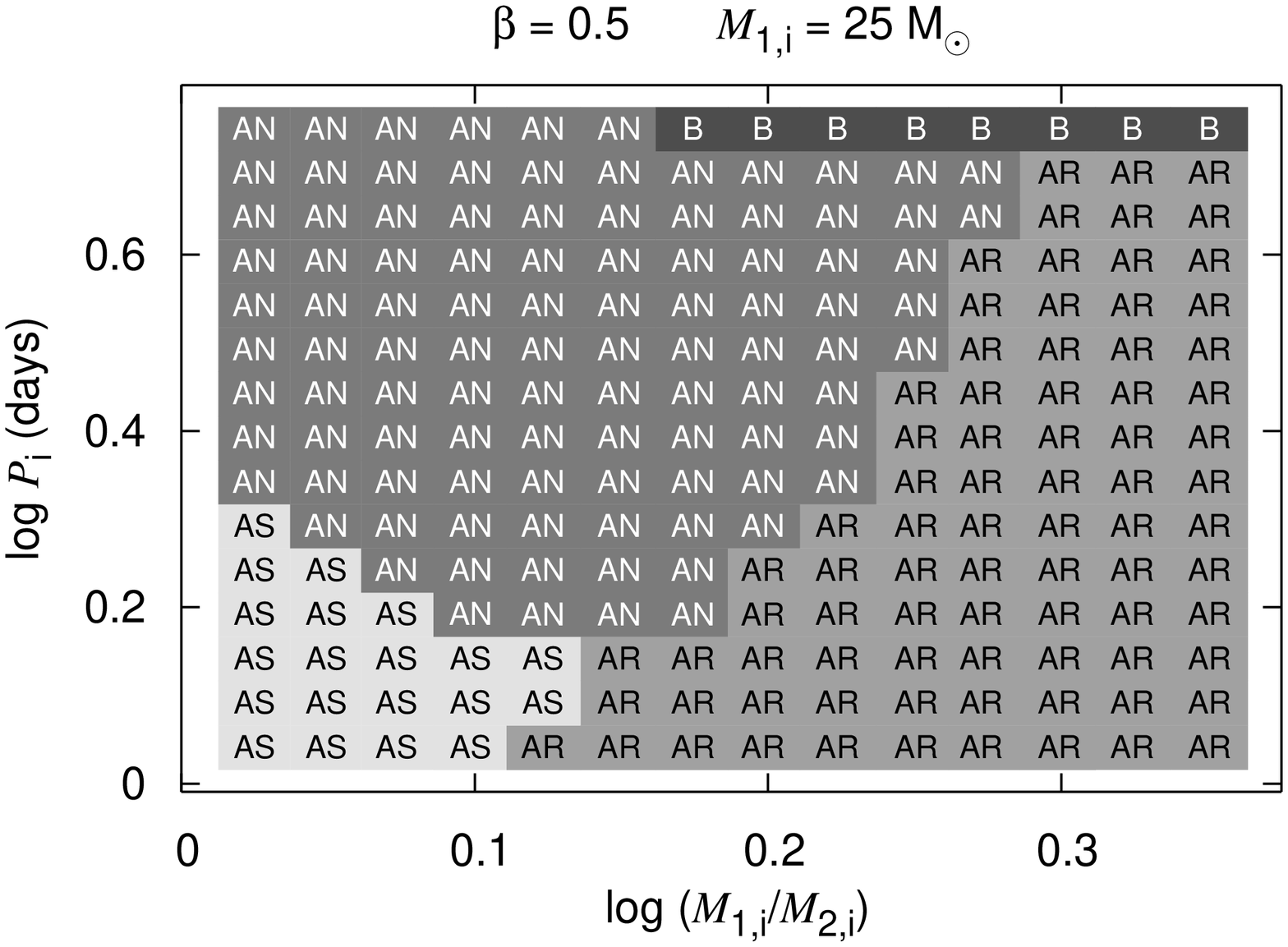}
	   \caption{ Classification of different types of binary
             evolution. In each diagram the sub-type is given as a
             function of initial orbital period and mass ratio for a
             fixed primary mass and fixed mass transfer efficiency
             $\beta$.  The initial primary mass is $5 \Msun$ for the
             figures on the left and $25 \Msun$ for the figures on the
             right. In the two top figures conservative evolution is
             assumed, $\beta = 1.0$, and the two lower figures $\beta
             = 0.5$.  The subtypes abbreviations are described in the
             text.  } \label{classification}
           \end{figure*}

    We have calculated four data cubes of binary evolution
    tracks\footnote{Our models are available on-line from the
    following website http://www.astro.uu.nl/data/stars/} 
    with different initial primary masses
    $M_{1,i}$, mass ratios $q_{i} =
    M_{1,i}/M_{2,i}$ and separations or orbital
    periods $P_i$. The subscript $i$ is used to indicate that
    these values are the initial values, i.e.  at the moment both
    stars start burning hydrogen on the main sequence. These
    parameters are spaced at approximately equal logarithmic
    intervals,
    \begin{equation}
    \begin{array}{cccccc}
    \log M_{1,i}             &=& 0.55, &0.60,& \ldots,& 1.55, \\
    \log q_{i}               &=& 0.025,& 0.050,&  \ldots, &0.350, \\
    \log (P_{i}/P_{\mathrm{ZAMS}}) &=& 0.05, &0.1, &\ldots,& 0.75, \\
    \end{array} \nonumber
    \end{equation}
    with $P_{\rm ZAMS}$ an approximation of the orbital period at which
    the initially more massive component would fill its Roche lobe on the
    zero-age main sequence for a system with equal masses,
    \begin{equation}
    P_{\rm ZAMS} \approx \frac {0.19 M_{1,i} + 0.47M_{1,i}^{2.33}}{1 +
      1.18M^2_{1,i}}.
    \end{equation}
    One data cube of binary tracks is calculated assuming conservative
    mass transfer $(\beta =1)$. Three additional data cubes are
    calculated assuming non-conservative mass transfer $(\beta = 0.75,
    0.5 \ \mathrm{ and } \  0.25)$.
    Each binary evolution track is terminated when one of the following
    conditions is met:
    \begin{itemize}
    \item 4000 time steps are taken,
    \item The stellar radius exceeds the Roche lobe radius by more than
    10\%, which happens if deep contact is established, a situation our code
    is not designed for.
    \item The mass loss rate becomes larger than $3\times10^2 M
     /\tau_{\mathrm{KH}}$, where $M$ is the mass of the initially more
     massive star and $\tau_{\mathrm{KH}}$ its Kelvin Helmholtz
     timescale. These systems are expected to come into contact.
   \item The code fails to converge. This occurs for example during
    advanced burning stages and in models for which the mass loss rate
    becomes very high. These models are expected to come into contact
    shortly after.
   \end{itemize}

     \subsection{ Subtypes of case A evolution \label{sec:subtypes}}

      Our conservative grid of models is very similar to the survey of
      case A binary evolution at solar metallicity published by
      \cite{2001ApJ...552..664N}.  In line with their definitions we
      identify different subtypes of case A binary evolution. As an
      example the subtypes are plotted in Figure~\ref{classification}
      for two different primary masses and for different assumptions
      about the mass transfer efficiency $\beta$.
     \begin{itemize}
     \item {Case B:} In our widest systems the initially
     most massive star will not fill its Roche lobe before depleting
     hydrogen in the core.  A system is classified as case B if it
     fills it Roche lobe after leaving the main sequence, which we
     define as the moment the central hydrogen abundance drops 
     below $0.001$. 
     \item {Case AR \emph{Contact during rapid mass transfer}: } In
     systems with large mass ratios, the mass-accreting star reacts to
     the mass exchange by expanding rapidly as it is driven out of
     thermal equilibrium. This leads to a contact situation during the
     first rapid phase of case A mass transfer. If the mass transfer
     rate $\dot{M}$ just before contact occurs is larger than $2\%$ of
     the typical thermal mass transfer rate $M/\tau_{\rm KH}$ a system 
     is classified as case AR. 
%
%
     \item {Case AS \emph{Contact during slow mass transfer}: }
     Binaries with small periods and moderately equal masses stay
     semi-detached during thermal-timescale mass transfer, but come into
     contact during the subsequent slow nuclear-timescale mass transfer
     phase.
     \item {Case AO \emph{Contact by overtaking}:} Mass transfer
     accelerates the evolution of the mass-accreting star. Especially
     in lower-mass systems there is a significant range of initial
     separations for which the mass-accreting star depletes its
     central hydrogen first.  Most systems come into contact shortly
     afterwards as the mass-accreting star expands in the Hertzsprung
     gap. This subtype corresponds to case AE in
     \cite{2001ApJ...552..664N}, with the small difference that we
     explicitly check if the system comes into contact.
     \item {Case AN \emph{No contact on the MS}:} Wider systems with
    moderately equal mass can avoid contact while both components are
    still on the main sequence. This subtype comprises cases
    AG, AL, AN and AB in \cite{2001ApJ...552..664N} between which we
    do not distinguish here.
     \end{itemize}

   \begin{figure} \centering
        \includegraphics[angle=-90,width=0.5\textwidth]{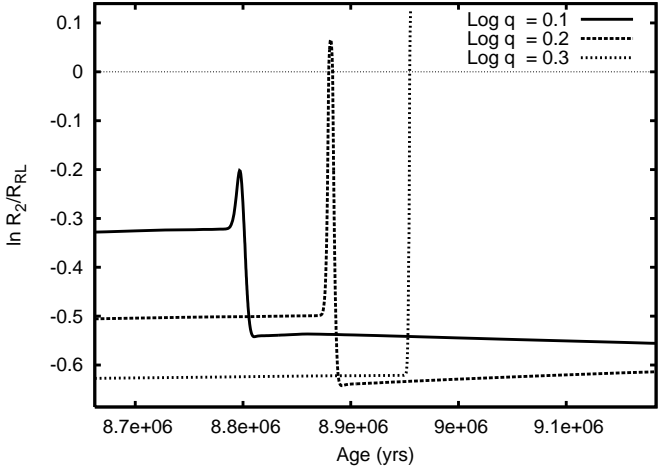}
        \caption{The natural logarithm of the radius over the Roche
          lobe radius as function of time for the mass-accreting star
          in binary systems with initial primary mass $M_{1,i} =
          18 M_{\odot}$, initial orbital period $P_i = 2.2$~days
          for three different mass ratios. The primary star of the
          system with $\log q = 0.2$ fills its Roche lobe at an age of
          $\approx 8.87$~Myrs. During the phase of rapid mass transfer
          the mass-accreting star expands so much that contact is
          established. Shortly afterwards it restores its thermal
          equilibrium and shrinks inside its Roche lobe again. The
          mass-accreting star in the system with $\log q_i = 0.1$
          stays well inside its Roche-lobe.  The accreting star in the
          $\log q_i = 0.3$ system excess its Roche Lobe by more than
          10\%. We stop the evolutionary calculations as our code is
          not able to handle deep contact configurations. Systems with
          more extreme initial mass ratios are likely to merge during
          rapid mass transfer.  }
        \label{FigVib} \end{figure}
%

     \subsection{Comparison to \cite{2001ApJ...552..664N} }

     In contrast to \cite{2001ApJ...552..664N}  our models are calculated
     at $Z=0.004$ instead of $Z=0.02$.  Due to lower opacity
     the stars are more compact, which causes the different subtypes of
     case A binary evolution to shift to smaller orbital periods.

     Another difference between our survey and
     \cite{2001ApJ...552..664N} is that in our version of the code the
     equations for both stars are solved simultaneously instead of
     first evolving the primary star, keeping track of the mass
     transfer rate and then evolving the secondary applying the stored
     mass transfer rate.  This enables us to model reverse mass
     transfer, although this does not occur often and it is not very
     relevant for the comparison with observations we undertake. A
     second advantage is that we continue our evolutionary
     calculations when the stars reach contact, see
     Figure~\ref{FigVib}. Therefore we find a subset of type AR
     systems that come into contact during rapid mass transfer, but in
     which the mass-accreting star quickly restores its thermal
     equilibrium and shrinks inside its Roche lobe again. This is
     followed by a slow mass-transfer phase like in systems that avoid
     contact altogether during rapid mass transfer. Some of the
     observed systems may have followed this type of
     evolution. Further evolution is either as case AS, AO or AN.

   \begin{figure*}
   \centering
   \includegraphics[ bb=80 50 540 750,clip,angle=-90,width=0.5\textwidth]%
    {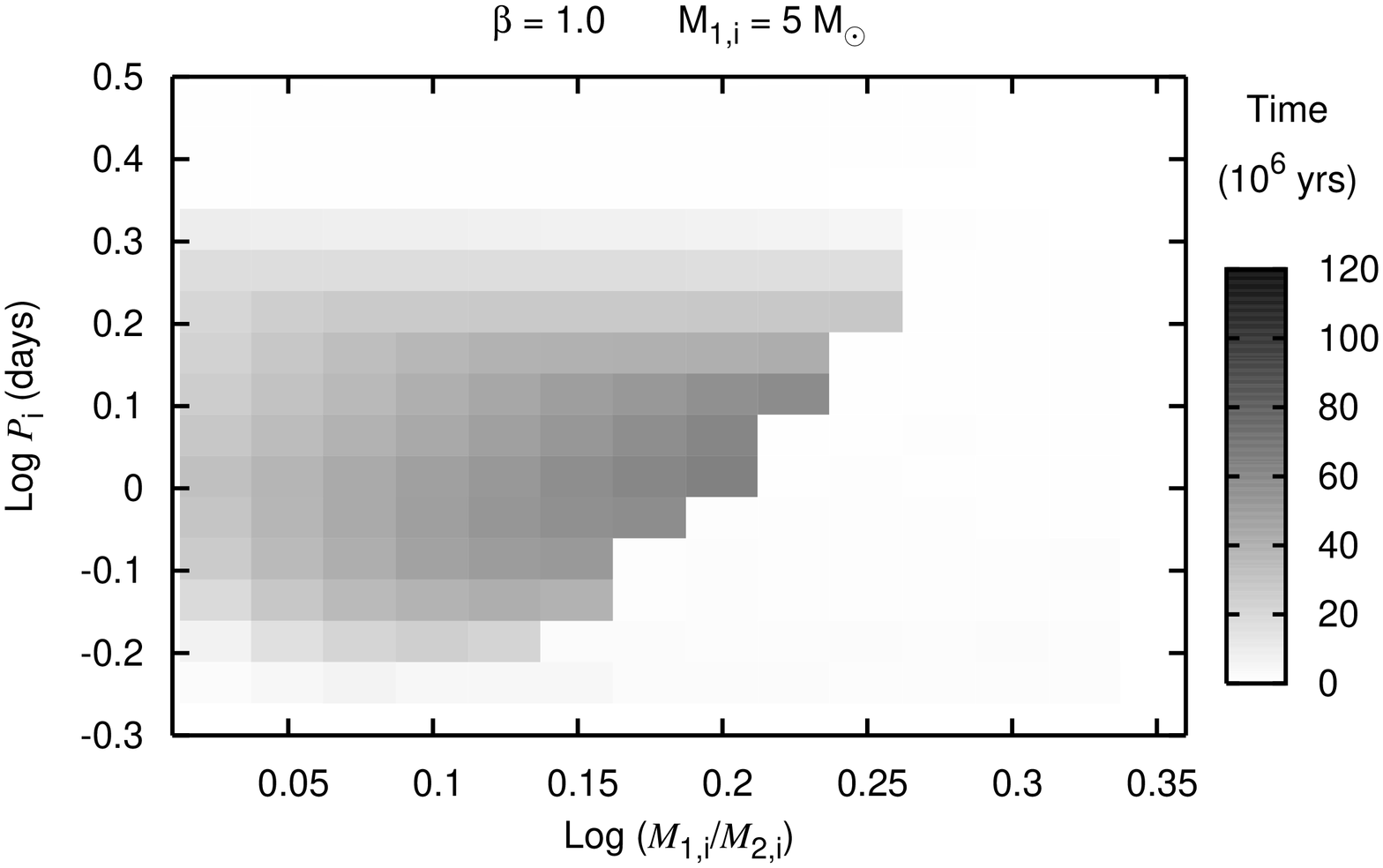}%
   \includegraphics[ bb=80 50 540 750,clip,angle=-90,width=0.5\textwidth]%
    {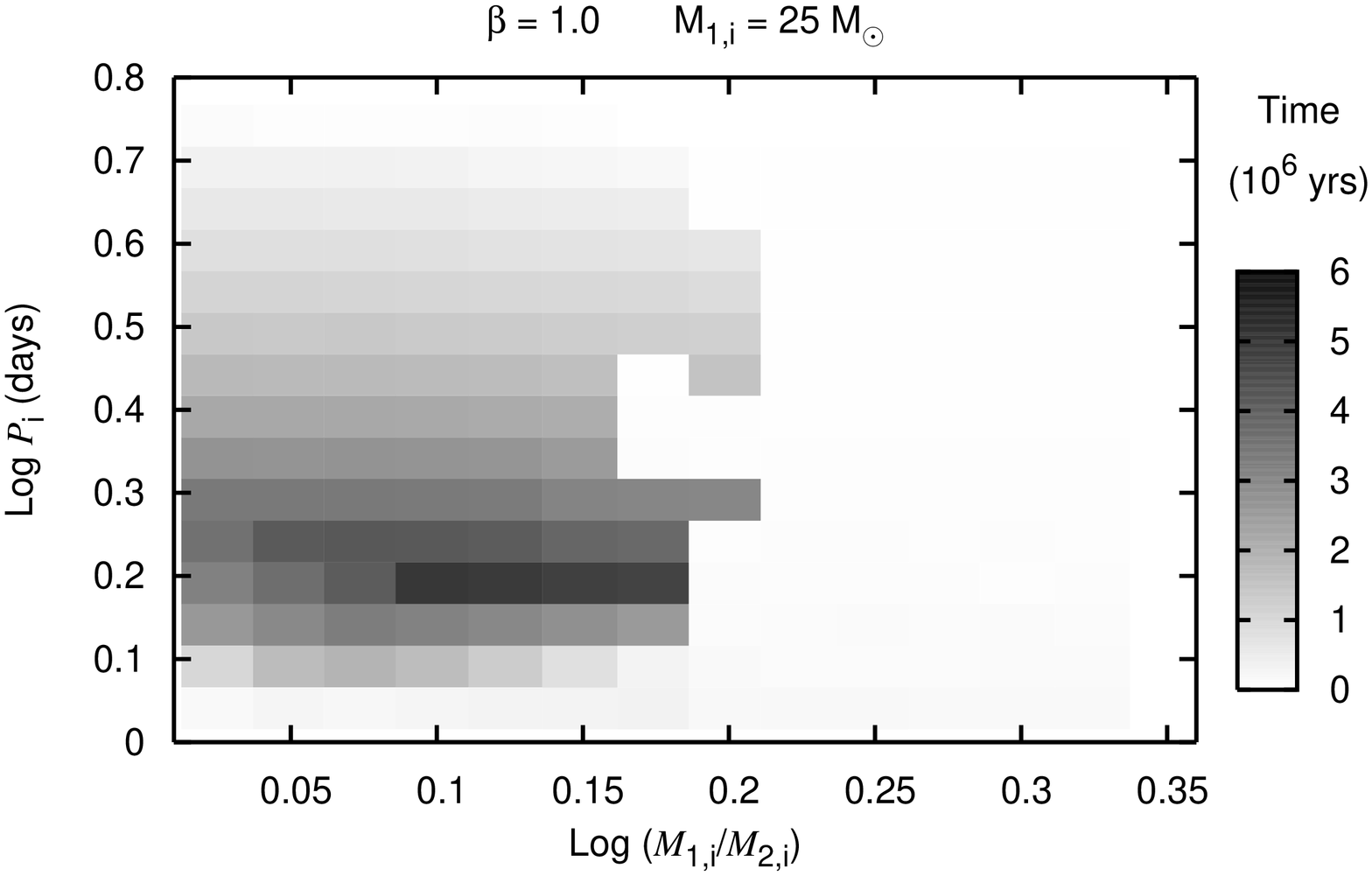}%

   \includegraphics[ bb=80 50 540 750,clip,angle=-90,width=0.5\textwidth]%
   {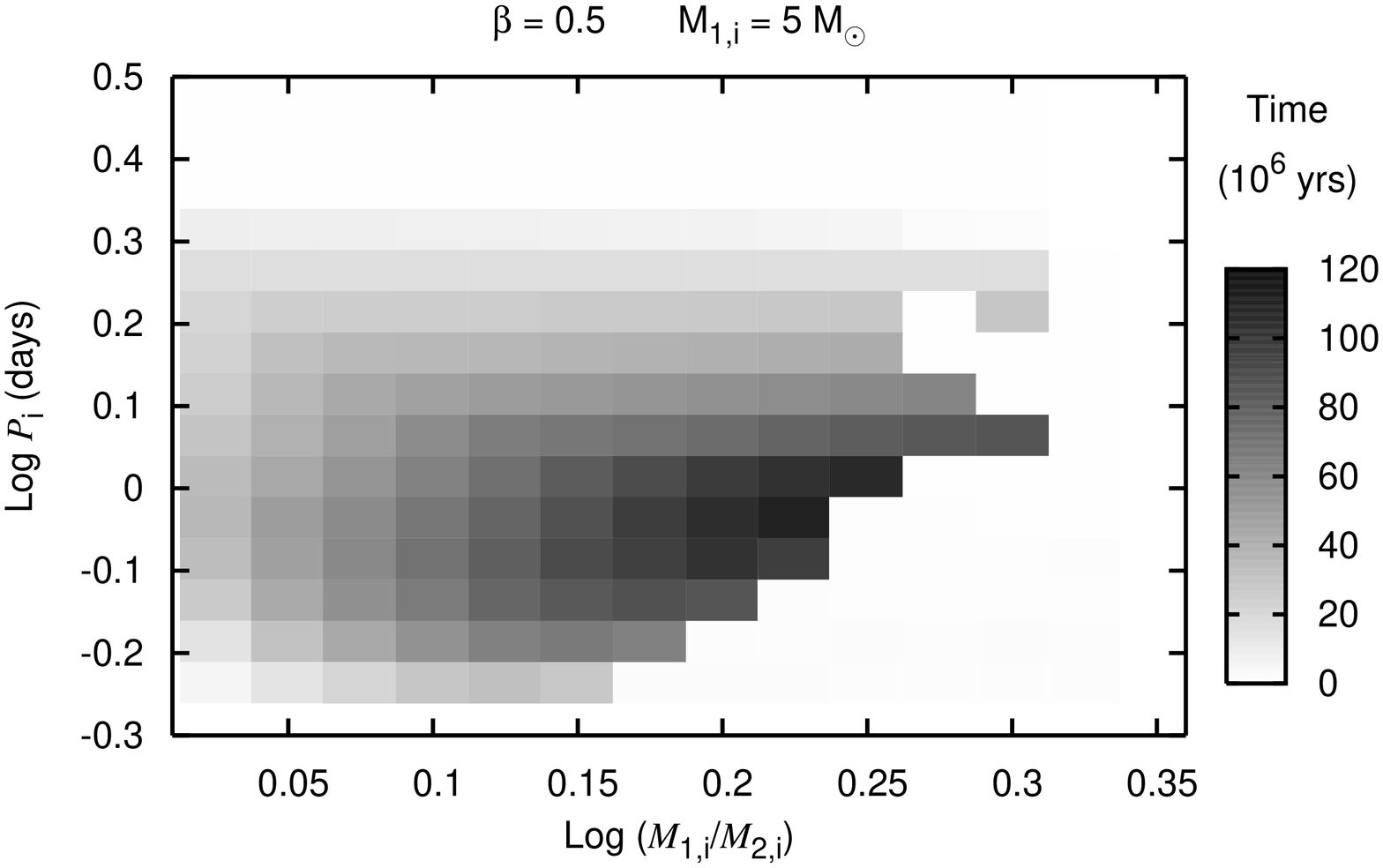}%
   \includegraphics[ bb=80 50 540 750,clip,angle=-90,width=0.5\textwidth]%
   {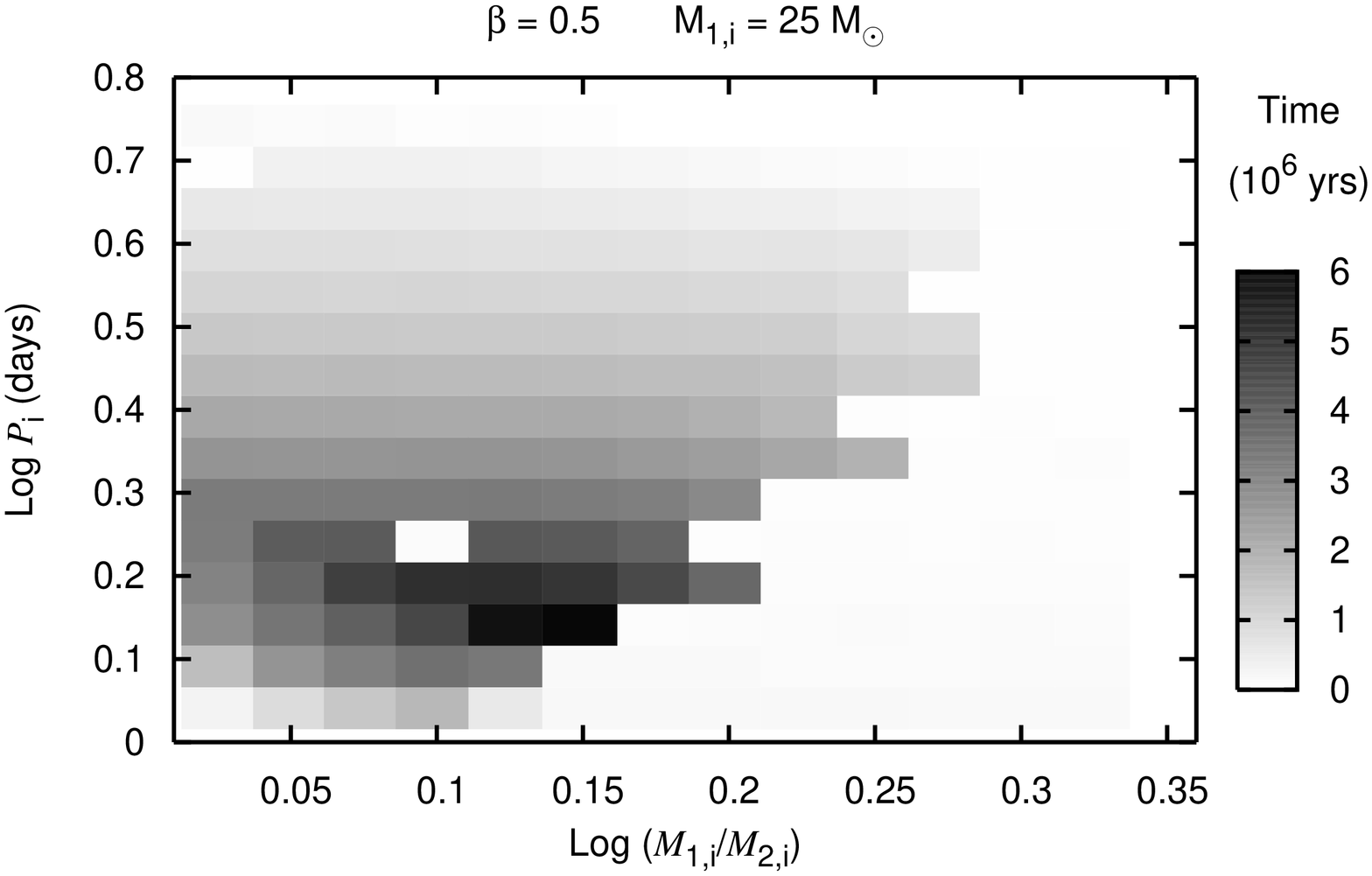}%

   \caption{ Time in years (color shading) spent in a semi-detached
     configuration for binary systems with different initial
     parameters. For a description of the difference between the four
     diagrams and their axis see Fig.~\ref{classification}.  The
     duration of the semi-detached phase for non-conservative models
     is longer than for conservative models and more systems avoid
     contact during rapid mass transfer, see also
     Fig.~\ref{classification}. This trend is seen for all different
     primary masses in our grid of binary models over the full range
     of $\beta$}
     \label{time} \end{figure*}

     \subsection{The influence of the mass transfer efficiency}
    
     An important effect of lowering the mass transfer efficiency is
     that more systems avoid contact during rapid mass transfer.  The
     critical mass ratio, which defines the border between type AN and
     type AR systems is up to $25\%$ larger if we assume $\beta = 0.5$
     than for the conservative tracks, see
     Fig~\ref{classification}. Since contact during rapid mass
     transfer potentially has catastrophic consequences for the
     further evolution of the binary, we expect that non-conservative
     mass transfer results in a larger number of binaries reaching a
     long-lived semi-detached phase.

     The mass transfer efficiency also strongly influences how much
     time a binary systems spends in a semi-detached phase, see
     Fig~\ref{time}. The accreting star will be less massive in
     non-conservative models and therefore it will evolve more
     slowly. The duration of the slow mass transfer phase can be twice
     as long for non-conservative models ($\beta = 0.5$) compared to
     the conservative models.

     The implication of non-conservative mass transfer that more
     binary systems become semi-detached and the duration of the
     semi-detached phase is longer, can in principle give a strong
     test of the mass transfer efficiency if the number of
     semi-detached systems is compared to the number of detached
     systems. However, a good understanding of the selection effects
     and knowledge of the initial mass-ratio distribution are critical
     in performing such a test.

\section{Observed systems} \label{sec:data}

 \begin{table*} \caption{\label{tab:data} Parameters of 50
 double-lined eclipsing binaries in the SMC. Details are given in
 Section~\ref{sec:data}.}
\centering
\begin{tabular}{ll|r|r|r|r|r|r|r}

 \hline\hline
 id & OGLE-id &
 \multicolumn{1}{c}{$ M_p (M_{\odot})$}&
 \multicolumn{1}{c}{$ M_p / M_s $}&
 \multicolumn{1}{c}{$ R_p (R_{\odot})$}&
 \multicolumn{1}{c}{$ R_s (R_{\odot})$}&
 \multicolumn{1}{c}{$ T_p (kK)$}& 
 \multicolumn{1}{c}{$  T_p/T_s$} & 
 \multicolumn{1}{c}{$ P$ (d)} \\
 \hline
 \multicolumn{2}{l}{}&\multicolumn{7}{l}{Detached systems}\\
 \hline 
01 &  04 056804 & $    13.0\pm     0.3$  & $   0.912\pm   0.019$  & $    4.11\pm     0.09$  & $    4.94\pm     0.08$  & $ 30.1\pm2.0$  & $   1.245\pm   0.017$  &    1.08987\\
02 &  04 103706 & $    17.5\pm     0.6$  & $   1.768\pm   0.020$  & $    5.36\pm     0.13$  & $    4.05\pm     0.12$  & $ 27.8\pm2.0$  & $   1.028\pm   0.017$  &    1.35585\\
03 &  04 163552 & $    13.3\pm     1.0$  & $   1.070\pm   0.052$  & $    5.31\pm     0.15$  & $    5.18\pm     0.15$  & $ 25.5\pm2.0$  & $   1.005\pm   0.007$  &    1.54581\\
04 &  05 038089 & $    17.1\pm     1.5$  & $   0.893\pm   0.058$  & $    6.27\pm     0.21$  & $    6.19\pm     0.26$  & $ 30.1\pm2.0$  & $   1.033\pm   0.036$  &    2.38946\\
05 &  05 095194 & $    20.3\pm     4.5$  & $   0.870\pm   0.149$  & $    8.06\pm     0.56$  & $    9.53\pm     0.66$  & $ 33.8\pm2.0$  & $   1.038\pm   0.002$  &    3.18742\\
06 &  05 140701 & $     6.9\pm     0.7$  & $   1.307\pm   0.064$  & $    7.88\pm     0.31$  & $    7.01\pm     0.31$  & $ 23.5\pm2.0$  & $   1.535\pm   0.025$  &    3.62544\\
07 &  05 180064 & $    10.7\pm     0.4$  & $   1.521\pm   0.019$  & $    5.57\pm     0.27$  & $    4.49\pm     0.29$  & $ 25.5\pm2.0$  & $   1.493\pm   0.107$  &    2.51491\\
08 &  05 255984 & $    11.6\pm     2.0$  & $   1.659\pm   0.133$  & $    4.16\pm     0.43$  & $    3.38\pm     0.34$  & $ 25.5\pm2.0$  & $   1.019\pm   0.038$  &    1.56384\\
09 &  05 305884 & $    17.6\pm     1.3$  & $   1.086\pm   0.059$  & $    7.76\pm     0.27$  & $    6.52\pm     0.37$  & $ 33.8\pm2.0$  & $   1.040\pm   0.007$  &    2.17648\\
10 &  05 311566 & $    12.9\pm     0.6$  & $   1.261\pm   0.034$  & $    4.38\pm     0.13$  & $    3.21\pm     0.19$  & $ 30.1\pm2.0$  & $   1.050\pm   0.024$  &    3.29139\\
11 &  06 011141 & $    15.1\pm     0.3$  & $   1.068\pm   0.016$  & $    5.05\pm     0.05$  & $    5.01\pm     0.06$  & $ 30.1\pm2.0$  & $   1.324\pm   0.027$  &    1.17737\\
12 &  06 180084 & $    16.2\pm     1.3$  & $   1.169\pm   0.073$  & $    5.85\pm     0.25$  & $    5.70\pm     0.25$  & $ 27.8\pm2.0$  & $   0.995\pm   0.011$  &    1.47523\\
13 &  06 215965 & $    16.0\pm     0.5$  & $   0.931\pm   0.027$  & $    9.91\pm     0.19$  & $   10.35\pm     0.19$  & $ 27.8\pm2.0$  & $   1.000\pm   0.037$  &    3.94304\\
14 &  06 221543 & $    11.9\pm     1.8$  & $   1.025\pm   0.103$  & $    5.27\pm     0.48$  & $    4.53\pm     0.52$  & $ 25.5\pm2.0$  & $   0.959\pm   0.009$  &    3.41678\\
15 &  07 120044 & $    12.5\pm     0.5$  & $   1.004\pm   0.028$  & $    4.78\pm     0.10$  & $    4.68\pm     0.12$  & $ 25.5\pm2.0$  & $   0.998\pm   0.016$  &    1.31081\\
16 &  07 255621 & $     9.3\pm     0.8$  & $   1.293\pm   0.049$  & $    5.04\pm     0.29$  & $    3.54\pm     0.56$  & $ 25.5\pm2.0$  & $   1.091\pm   0.043$  &    4.33063\\
17 &  08 087175 & $    12.0\pm     1.0$  & $   1.119\pm   0.073$  & $    4.48\pm     0.21$  & $    4.38\pm     0.22$  & $ 25.5\pm2.0$  & $   1.033\pm   0.021$  &    1.10226\\
18 &  08 104222 & $    13.1\pm     0.9$  & $   1.078\pm   0.055$  & $    5.25\pm     0.38$  & $    5.19\pm     0.38$  & $ 25.5\pm2.0$  & $   1.012\pm   0.033$  &    1.56384\\
19 &  10 037156 & $    19.5\pm     0.4$  & $   1.147\pm   0.014$  & $    7.19\pm     0.16$  & $    4.41\pm     0.46$  & $ 32.2\pm2.0$  & $   1.028\pm   0.012$  &    2.69834\\
20 &  10 110440 & $    10.8\pm     2.1$  & $   1.825\pm   0.131$  & $    3.97\pm     0.39$  & $    3.66\pm     0.34$  & $ 25.5\pm2.0$  & $   1.194\pm   0.027$  &    1.56384\\
21 &  11 057855 & $    12.4\pm     1.1$  & $   1.510\pm   0.061$  & $    5.22\pm     0.21$  & $    3.73\pm     0.24$  & $ 27.8\pm2.0$  & $   1.072\pm   0.039$  &    1.29695\\
 \hline
 \multicolumn{2}{l}{}&\multicolumn{7}{l}{Semi-detached systems and contact systems (47 and 49)}\\
 \hline
22 &  01 099121 & $    11.3\pm     0.8$  & $   1.704\pm   0.048$  & $    4.97\pm     0.19$  & $    6.66\pm     0.24$  & $ 27.8\pm2.0$  & $   1.758\pm   0.018$  &    2.45890\\
23 &  04 110409 & $    13.7\pm     0.8$  & $   1.544\pm   0.046$  & $    4.34\pm     0.20$  & $    8.36\pm     0.29$  & $ 25.5\pm2.0$  & $   1.660\pm   0.017$  &    2.97315\\
24 &  05 026631 & $    11.5\pm     0.6$  & $   1.021\pm   0.039$  & $    5.12\pm     0.11$  & $    5.65\pm     0.12$  & $ 25.5\pm2.0$  & $   1.489\pm   0.026$  &    1.41169\\
25 &  05 060548 & $    10.8\pm     0.4$  & $   1.238\pm   0.027$  & $    8.39\pm     0.14$  & $    9.63\pm     0.15$  & $ 30.1\pm2.0$  & $   1.722\pm   0.025$  &    3.63863\\
26 &  05 202153 & $    19.9\pm     1.1$  & $   1.592\pm   0.036$  & $    9.53\pm     0.26$  & $   12.84\pm     0.32$  & $ 32.2\pm2.0$  & $   1.374\pm   0.012$  &    4.60677\\
27 &  05 208049 & $    10.0\pm     0.2$  & $   2.093\pm   0.014$  & $    6.62\pm     0.08$  & $    6.82\pm     0.08$  & $ 25.5\pm2.0$  & $   1.791\pm   0.020$  &    3.02982\\
28 &  05 243188 & $    27.3\pm     1.5$  & $   1.466\pm   0.041$  & $    7.29\pm     0.19$  & $    7.89\pm     0.20$  & $ 35.5\pm2.0$  & $   1.104\pm   0.002$  &    1.87174\\
29 &  05 277080 & $    17.4\pm     0.9$  & $   1.540\pm   0.036$  & $    5.05\pm     0.14$  & $    6.82\pm     0.17$  & $ 25.5\pm2.0$  & $   1.607\pm   0.015$  &    1.93934\\
30 &  05 300549 & $    25.4\pm     1.0$  & $   1.458\pm   0.027$  & $    6.35\pm     0.11$  & $    6.15\pm     0.11$  & $ 30.1\pm2.0$  & $   1.879\pm   0.030$  &    1.33167\\
31 &  05 316725 & $    16.9\pm     1.5$  & $   1.916\pm   0.048$  & $    6.13\pm     0.25$  & $    7.66\pm     0.29$  & $ 33.8\pm2.0$  & $   1.072\pm   0.006$  &    2.55606\\
32 &  06 077224 & $    15.9\pm     1.0$  & $   1.218\pm   0.049$  & $    8.93\pm     0.26$  & $   11.52\pm     0.32$  & $ 32.2\pm2.0$  & $   1.253\pm   0.026$  &    3.82087\\
33 &  06 152981 & $    12.5\pm     0.4$  & $   1.533\pm   0.018$  & $    4.94\pm     0.08$  & $    6.26\pm     0.09$  & $ 25.5\pm2.0$  & $   1.268\pm   0.009$  &    2.00334\\
34 &  06 158118 & $    16.0\pm     0.7$  & $   2.029\pm   0.023$  & $    7.61\pm     0.15$  & $    7.38\pm     0.15$  & $ 29.2\pm2.0$  & $   1.589\pm   0.020$  &    2.57832\\
35 &  06 251047 & $     8.1\pm     0.2$  & $   1.473\pm   0.016$  & $    4.62\pm     0.11$  & $    6.40\pm     0.12$  & $ 23.5\pm2.0$  & $   1.738\pm   0.012$  &    2.51323\\
36 &  06 311225 & $    21.2\pm     0.4$  & $   1.773\pm   0.012$  & $    6.55\pm     0.06$  & $    6.69\pm     0.06$  & $ 30.1\pm2.0$  & $   1.282\pm   0.015$  &    1.84606\\
37 &  06 319960 & $    10.6\pm     0.8$  & $   1.569\pm   0.057$  & $    4.47\pm     0.18$  & $    9.38\pm     0.34$  & $ 25.5\pm2.0$  & $   1.567\pm   0.014$  &    4.05690\\
38 &  07 066175 & $    19.6\pm     1.8$  & $   1.703\pm   0.059$  & $    7.88\pm     0.31$  & $   10.38\pm     0.40$  & $ 32.2\pm2.0$  & $   1.271\pm   0.021$  &    3.62638\\
39 &  07 142073 & $    12.6\pm     1.2$  & $   2.000\pm   0.053$  & $    9.57\pm     0.35$  & $    7.75\pm     0.28$  & $ 30.1\pm2.0$  & $   1.435\pm   0.019$  &    3.18871\\
40 &  07 189660 & $    15.3\pm     1.2$  & $   1.498\pm   0.046$  & $    5.35\pm     0.16$  & $    5.98\pm     0.18$  & $ 25.5\pm2.0$  & $   1.510\pm   0.014$  &    1.66784\\
41 &  07 193779 & $    11.6\pm     1.0$  & $   1.966\pm   0.053$  & $    5.79\pm     0.21$  & $    4.93\pm     0.18$  & $ 25.5\pm2.0$  & $   1.486\pm   0.019$  &    1.67293\\
42 &  07 243913 & $    18.6\pm     1.1$  & $   1.768\pm   0.034$  & $    8.23\pm     0.20$  & $    8.28\pm     0.20$  & $ 32.2\pm2.0$  & $   1.242\pm   0.015$  &    2.63160\\
43 &  08 209964 & $    18.8\pm     0.9$  & $   1.300\pm   0.031$  & $    7.67\pm     0.20$  & $   10.69\pm     0.24$  & $ 36.3\pm2.0$  & $   1.371\pm   0.014$  &    3.31475\\
44 &  09 010098 & $    17.8\pm     1.8$  & $   1.304\pm   0.075$  & $    5.08\pm     0.22$  & $    5.06\pm     0.22$  & $ 33.8\pm2.0$  & $   1.062\pm   0.003$  &    1.11208\\
45 &  09 047454 & $    12.6\pm     2.3$  & $   1.362\pm   0.107$  & $    4.20\pm     0.26$  & $    5.58\pm     0.35$  & $ 27.8\pm2.0$  & $   1.644\pm   0.029$  &    1.57397\\
46 &  09 064498 & $     8.4\pm     0.7$  & $   3.092\pm   0.044$  & $    5.40\pm     0.24$  & $    5.10\pm     0.23$  & $ 25.5\pm2.0$  & $   1.496\pm   0.032$  &    2.63512\\
47 &  09 175323 & $    23.5\pm     1.6$  & $   1.451\pm   0.041$  & $   10.16\pm     0.26$  & $    8.48\pm     0.23$  & $ 39.3\pm2.0$  & $   1.021\pm   0.006$  &    2.20596\\
48 &  10 094559 & $    12.0\pm     1.0$  & $   1.208\pm   0.065$  & $    4.68\pm     0.21$  & $    6.19\pm     0.26$  & $ 30.1\pm2.0$  & $   1.262\pm   0.019$  &    1.75193\\
49 &  10 108086 & $    16.9\pm     1.2$  & $   1.183\pm   0.057$  & $    5.67\pm     0.18$  & $    5.32\pm     0.16$  & $ 33.8\pm2.0$  & $   1.130\pm   0.038$  &    0.88308\\
50 &  11 030116 & $    14.3\pm     1.9$  & $   1.868\pm   0.086$  & $    7.42\pm     0.42$  & $    8.05\pm     0.46$  & $ 27.8\pm2.0$  & $   1.368\pm   0.016$  &    2.95427\\
 \hline
 \multicolumn{2}{l}{}&\multicolumn{7}{l}{Alternative solutions for 
  system 1 (51) and system 6 (52)}\\
 \hline
51 &  04 056804 & $    16.8\pm     2.1$  & $   1.724\pm   0.088$  & $    4.46\pm     0.25$  & $    4.48\pm     0.25$  & $ 30.1\pm2.0$  & $   1.250\pm   0.009$  &    1.08987\\
52 &  05 140701 & $     7.0\pm     0.7$  & $   1.307\pm   0.064$  & $    6.00\pm     0.20$  & $    8.30\pm     0.30$  & $ 24.3\pm2.0$  & $   1.493\pm   0.007$  &    3.62544\\

\hline
\end{tabular}
 \end{table*}

    The stellar masses, radii and temperatures of the components of 50
    double-lined eclipsing OB-type binaries in the SMC were presented
    by \citetalias{2003MNRAS.339..157H} and
    \citetalias{2005MNRAS.tmp...20H}.  In this section we will
    summarize their work. The systems were initially taken from 1400
    eclipsing binaries detected in the Optical Gravitational Lensing
    Experiment (OGLE) survey. 169 systems were chosen for
    spectroscopic follow-up observations by selecting systems brighter
    than $B<16$, to allow a sufficient signal-to-noise ratio, and
    orbital period $P_{\rm orb}<5$, to ensure adequate phase coverage.
    About 100 systems had spectroscopic observations obtained near
    quadrature phases calculated from the adopted photometric
    ephemeris \citep{2004AcA....54....1W, 1998AcA....48..563U}. A
    spectral disentangling procedure was used to establish the orbital
    parameters directly, together with the average disentangled
    spectrum of each binary component. Orbital solutions were
    determined for a total of 50 systems, and were combined with
    analysis of the OGLE $I$-band light curves to yield complete
    astrophysical parameters. The 50 systems are spread over the 10
    OGLE fields which cover the central 2.4 square degree area of the
    Small Magellanic Cloud.

    Twenty-one systems are in a detached configuration, two are
    contact systems. The remaining 27 systems are semi-detached and
    believed to be undergoing their first phase of mass transfer. 
    In Table~\ref{tab:data} the observed stellar parameters are
    summarized. For two of the detached systems, 1 and 6, alternative
    semidetached solutions were found after the publication of
    \citetalias{2005MNRAS.tmp...20H}. These are listed as system 51
    and 52 respectively. 

    The light curves show no evidence for departures from circular
    orbits, except for the following detached systems: 10 ($e = 0.063
    \pm 0.002$), 14 ($e = 0.113 \pm 0.004$), 16 ($e = 0.190 \pm
    0.006$) and 19 ($e = 0.013 \pm 0.009$). We can therefore assume
    that the spin periods of the stars in each system are synchronized
    with the orbital period. The orbital periods are taken from
    \cite{2004AcA....54....1W} 
    and
    \cite{1998AcA....48..563U}. 
    The mass of the brightest component, the stellar radii, their
    errors and the temperature of the brightest component are taken
    directly from
    \citetalias{2003MNRAS.339..157H} and 
    \citetalias{2005MNRAS.tmp...20H}. 
    We derived the mass ratio $q = M_\mathrm{p}/M_\mathrm{s}$ (where
    $M_\mathrm{p}$ is the mass of the primary, i.e. the star eclipsed
    during the deepest eclipse which is usually the most luminous
    component, and $M_\mathrm{s}$ is the mass of the companion) and
    its error directly from the radial velocity semi-amplitudes
    $K_{\mathrm s}$ and $K_{\mathrm p}$ and their errors, given in the
    same papers ($q = K_\mathrm{s}/K_\mathrm{p}$).  The three authors
    independently classified the spectral type of the primary
    component to determine the effective temperature. They estimate
    the error on the temperature to be in the order of the difference
    in temperature between two spectral subtypes\footnote{The errors
    in the secondary temperatures quoted in
    \citetalias{2003MNRAS.339..157H} and
    \citetalias{2005MNRAS.tmp...20H} only reflect the uncertainty in
    the flux ratio and not the uncertainty induced by the error on the
    temperature of the primary star.} An error of 1000\,K is given in
    \citetalias{2003MNRAS.339..157H} while an error of 1500\,K is
    given in \citetalias{2005MNRAS.tmp...20H}. However, the average
    temperature difference between spectral subtypes in the range
    O6-B2 is $\sigma_{T_{\rm eff}} \approx 1800$\,K.
%
%
    In addition there are significant uncertainties in the conversion
    of spectral type to temperature at the metallicity of the
    SMC. Therefore we decided to adopt an error of $\Delta T_{\rm eff}
    = 2000$\,K to allow for some systematic uncertainty.
    The temperature ratio and its error were determined by
    \citetalias{2005MNRAS.tmp...20H} and
    \citetalias{2003MNRAS.339..157H} from the $I$-band flux
    ratio. Unfortunately this band is not very sensitive to
    temperature for OB stars as it covers the Raleigh-Jeans tail of
    the spectrum. Nevertheless, the formal errors on the temperature
    ratio as quoted in Table~\ref{tab:data} are often very
    small. Photometric observations in additional bands or higher
    resolution spectra would improve the determination of the primary
    temperature, the temperature ratio and their errors.



     \section{Fit method} \label{sec:method}

     For each system in the observed sample we determine the best
     fitting binary evolution track using a least-squares fit.  Our
     four fitting parameters are the three initial binary parameters that
     determine the binary evolution track -- the initial primary
     mass~$M_{1,i}$, the initial mass ratio~$q_{i}$ and
     the initial orbital period~$P_{i}$ -- and the age~$t$ of the best
     fitting model on that track.

     There is freedom in the set of observed parameters we can use for
     fitting. To avoid propagation of errors, we preferentially use
     observables that are determined directly from the light curve and
     the spectra.
     For a proper \chisq-test the observables should be independent.
     In case of the semi-detached systems the radius of the Roche-lobe
     filling star is determined from the light curve, but it is not
     independent of the mass ratio and the orbital period. Therefore
     we cannot use all three observables at the same time. In detached
     binaries the orbital period does not change significantly during
     the evolution before the onset of mass transfer.
     Fitting the period within the tiny observed error is pointless
     given the much larger uncertainties in the other observables. We
     therefore choose a definition of $\chisq$ which does not include
     the orbital period. We do demand that a fit solution matches the
     period within the accuracy of our grid.

     We use the following set of six observed parameters, listed in
     Table~\ref{tab:data}: the mass of the most luminous star $\log
     M_{\mathrm{p}}$, the mass ratio $\log q$, the radius of
     the most luminous star $\log R_\mathrm{p}$, the radius of the
     companion star $\log R_\mathrm{s}$, the temperature of the most
     luminous star $\log T_\mathrm{p}$ and the temperature ratio $\log
     T_\mathrm{p}/T_\mathrm{s}$.
     With six independent observables and four model parameters we
     have two degrees of freedom ($\nu = 2$) when we fit semi-detached
     systems. In the case of detached systems one of our model
     parameters, the initial orbital period, is degenerate because
     stars evolve essentially as if they were single until the onset
     of mass transfer. In practice we therefore have three model
     parameters which yields $\nu = 3$ for the detached systems.

     The cube of models only has discrete values for the model
     parameters $M_{1,i}$, $q_i$ and $P_i$. Ideally we want to vary
     the model parameters continuously, but our grid spacing is not
     fine enough for some of the more accurately determined
     observables.
     Following \cite{2001ApJ...552..664N} we therefore introduce `theoretical
     errors'~$\sigma_{\rm th}$ to account for the discreteness of the
     model grid. For the total error~$\sigma_j$ on the $j$-th
     observed parameter we add the observational error~$\sigma_{\rm
     obs}$ and the model error~$\sigma_{\rm th}$ in quadrature,
     \begin{equation}
     \sigma_{ j}^2 = \sigma_{\rm obs}^2 + \sigma_{\rm th}^2.
     \end{equation}
     This leads to the following definition of \chisq,
     \begin{equation}    
        \chi^2\,(M_{1,i},\, q_{i},\, P_{i},
        \,t\,) =   
	 \frac{1}{\nu}\sum_{j} \frac{[\, X_{\mathrm{obs},j} -
       X_{\mathrm{th},j}\,(\,M_{1,i},\, q_{i},\,
       P_{i},\,t)\,]^2} {\sigma_j^2}. 
     \end{equation}
     The sum is taken over all observed parameters
     $X_{\mathrm{obs},j}$ and the corresponding values in our models
     $X_{\mathrm{th},j}$.
     We determine the best fitting evolution track
     $(M^*_{1,i},\, q^*_{i},\, P^*_{i})$ and
     the best fitting age ($t^*$) by minimizing $\chi^2$ with a
     systematic grid search of the fitting parameters. We define
     \begin{equation}
     \chi^2_{\min} = \chi^2\,(M^*_{1,i},\, q^*_{i},\, P^*_{i}, \,t^*). 
     \end{equation}
     We only allow fit solution in which the observed period is
     matched within $|\log P_{\mathrm{obs}} - \log P_{\mathrm{th}}| <
     0.05$.  
     For the theoretical errors we take half the initial grid spacing,
     i.e.  for $\log M_{1, \mathrm{i}}$ and $\log P_\mathrm{i}$,
     $\,\sigma_\mathrm{th} = 0.025$ and for $\log q_\mathrm{i}$,
     $\sigma_\mathrm{th} = 0.0125$. We estimate the initial grid
     spacing for the radii and the effective temperature for each
     system individually from the two initial models with primary
     masses closest to the observed mass of the primary.

     The initial grid spacing for the logarithm of the temperature
     ratio scales approximately linearly with the spacing of the
     logarithm of the mass ratio, with no dependence on the orbital
     period and a weak dependence on the primary mass. For a typical
     primary mass of $10\,\Msun$ we fit a straight line through $\log
     T_1/T_2$ as function of $\log M_1/M_2$, which results in an
     estimate for the theoretical error on the logarithm of the
     temperature ratio of $0.0065$.

     We realize that the introduction of theoretical errors in the
     definition of \chisq\ compromises a rigorous statistical
     interpretation of the obtained \chirmin\ values. Furthermore,
     using the initial grid spacing is not ideal, especially for
     post-mass transfer systems, because the real grid spacing will
     vary during the evolution.
     Therefore we compared different definitions of \chisq\ and of
     $\sigma_\mathrm{th}$. We tested the effect of using the orbital
     period as an observable instead of the radius of the Roche-lobe
     filling star, and we varied the magnitude of $\sigma_\mathrm{th}$
     between zero and the full initial grid spacing. The latter
     definition was used by \cite{2001ApJ...552..664N}. The magnitude
     $\chi_{\min}^2$ for a certain system depends directly the chosen
     definition; in particular, overestimating $\sigma_\mathrm{th}$
     leads to artificially low values of \chisq.
     On the other hand, $\sigma_\mathrm{th} = 0$ can yield artificially
     high values of \chisq\ due to incomplete sampling of the model
     parameters.
     Therefore, although we attempted to choose a reasonable definition,
     $\chi^2_{\min}$ can not be used as an objective criterion to
     determine whether a fit is good or bad, only as a relative
     estimate.
     Nevertheless, we found that the $\chi_{\min}^2$ values obtained
     with different methods are well correlated; in other words, the
     relative quality of a fit is independent of the chosen fit
     method.  The best fitting parameters are not sensitive to the
     chosen definition, i.e. do not vary more than one step in our
     grid of models except for a few systems for which no good fit
     could be obtained at all.

     For the detached systems we only take detached solutions,
     i.e. the radii of both stars should be smaller than 98$\% $ of
     their Roche-lobe radii. This definition is consistent with the
     definition used for the observed systems
     \citepalias{2005MNRAS.tmp...20H}.  For the semi-detached systems
     we only accept solutions during which the evolution is slow, on a
     nuclear timescale.  Case A mass transfer starts with a phase of
     rapid mass transfer on the thermal timescale, $\tau_{\rm
     KH}$. After the mass ratio reverses, the orbit widens and a phase
     of slow mass transfer follows on the nuclear timescale,
     $\tau_{\rm nuc}$, of the mass-donating star. As $\tau_{\rm KH}
     \approx 0.01 \tau_{\rm nuc}$ one would expect only 1 out of 100
     systems to be in the rapid mass transfer phase, if no other
     biases play a role.  We therefore assume that all observed
     semi-detached systems are currently in the slow mass transfer
     phase, and we only fit models in which:

     \begin{itemize}
     \item{the donor star is filling its Roche lobe, such that the
         radius of the initially more massive star is larger than $98\%$
         of the Roche-lobe,}

     \item{the stars are in thermal equilibrium, i.e. the thermal
     luminosity of the mass donor is less than $1\%$ of its
     total luminosity, }
 
     \item{the mass transfer rate is smaller than $1\%$ of the typical
     thermal-timescale mass transfer rate of the donor,
     $M_\mathrm{D}/\tau_{1, \rm{KH}}$, and}

     \item{the donor star has become less massive than the
     accreting star, $M_\mathrm{D} <M_\mathrm{A}$.}

     \end{itemize}

     To determine confidence limits on the fitted model parameters one
     has to determine for which values of the parameters the condition
     $\chi^2 (M_i, q_{i}, P_{i}, t) < \chi^2_{\min} + c$ is satisfied,
     varying each parameter in turn, where $c=1/\nu$ to determine
     $1\sigma$ boundaries for a proper $\chi^2$-test.  To compensate
     for any systematic over- or underestimation of the observational
     and theoretical errors we take $c = \chi^2_{\min}/\nu$. As our
     model grid is discrete we determine the upper and lower limit on
     each parameter by the maximum and minimum values for which the
     condition is satisfied, plus or minus half of the grid spacing
     for that parameter.
     To estimate the range of $\beta$ values which fit well for a certain
     model, we first determine the best fitting model for each of the
     four assumptions of $\beta$, and the corresponding \chisqmin.
     To obtain $\chi^2_{\rm min}$ as a function of $\beta$ on the full
     range $0.25 \le \beta \le 1.0$ we interpolate linearly between
     the four values.  Let $\beta^*$ denote the best-fitting value of
     $\beta$. To estimate the confidence limits on $\beta$ we
     determine for which values $\chi_{\min}^2 (\beta) <
     \chi_{\min}^2(\beta^*) +\chi_{\min}^2(\beta^*) /\nu$.

\section{Results} \label{sec:results}

    The best fitting model parameters for the detached systems are
    given in Table~\ref{tab:det1} and for the semi-detached systems in
    Table~\ref{tab:sem1}.

 \subsection{Detached Systems }

   \begin{figure*} \centering
   \includegraphics[width=0.85\textwidth]{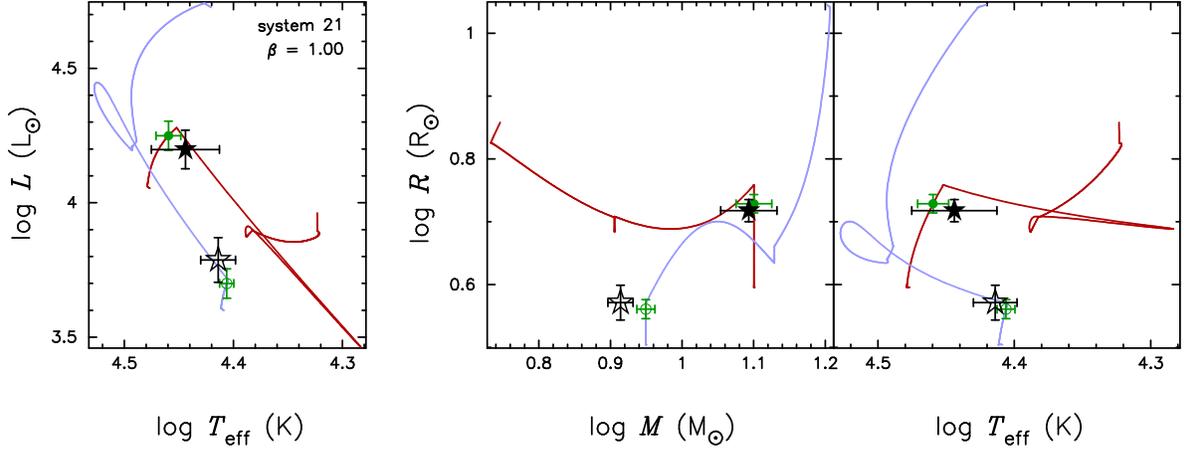}
   \caption{ Example of a fit for a detached system. The star-shaped
   symbols indicate the observed parameters for system 21, the circles
   (green in on-line version) indicate the best-fitting model. The
   conservative evolution track of the best-fitting model is plotted
   in dark gray (red) for the star indicated as the primary, see
   Table~\ref{tab:data}, and in lighter gray (blue) for the
   companion. On the left a Hertzsprung-Russel diagram is given for
   reference, although we do not use the luminosity as a fit
   parameter. The error bars on the filled star symbols indicate the
   error on the primary mass, temperature and radius. In contrast the
   error bars on the open star symbols indicate the error on the mass
   ratio, the temperature ratio and the secondary radius. The error
   bars on the circles indicate the adopted theoretical errors on the
   same quantities (see Sect.~\ref{sec:method}), which correspond to
   half the initial grid spacing of the models.} \label{fig:detfitex}
   \end{figure*}

 \begin{table*} \caption{\label{tab:det1} Fit-results for the detached
 systems: The best fitting initial primary mass $M_{1,i}$, initial
 mass ratio $q_{\mathrm i} = M_{1,{\mathrm i}}/M_{2,{\mathrm i}}$ and
 initial orbital period $P_{\mathrm i}$ are given together with
 estimates of the 1$\sigma$ confidence limits (for details see
 the last paragraph of Section~\ref{sec:method}). The last two
 columns give the best fitting age on the best fitting evolution track
 and, for systems with $\chi^2 <10$, the evolutionary subtype
 assuming conservative mass transfer, see Section~\ref{sec:subtypes}.
 The notation AR $\rightarrow$ ... indicates that the
 best-fitting model experiences a temporary contact phase during
 rapid mass transfer after which the evolution continues as
 indicated.}
\centering
\begin{tabular}{lr|ccc|ccc|ccc|cc}
 \hline\hline
 id & $\chi^2_{\rm min}$ & 
 \multicolumn{3}{|c|}{$M_{1,i}$ ($M_{\odot}$) }&
 \multicolumn{3}{|c|}{$q_{\mathrm i} $}&
 \multicolumn{3}{|c|}{$P_{\mathrm i}$ (days)} & 
   $t$ (Myrs) & case
   \\
 \hline 

 1 &  43.0 & 	14.1 & -4.7 & +4.7   &     1.05 & -0.05 & +0.24   &	 1.12 & -0.21 & +0.18	&  3.1& -  \\
 2 &  21.8 & 	15.8 & -2.5 & +5.3   &     1.58 & -0.21 & +0.15   &	 1.31 & -0.16 & +0.32	&  6.8& -  \\
 3 &   1.2 & 	12.6 & -0.7 & +0.7   &     1.06 & -0.06 & +0.03   &	 1.51 & -0.08 & +0.29	& 10.6& AS  \\
 4 &   1.7 & 	17.8 & -2.8 & +1.0   &     1.06 & -0.06 & +0.03   &	 2.18 & -0.12 & +0.60	&  6.2& AN  \\
 5 &   1.0 & 	20.0 & -1.2 & +3.7   &     1.13 & -0.13 & +0.02   &	 3.21 & -0.39 & +0.60	&  8.1& AN  \\
 6 & 114.0 & 	10.0 & -3.3 & +27.6  &     1.06 & -0.06 & +1.31   &	 3.47 & -0.42 & +1.20	& 23.7& -  \\
 7 &  10.7 & 	11.2 & -1.8 & +3.8   &     1.49 & -0.12 & +0.14   &	 2.57 & -0.41 & +0.59	& 14.8& -  \\
 8 &   4.5 & 	10.0 & -1.6 & +3.3   &     1.33 & -0.11 & +0.21   &	 1.55 & -0.18 & +0.40	& 15.3& AN  \\
 9 &   1.2 & 	20.0 & -3.2 & +1.1   &     1.13 & -0.04 & +0.02   &	 2.27 & -0.36 & +0.47	&  6.3& AN  \\
10 &   2.3 & 	12.6 & -0.7 & +0.7   &     1.26 & -0.11 & +0.04   &	 3.02 & -0.17 & +0.78	&  3.9& AN  \\
11 &  30.9 & 	15.8 & -3.9 & +5.3   &     1.18 & -0.18 & +0.19   &	 1.31 & -0.30 & +0.20	&  4.3& -  \\
12 &   2.0 & 	14.1 & -0.8 & +2.7   &     1.05 & -0.05 & +0.10   &	 1.58 & -0.25 & +0.27	&  9.4& AS  \\
13 &   3.6 & 	17.8 & -2.8 & +1.0   &     1.06 & -0.06 & +0.03   &	 4.34 & -0.84 & +0.53	& 10.0& B  \\
14 &   2.1 & 	11.2 & -1.8 & +2.1   &     1.06 & -0.06 & +0.10   &	 3.63 & -0.71 & +0.63	& 12.7& B  \\
15 &   2.0 & 	12.6 & -2.0 & +0.7   &     1.06 & -0.06 & +0.03   &	 1.20 & -0.07 & +0.42	&  8.0& AS  \\
16 &   0.3 & 	10.0 & -0.6 & +0.6   &     1.27 & -0.04 & +0.11   &	 3.89 & -0.22 & +0.48	& 17.1& B  \\
17 &   0.9 & 	11.2 & -0.6 & +0.7   &     1.06 & -0.06 & +0.10   &	 1.15 & -0.18 & +0.14	&  9.7& AS  \\
18 &   0.9 & 	12.6 & -0.7 & +0.7   &     1.06 & -0.06 & +0.03   &	 1.51 & -0.08 & +0.39	& 10.7& AS  \\
19 &   3.1 & 	17.8 & -1.0 & +3.3   &     1.19 & -0.10 & +0.03   &	 2.44 & -0.14 & +0.77	&  7.2& AN  \\
20 &   3.7 & 	11.2 & -1.8 & +2.1   &     1.58 & -0.12 & +0.15   &	 1.45 & -0.08 & +0.51	&  9.6& AR  \\
21 &   1.4 & 	12.6 & -0.7 & +0.7   &     1.42 & -0.04 & +0.12   &	 1.35 & -0.21 & +0.16	& 10.0& AR $\rightarrow$ AS  \\
\hline
\end{tabular}
 \end{table*}

    For twelve of the twenty-one detached systems we find good or
    reasonable fit solutions with a $\chirmin \leq 3$, see
    Table~\ref{tab:det1}. Inspection by eye confirms that the fits for
    these systems are indeed reasonably good, see
    Figure~\ref{fig:detfitex} for an example.  An additional set of
    four systems have $3 < \chirmin < 5$, which we still regard as
    acceptable fits on the basis of eye inspection. For five systems
    no acceptable fit could be obtained ($\chirmin > 10$). In
    Sect.~\ref{sec:detfits:indiv} we comment on the individual fit
    solutions of these systems. Figures similar to
    Fig.~\ref{fig:detfitex} for all systems are given in the on-line
    Appendix.  Here we first discuss the overall properties of the
    results as presented in Table~\ref{tab:det1}.

    If we eliminate system 1 because of the existence of a well
    fitting alternative semi-detached solution and system 7 because of
    bad quality of the data (see Sect.~\ref{sec:detfits:indiv}) we are
    left with good or acceptable fits for 16 out of 19 (84\%) of the
    detached systems. We may thus conclude that on the whole our
    models are able to reproduce the detached binaries in the sample
    quite well.

    In some cases the best fitting solution lies on the border of our
    grid. This happens often for systems with a mass ratio near 1 as our
    grid only holds models with initial mass ratios down to 1.059. In
    two cases the observed orbital period is too high for our
    grid. This is the case for system 6, for which no good fit could
    be obtained for other reasons than the period alone, and for
    system 16 which nevertheless gives a good fit. The boundaries for
    the masses, the lower boundary for the period and the upper
    boundary for the mass ratio are never reached.

    Out of the 16 systems for which we obtained good or acceptable
    fits three systems are so wide that our conservative models
    predict that mass transfer will not start before the most massive
    star enters the Hertzsprung gap, six systems will start mass
    transfer on the main sequence but avoid contact, at least until
    the primary reaches the Hertzsprung gap, and six systems will come
    into contact during slow mass transfer, see the last column of
    Table~\ref{tab:det1}. One system will reach contact already during
    the rapid mass transfer phase.

      \subsubsection{Comments on individual fits} \label{sec:detfits:indiv}

    We first give brief notes on the systems that fit well and then
    more detailed notes on the systems for which no good fits could be
    obtained.  

    Several systems have lower observed temperatures than our models
    predict: 3, 4, 15 and 18 and to a lesser extent 12 and 17. Good
    fits with small values of \chisqmin\ were nevertheless obtained,
    because of the large adopted error on the effective temperature of
    the primary.  However, the discrepancy is systematic and the
    opposite situation does not occur. In addition, the components of
    system 12 have very equal temperatures, while the masses are
    different.

    Both stars of system 10 are very close to the zero-age main
    sequence. \citetalias{2005MNRAS.tmp...20H} note that the orbit of
    this system is slightly eccentric, see Section~\ref{sec:data},
    which is consistent with the young age and the fact that both
    stars are well inside their Roche lobes.  Also systems 14 and 16
    have significant eccentricities. Although the stars in these
    binaries are more evolved than in system 10, both systems have
    wide orbits and are well detached.

    For  four systems the fits we obtained were only marginally
    acceptable, $3 \leq  \chirmin \leq 5 $:  
    \begin{itemize}
        \item[8]{For this system we cannot fit the small temperature
        ratio well at the same time as the moderately large mass
        ratio, resulting in $\chirmin = 4.5 $.}
        \item[13]{In this wide and evolved binary, the equal
        temperatures and the large and almost equal radii of the stars
        suggest that the masses must also be nearly equal. The best
        fit ($\chirmin = 3.6$) is obtained with the smallest mass
        ratio available in our grid, $q = 1.06$. We expect that if a
        model with equal masses were available, we would have found a
        very good fit for this system.}
        \item[19]{The stars in this system have fairly equal masses,
        while the radii are very different. The less massive component
        has the size of a zero-age main-sequence star.
        \citetalias{2005MNRAS.tmp...20H} note that the system may be
        slightly eccentric. One possibility is that the larger star is
        still a pre-main sequence star, but this is also the more
        massive star which should have contracted towards the zero-age
        main sequence first. Another speculative possibility is that
        the stars were not born at the same time and that they became
        part of a binary system after the larger star started burning
        hydrogen. However, given the moderate value of $\chirmin=3.1$
        such exotic scenarios are probably not required.}
        \item[20]{ This system has a large secondary radius compared
        to its mass (the radii are nearly equal whereas the masses are
        not) which cannot be explained by our detached models. As a
        result the overall fit is not very good ($\chirmin = 3.7$).
        An alternative fit solution was found for this system during a
        post-mass transfer phase.  After a rapid phase of mass
        transfer the donor star briefly detaches when it restores its
        thermal equilibrium. This alternative fit was not
        significantly better and the probability of observing a system
        in the short post-mass transfer phase is small compared to the
        pre-mass transfer solution.}
    \end{itemize} 
    \noindent
    For five systems no good fit could be obtained:
    \begin{itemize}
        \item[1]{ This system has nearly equal masses, which is not
        consistent with the large temperature ratio and the very
        different radii. The brightest star is the less massive
        component of the system which fits well to the evolution track
        of a star that has just arrived on the main sequence. However,
        the less bright and more massive component is too big and too
        cool compared to our models.

        For this system an alternative semi-detached light curve
        solution was found. This alternative solution can be fitted
        well against the semi-detached models with a reduced $\chirmin$
        of 1.6. We propose to reject the detached solution published
        in \citetalias{2003MNRAS.339..157H} and
        \citetalias{2005MNRAS.tmp...20H} and adopt the new
        semi-detached solution indicated as system 51, see
        Sect.~\ref{sec:ressemi}.}
        \item[2]{ The mass ratio of the components in this system is
        large ($M_{\mathrm p}/M_{\mathrm s}=1.8$) while the $I$-band
        flux ratio indicates that the effective temperatures of both
        stars are very similar. The less massive component is
        consistent with a single star model, while the more massive
        component is too cool and too small for its mass compared to
        the models.}
        \item[6]{ The large radii of both components in this systems
        suggest that the stars are close to hydrogen exhaustion in the
        core, at the end of the main sequence or even at the beginning
        of the Hertzsprung gap. This implies that, if no mass transfer
        has taken place yet, the masses of the stars must be nearly
        equal, approximately $6\,\Msun$. However, one would expect
        approximately equal effective temperatures for both stars,
        which is not the case.  An alternative semi-detached solution
        is available for this system, but no acceptable fit against
        semi-detached models was obtained either (see
        Section~\ref{sec:ressemi}). }
        \item[7]{ The more massive component of this system can be
        fitted to the evolution track of a $11\,\Msun$ star, but the
        less massive component is cooler and bigger than one would
        expect for its mass. However, the light curve has a very small
        amplitude (0.06 mag) and is poorly defined due to large
        scatter (H05). Therefore the light curve solution is not very
        reliable and this system probably should not have been
        included in the sample.}
        \item[11]{ The equal masses of this system together with the
         extreme temperature ratio make this system difficult to
         fit. The more massive component is consistent with a single
         star of about 15\,\Msun, but the less massive star is too cool
         compared to the models.}  
\end{itemize}
\noindent

 \subsection{Semi-detached Systems \label{sec:ressemi}}

   \begin{figure*} \centering
   \includegraphics[width=0.85\textwidth]{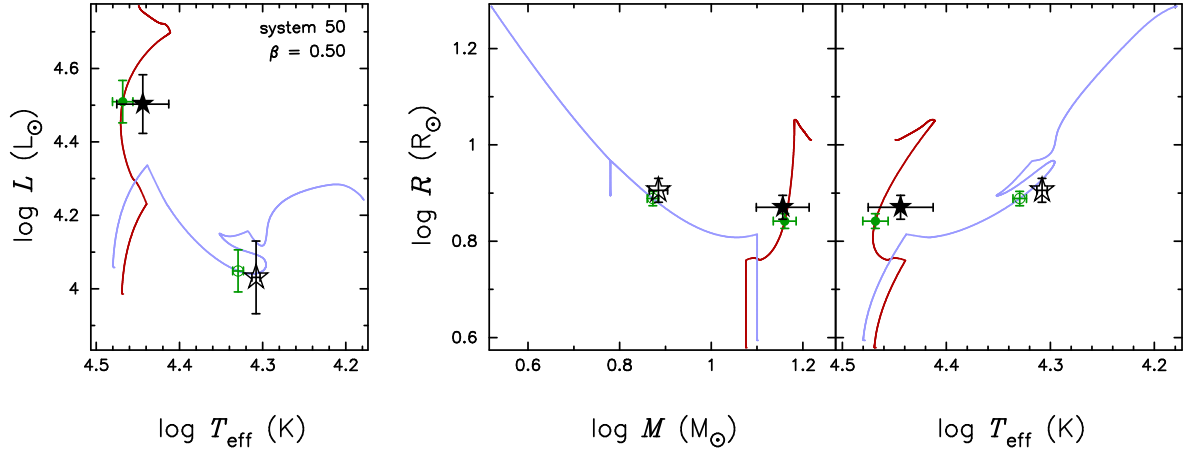}%
   \caption{ Example of a fit for a semi-detached system.  The
   observed parameters for system 50 are plotted together with the
   best fitting model assuming a mass transfer efficiency of 0.5. See
   Fig~\ref{fig:detfitex} for description of the symbols. According to
   the best fitting model this system consisted initially of two stars
   with masses of about 13 and 12 \Msun\ and an orbital period of 1.7
   days. The most massive star, which lost about 5 \Msun\ by Roche-lobe
   overflow, is now close to the end of its main sequence. This system
   will come into contact when the mass donor enters the
   Hertzsprung gap.\label{fig:semfitex} 
  } 
   \end{figure*}

 \begin{table*} \caption{\label{tab:sem1} Fit-results for semi-detached
 and contact systems. See the caption of Table~\ref{tab:det1} for a description. }
 \centering
\begin{tabular}{lrr|ccc|ccc|ccc|cc}
 \hline\hline
 id & $\chi^2_{\rm min}$ & $\beta$ & 
 \multicolumn{3}{|c|}{$M_{1,i}$ ($M_{\odot}$) }&
 \multicolumn{3}{|c|}{$q_{i}$} &
 \multicolumn{3}{|c|}{$P_{i}$ (days)} & 
   $t$ (Myrs) & case 
   \\
 \hline

22 &  16.9 &  1.00 &      7.9 & -1.3 & +1.5   &     1.06 & -0.06 & +0.16   &	 1.99 & -0.32 & +0.22	& 33.9& -  \\
23 &  17.4 &  0.50 &     10.0 & -1.6 & +1.9   &     1.19 & -0.19 & +0.26   &	 2.46 & -0.48 & +0.27	& 23.2& -  \\
24 &  50.0 &  0.75 &     10.0 & -3.3 & +5.0   &     1.12 & -0.12 & +0.33   &	 1.74 & -0.58 & +0.43	& 20.6& -  \\
25 &  97.5 &  0.25 &     11.2 & -2.8 & +7.6   &     1.06 & -0.06 & +0.24   &	 2.89 & -1.09 & +1.04	& 19.7& -  \\
26 &   4.1 &  0.50 &     17.8 & -1.0 & +3.3   &     1.06 & -0.06 & +0.03   &	 3.45 & -0.42 & +0.36	& 10.2& AN \\
27 &  26.4 &  0.50 &      7.1 & -0.4 & +2.4   &     1.05 & -0.05 & +0.10   &	 1.34 & -0.16 & +0.87	& 50.1& -  \\
28 &   1.8 &  0.75 &     25.1 & -1.4 & +4.8   &     1.33 & -0.17 & +0.13   &	 1.75 & -0.28 & +0.18	&  5.6& AN \\
29 &  31.1 &  1.00 &     10.0 & -1.6 & +3.3   &     1.12 & -0.12 & +0.25   &	 1.95 & -0.39 & +0.59	& 21.6& -  \\
30 & 141.5 &  1.00 &     11.2 & -3.7 & +7.6   &     1.06 & -0.06 & +0.57   &	 2.04 & -0.96 & +1.08	& 18.2& -  \\
31 &   1.8 &  0.50 &     20.0 & -1.2 & +1.1   &     1.69 & -0.06 & +0.03   &	 2.02 & -0.11 & +0.12	& 10.7& AR $\rightarrow$ AN  \\
32 &   5.7 &  0.25 &     17.8 & -1.0 & +3.3   &     1.06 & -0.06 & +0.10   &	 3.07 & -0.49 & +0.32	&  9.9& AN \\
33 &   1.4 &  0.50 &     12.6 & -0.7 & +0.7   &     1.26 & -0.04 & +0.04   &	 1.51 & -0.08 & +0.09	& 14.7& AN \\
34 &  15.3 &  1.00 &     12.6 & -2.0 & +2.4   &     1.06 & -0.06 & +0.10   &	 1.91 & -0.48 & +0.36	& 15.8& -  \\
35 &  41.4 &  0.50 &      7.1 & -1.8 & +2.4   &     1.12 & -0.12 & +0.18   &	 1.90 & -0.38 & +0.47	& 42.7& -  \\
36 &   3.7 &  1.00 &     15.8 & -0.8 & +3.0   &     1.11 & -0.11 & +0.26   &	 1.48 & -0.08 & +0.28	&  9.9& AS \\
37 &   3.1 &  0.25 &     10.0 & -0.6 & +0.6   &     1.19 & -0.04 & +0.11   &	 2.46 & -0.14 & +0.15	& 23.7& AN \\
38 &   0.7 &  0.25 &     17.8 & -1.0 & +1.0   &     1.13 & -0.04 & +0.03   &	 2.18 & -0.12 & +0.13	& 10.5& AN \\
39 &  10.4 &  0.25 &     11.2 & -0.6 & +2.1   &     1.06 & -0.06 & +0.03   &	 1.45 & -0.08 & +0.27	& 22.7& -  \\
40 &  17.7 &  1.00 &     10.0 & -2.5 & +1.9   &     1.06 & -0.06 & +0.23   &	 1.95 & -0.56 & +0.21	& 21.6& -  \\
41 &   3.0 &  1.00 &      8.9 & -1.4 & +0.5   &     1.06 & -0.06 & +0.09   &	 1.32 & -0.21 & +0.08	& 26.9& AO \\
42 &   0.6 &  0.50 &     17.8 & -1.0 & +1.0   &     1.06 & -0.06 & +0.03   &	 1.73 & -0.10 & +0.10	&  9.6& AN \\
43 &  14.1 &  0.50 &     17.8 & -2.8 & +3.3   &     1.13 & -0.13 & +0.17   &	 2.74 & -0.26 & +0.66	&  9.7& -  \\
44 &   1.2 &  1.00 &     17.8 & -2.8 & +1.0   &     1.19 & -0.19 & +0.18   &	 1.09 & -0.17 & +0.06	&  3.4& AS \\
45 &  11.6 &  1.00 &      7.9 & -2.0 & +2.7   &     1.12 & -0.12 & +0.33   &	 1.77 & -0.44 & +0.29	& 32.7& -  \\
46 &   6.5 &  1.00 &      7.1 & -0.4 & +1.3   &     1.68 & -0.22 & +0.05   &	 1.34 & -0.21 & +0.15	& 63.5& AR $\rightarrow$ AN  \\
47 &   2.4 &  0.25 &     28.2 & -1.6 & +5.3   &     1.19 & -0.10 & +0.11   &	 1.44 & -0.08 & +0.15	&  6.1& AS \\
48 &   3.5 &  0.50 &     11.2 & -1.8 & +2.1   &     1.18 & -0.09 & +0.19   &	 1.82 & -0.39 & +0.20	& 17.2& AN \\
49 &   0.9 &  1.00 &     15.8 & -0.8 & +1.0   &     1.05 & -0.05 & +0.04   &	 1.17 & -0.07 & +0.07	&  5.9& AS \\
50 &   1.2 &  0.50 &     12.6 & -0.7 & +2.4   &     1.06 & -0.06 & +0.03   &	 1.70 & -0.09 & +0.18	& 16.3& AN \\
51 &   2.6 &  1.00 &     12.6 & -2.0 & +0.7   &     1.42 & -0.19 & +0.04   &	 1.07 & -0.11 & +0.36	& 12.0& AR $\rightarrow$ AS \\
52 &  14.0 &  0.25 &      8.9 & -1.4 & +1.7   &     1.06 & -0.06 & +0.09   &	 2.34 & -0.46 & +0.26	& 28.4& - \\

\hline
\end{tabular}
 \end{table*}

     In total there are 29 semi-detached systems and 2 contact systems
     in the sample, including the two detached systems (system 1 and
     6) for which we have alternative semi-detached solutions. The
     best fitting parameters for these systems are listed in
     Table~\ref{tab:sem1}.  An example of a semi-detached system for
     which we obtained an excellent fit is given in
     Figure~\ref{fig:semfitex}.

     Our models provide a good or reasonable fit to 11 out of the 31
     systems, with $\chirmin \leq 3$, for at least one value of
     $\beta$. This includes the two contact binaries. Four systems
     have $3 < \chirmin < 5$, and another two have $\chirmin \leq
     6.5$, and we consider these fits as (marginally) acceptable. The
     remaining 14 semi-detached systems cannot be fitted well
     ($\chirmin > 10$).
     As we have mentioned before the absolute value of \chirmin\ is not
     a very good measure of the goodness-of-fit. The distinction we
     make here, and the decision to accept fits with $\chirmin
     \leq 6.5$, is primarily guided by eye inspection of the
     solutions. In Sections~\ref{sec:semfitok:indiv}
     and~\ref{sec:semfitbad:indiv} we briefly discuss the individual
     fit solutions, for good or acceptable and bad fits, respectively.

     There are several common problems that occur in fitting our
     models to semi-detached binaries. We distinguish three types of
     conflicts between the models and the observations:
     \begin{itemize}
     \item \emph{q-conflict:\,\,}
      The most ubiquitous discrepancy, occurring in more or
     less extreme form in all systems with $\chirmin > 5$ except one,
     is the combination of a relatively small mass ratio (close to
     unity) with a large or very large temperature ratio. Such a
     combination cannot be achieved by our models for any value of
     $\beta$. We discuss possible causes for this problem in
     Sect.~\ref{sec:discussion}. 
     \item \emph{R-conflict:\,\,} 
     A second discrepancy that plagues nearly half
     the systems with $\chirmin > 10$, is that the ratio of donor
     radius over accretor radius, $\Rdon/\Racc$, is smaller than shown
     by our models, in particular the radius of the accreting star is
     often under-predicted.
     \item \emph{MT-conflict:\,\,} 
     A third problem affecting several
     systems, is the combination of a large primary mass with a
     relatively low temperature which cannot be matched simultaneously
     by our models. The inverse (small mass together with high
     temperature) also occurs in a few systems.
     \end{itemize}
     We refer to these conflicts when discussing the individual
     systems in Sect.~\ref{sec:semfitbad:indiv}.

     The effect of varying the mass transfer efficiency on the fit
     solutions is often rather obscure, but we can make a number of
     general statements. The most obvious effect is that lowering
     $\beta$ results in a smaller (less extreme) post-mass transfer
     mass ratio, for given initial masses and period and at a given
     age. The same is true, although somewhat less obvious, for the
     temperature ratio. Since we are trying to fit a certain observed
     mass (and temperature) ratio, this means that models with smaller
     $\beta$ require either a larger (more extreme) initial mass
     ratio, or a larger amount of mass to be transferred, i.e.\ a more
     advanced evolution stage. A consequence of the latter is in many
     cases a larger donor radius compared to the radius of the
     accretor. This effect is often (though not always) seen in our
     best-fitting solutions for a certain system: a smaller $\beta$
     leads to a larger ratio of donor to accretor radius, as well as
     (or sometimes instead of) a less extreme mass ratio.

     A few systems can be fitted well for different assumptions of
     $\beta$, such that we cannot determine, on basis of the available
     observables, if the system has gone through conservative or
     non-conservative mass transfer. However, in some of these cases
     (i.e.\ systems 31, 33, 38 and 51) the surface abundances of carbon
     and nitrogen differ substantially between the best-fitting
     models for different $\beta$. This is a consequence of the fact
     that a smaller $\beta$ yields a fit at a more advanced evolution
     stage, when the donor star is peeled off to a larger extent.
     Abundance determinations of carbon and nitrogen would therefore
     enable us in principle to constrain the mass transfer
     efficiency. Unfortunately these are not available at this moment,
     because the resolution of the spectra is not high enough for
     abundance determinations. We limit ourselves to qualitative
     statements as the absolute abundances in our models are sensitive
     to uncertainties in the initial composition and the chemical
     profile inside the star.

      \subsubsection{Comments on individual fits: $\chirmin<6.5$} 
            \label{sec:semfitok:indiv}

      The seventeen systems for which we found fits that we deem
      acceptable are discussed individually below. We focus the
      discussion on the characteristics that distinguish the
      best-fitting mass transfer efficiencies.

\begin{itemize}

       \item[26] This is one of the brightest systems, with an O9.5
        primary. \citetalias{2003MNRAS.339..157H} mention the high
        quality of the spectra. The asymmetric light curve shows
        evidence for an accretion stream.

        Our non-conservative models ($\beta\leq0.5$) give better
        fits to this system. In fits to conservative models, the
        primary mass and temperature prefer tracks with small initial
        masses, but they cannot be fitted at the same time as the mass
        ratio and accretor radius, which are better explained by
        models with higher initial masses. The substantial value of
        \chirmin\ is dominated by the contribution of the accretor
        radius, which is somewhat under-predicted in all models. 

	This is the semi-detached system with the widest orbit, both
	currently and initially. As a consequence it experienced mass
	transfer in a late stage of the primary's main-sequence phase,
	and the current slow mass-transfer phase will be relatively
	short.

        \item[28] This system, consisting of two O stars, is the most
        massive in the whole sample.  It fits well to models with
        $\beta= 0.5-1.0$ and an initial mass ratio of about 1.3,
        although the effective temperature is overestimated by more
        than one sigma assuming conservative evolution.
        According to these models the system is currently seen in
        the early stages of the slow mass-transfer phase.  For
        $\beta=0.25$ the observed primary mass (which requires a model
        with an initially massive primary) cannot be fitted well
        together with the primary radius and the mass ratio, which
        prefer models with a lower initial primary mass and more
        extreme mass ratio. 

        \item[31] The best fit to this systems is for $\beta=0.5$ with
        a large initial mass ratio (1.7). This model narrowly missed
        contact during rapid mass transfer, and predicts the current
        system to be near the end of the slow mass-transfer phase, not
        long before core H-exhaustion of the initial primary.
        Conservative ($\beta\geq0.75$) models also require a large
        initial mass ratio to fit the currently observed one, but fail
        to account for the observed temperature ratio. These
        conservative models undergo temporary contact during rapid
        mass transfer, and it is not excluded that this system evolved
        fairly conservatively through a deeper temporary contact
        phase, which we cannot model properly.

	As a result of the advanced evolution stage, our models
	predict that the C and N abundances have reached equilibrium
	at the surface but oxygen has not, and the O abundance depends
	quite strongly on the adopted value of $\beta$.

        \item[32] This system has a fairly wide orbit, and the
        components have very similar masses. The only acceptable fit
        for this binary is for highly non-conservative evolution
        ($\beta = 0.25$). The fit is not very good ($\chirmin=5.7$) as
        the models over-predict the mass ratio and under-predict the
        primary mass, but these problems are much worse for higher
        mass-transfer efficiencies.  It is possible that a value of
        $\beta<0.25$ would fit this system better.

        \item[33] This system fits best for $0.5 \leq \beta \leq
        0.75$, although the best fitting models overestimate the
        effective temperature by more than 1 sigma. Conservative
        models overestimate the temperatures even more, whereas models
        with $\beta=0.25$ under-predict the masses, but these models
        are not strongly excluded.

	For decreasing values of $\beta$ the best-fitting models start
	with a more massive primary and a more extreme mass ratio, so
	that the initial secondary mass is always $\approx 10\,\Msun$.
	The less conservative models fit this system in a later phase
	of evolution, when more mass has been transferred.  Abundance
	measurements of C and N could possibly distinguish between a
	conservative and non-conservative solution, since the
	predicted model abundances for this system depend quite
	strongly on $\beta$.

        \item[36] This system has a fairly large mass ratio.  The
        best-fitting models assume conservative mass transfer $\beta
        \geq 0.75$. However, the model temperatures are larger than
        observed while the masses are smaller (a problem that also
        occurs in many of the poorly-fitting systems,
        Sect.~\ref{sec:semfitbad:indiv}).  The main problem with the
        non-conservative models is that the masses are under-predicted
        even more by the models, by about $2 \Msun$

        \item[37] With an orbital period of 4 days, this system is one
        of the wider systems in the sample. Only very non-conservative
        models ($\beta < 0.5$) can fit both the large radius of the
        Roche-lobe filling star and the much smaller radius of its
        companion at the same time as the mass ratio of the system.
        Like system 26, this binary underwent mass transfer fairly
        late on the main sequence and the primary is currently close
        to hydrogen exhaustion in its core.

        \item[38] This system fits well to non-conservative models
        ($\beta \leq0.53$) not long before the donor will leave the
        main sequence.  For the more conservative models, the observed
        masses and temperatures would fit better against initially
        less massive models, while the radii and temperature ratio
        would fit better against more massive models. Also for this
        system the best-fitting models suggest that abundance
        measurements of C and N (and perhaps O) could possibly
        constrain $\beta$ further.

        \item[41] A good fit to this system, which has a large mass
        ratio and temperature ratio, is obtained for fairly
        conservative evolution ($\beta\geq0.75$), assuming that the
        system started with a mass ratio near unity. The
        non-conservative models cannot fit the observed masses (which
        prefer models with larger initial masses) at the same time as
        the temperature ratio and the radius of the Roche-lobe filling
        star.

        \item[42] This system can be fitted well assuming
        non-conservative evolution, particularly for $\beta=0.5$. The
        main problem with fitting a conservative model is the large
        radius of mass-accreting star (nearly equal to that of its
        companion) in comparison with its mass and temperature. This
        system is very similar to system 38 in terms of its inferred
        initial mass and mass ratio, but started with a smaller
        period.

        \item[44] This system is fitted well by our models without a
        clear preference for a certain mass transfer efficiency. It
        has one of the shortest periods in the sample, and in all
        models mass transfer started soon after the onset of hydrogen
        burning and will eventually lead to contact (case AS
        evolution). The smaller the value of $\beta$, the closer to
        contact the system currently is, according to our models.

        \item[46] This is the least massive of the semi-detached
        systems and it has the most extreme mass ratio (3.1) of the
        sample.  The non-conservative models cannot explain this
        extreme mass ratio, to which we obtain the best fit with a
        completely conservative model with a large initial mass
        ratio. This model overestimates the temperature ratio,
        however, and this is the major contribution to the rather poor
        $\chirmin=6.5$. The primary temperature is also not perfectly
        reproduced but the overall fit is acceptable. This model went
        through a brief contact phase during rapid mass transfer. The
        Roche-lobe filling star is fitted right at the end of the slow
        mass transfer phase, and also the accreting star is very close
        to the end of the main sequence according to this model. The
        system is on the verge of detaching, before entering another
        phase of case AB mass transfer.

        \item[47] This is one of the more massive and hottest systems
        in the sample, and the light-curve solution indicates it is a
        near-contact binary in which the primary star fills 99 per
        cent of its Roche lobe. We find a fairly good fit to a
        non-conservative ($\beta=0.25$) model in contact, which
        however somewhat overestimates the masses and underestimates
        the temperatures. This solution has a very large initial mass
        of 28\,\Msun.  For the more conservative models the observed
        masses, which prefer initially even more massive models,
        cannot be fitted simultaneously with the other observables.

        \item[48] This system can be fitted well assuming $0.5 \leq
        \beta \leq 0.75$. More conservative models overestimate the
        observed small mass ratio.  The system is currently seen
        shortly after the end of the rapid mass transfer phase.

        \item[49] This is the shortest-period binary in the sample,
        for which the light-curve solution indicates a deep-contact
        system. A good fit solution is obtained for all values of
        $\beta$; only the most non-conservative grid models cannot
        explain the temperature ratio well. In all best-fitting models
        the system is still semi-detached but the accreting star is
        very close to filling its Roche lobe. However, contact is not
        reached until substantially more mass is transferred.

        \item[50] The system can be fitted well assuming
        non-conservative evolution ($\beta\leq0.5$) and an initial
        mass ratio near unity. If conservative mass transfer is
        assumed the primary mass and temperature cannot be fitted well
        at the same time as the very large radius of the
        mass-accreting star. On the basis of the best-fitting model we
        interpret this system as being in the last stage of the slow
        mass-transfer phase. It is very similar to system 42 in terms
        of its initial parameters and current evolution state, except
        for a smaller initial primary mass (and consequently larger
        age).

        \item[51] The alternative detached light-curve solution of
        this system does not fit well to the models, as explained in
        Sect.~\ref{sec:detfits:indiv}. Conservative models with $\beta
        \geq 0.75$ give good fits to the semi-detached configuration,
        although the masses are underestimated and the radius of the
        accreting star is overestimated.  This is worse for
        non-conservative models, which also overestimate the radius of
        the accreting star. The best fit is found for a model with a
        large initial mass ratio, which went through a contact phase
        during rapid mass transfer and is currently undergoing case AB
        mass transfer. The C/N ratio of
        the best-fitting models decreases with $\beta$, so that its
        determination might further constrain the mass transfer
        efficiency.

\end{itemize}
     \subsubsection{Comments on individual fits: $\chirmin>10$} 
            \label{sec:semfitbad:indiv}

     For the systems, for which we find no acceptable fits, we
     indicate which of the three common conflicts occur (see the start
     of this subsection).  We divide the systems in two groups, based
     on their \chir\ values. We start with the eight systems that have
     $10 < \chirmin < 20$:

\begin{itemize}

	\item[22]  This system shows the $q$-conflict.
	Only the conservative models can match the very large
	temperature ratio of this system, as well as the moderately
	small value of $\Rdon/\Racc$. The mass ratio is then
	over-predicted by the models.

	\item[23] This system exhibits the $q$- and $MT$-conflict.
	The best fitting model, obtained assuming $\beta = 0.5$, fits
	the observed radii but underestimates the mass of the donor
	star by about 3 \Msun.

	\item[34] The $R$-conflict and a moderate form of the
	$q$-conflict occur for this system.  The best fit is obtained
	with conservative models.

	\item[39] The $q$- and $R$-conflicts are present as well as a
	moderate inverse $MT$-conflict.

	\item[40] A combination of all three conflicts occur for this system.

	\item[43] This system, which shows the $q$-conflict and a moderate
	inverse $MT$-conflict, has one of the highest effective
	temperatures. In the best fit, assuming $\beta = 0.5$, the
	temperature ratio is under-predicted and the mass ratio is
	over-predicted by the models.

	\item[45] For this system with a $q$-conflict, the
	conservative models provide a reasonably good fit to nearly
	all observed parameters for this system except the mass ratio.

	\item[52] The alternative detached solution for this model
	could not be fitted well with our models. The same is the case
	for the semi-detached solution. The system shows the $q$- and
	inverse $MT$-conflict. The radius of the accretor is also
	under-predicted.

\end{itemize}

        Six systems have very poor fit solutions, with $\chirmin > 25$:

\begin{itemize}

     	\item[24] This systems shows a severe form of the
     	$q$-conflict and cannot be matched for any assumption of
     	$\beta$.

     	\item[25] This system suffers from a severe $q$-conflict as
     	well as the $R$-conflict and the inverse $MT$-conflict.

	\item[27] This system shows the $q$- and $R$-conflict.

	\item[29]
	  This system exhibits the $q$- and $MT$-conflict.

	\item[30] This system shows a severe form of the $q$-conflict.
	The large \chirmin\ is dominated by the contribution of the
	extreme temperature ratio, which is the largest in the
	sample. The other two common discrepancies, the $R$- and
	$MT$-conflict, are also very serious in this system,
	$\Rdon/\Racc$ is even slightly below unity.
        The photometric flux ratio is incorrectly reported in
        H05 as 1.2; it should be 2.7. The spectral type of the
        secondary component is B2 or later.

	\item[35] This system is the least massive of all
	semi-detached systems and cannot be fitted well because of the
	$q$-conflict.

\end{itemize}

\subsection{Summary of fit results}

   \begin{figure} \centering
   \includegraphics[angle=-90,width=0.5\textwidth]{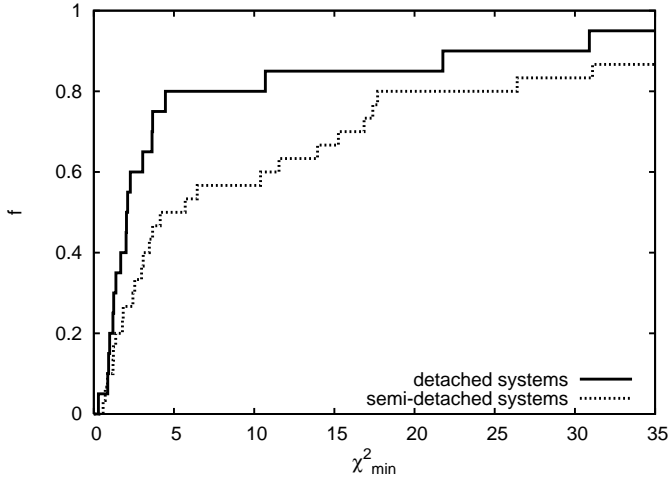}%
   \caption{ Fraction of detached (full line) and semi detached
   systems (dotted line) for which a fit solutions were obtained with
   \chisqmin lower than a certain value. We have excluded system 1
   from the detached systems and system 52 from the semi-detached
   systems.} \label{FigCumDist} \end{figure}


    As we have argued in Sect.~\ref{sec:method}, due to the nature of
    our fit method the absolute value of \chirmin\ is not a very good
    measure of the goodness-of-fit. The observables we have fitted are
    not truly independent, and their errors may not in all cases be
    representative of the true uncertainties. This applies especially
    to the temperature ratio, which is determined from a flux ratio
    that is not very sensitive to temperature for OB stars. Also the
    error on the primary temperature is only a rough estimate.
    Therefore we feel we can be fairly generous in accepting fit
    solutions with large \chirmin, and we have argued in the previous
    subsections why we regard fits with $\chirmin \leq 6.5$ as
    acceptable.

    In Fig.~\ref{FigCumDist} we show the cumulative distribution of
    \chisqmin for the detached and the semi-detached systems. For
    about $80\%$ of the detached systems and only about $55\%$ of the
    semi-detached systems we obtain fits with $\chisqmin \leq
    6.5$. Interestingly, both distributions show a break between this
    and larger values of \chir. There is no reason to expect that, if
    any of the observational errors were underestimated, this should
    occur preferentially for semi-detached binaries. It is noteworthy
    that the $q$-conflict, which afflicts a large fraction of the
    semi-detached fits, only occurs in one detached system (11). We
    must therefore conclude that our models do significantly worse at
    explaining post-mass transfer binaries than detached systems which
    have not yet undergone mass transfer.  We will discuss possible
    causes and the limitations of our models in
    Sect.~\ref{sec:discussion}.

	\section{Conclusions on the mass transfer efficiency}
	\label{sec:beta}

 \begin{table}
 \caption{\label{tab:chibeta} For each semi-detached system
  $\chi^2_{\min}$ is given as function of $\beta$.  For each system
  the value corresponding to the best fitting $\beta$ is marked. The
  last column indicates the estimated range of $\beta$ which fits well for a
  certain system. }
\centering
 \begin{tabular}{l|cccc|c}
 \hline\hline 
  id &  
  $\beta = 1.00$ & 
  $ 0.75$ & 
  $ 0.50$ & 
  $ 0.25$ & 
  range   \\
 \hline
 
22 &    \bf    16.9$^*$ &     18.0 &	 21.3 &     35.4 &  0.43$ - $ 1.00\\
23 &       31.0 &     22.5 &  \bf    17.4$^*$ &     25.9 &  0.25$ - $ 0.86\\
24 &       52.3 &  \bf    50.2$^*$ &	 50.9 &     57.2 &  0.25$ - $ 1.00\\
25 &      142.5 &    126.0 &	105.2 &  \bf	97.7$^*$ &  0.25$ - $ 1.00\\
26 &       15.2 &      6.9 &  \bf     4.1$^*$ &      4.2 &  0.25$ - $ 0.69\\
27 &       32.9 &     26.8 &  \bf    26.4$^*$ &     28.2 &  0.25$ - $ 1.00\\
28 &        2.3 &  \bf     1.9$^*$ &	  2.4 &      6.5 &  0.48$ - $ 1.00\\
29 &    \bf    31.1$^*$ &     33.0 &	 41.1 &     57.1 &  0.42$ - $ 1.00\\
30 &    \bf   141.4$^*$ &    153.8 &	176.1 &    212.2 &  0.25$ - $ 1.00\\
31 &       22.9 &      9.6 &  \bf     1.8$^*$ &      4.4 &  0.38$ - $ 0.63\\
32 &       34.5 &     22.1 &	 11.6 &  \bf	 5.7$^*$ &  0.25$ - $ 0.38\\
33 &        3.2 &      1.9 &  \bf     1.4$^*$ &      3.4 &  0.38$ - $ 0.78\\
34 &    \bf    15.2$^*$ &     15.9 &	 20.2 &     32.2 &  0.45$ - $ 1.00\\
35 &       63.1 &     43.8 &  \bf    41.4$^*$ &     43.1 &  0.25$ - $ 0.99\\
36 &    \bf	3.7$^*$ &      5.1 &	 10.8 &     27.5 &  0.73$ - $ 1.00\\
37 &       30.7 &     15.0 &	  4.9 &  \bf	 3.1$^*$ &  0.25$ - $ 0.46\\
38 &        4.8 &      2.5 &	  0.8 &  \bf	 0.7$^*$ &  0.25$ - $ 0.53\\
39 &       21.5 &     15.6 &	 12.5 &  \bf	10.6$^*$ &  0.25$ - $ 0.76\\
40 &    \bf    17.7$^*$ &     19.4 &	 24.1 &     33.3 &  0.43$ - $ 1.00\\
41 &    \bf	3.0$^*$ &      3.7 &	  7.7 &     10.4 &  0.70$ - $ 1.00\\
42 &        7.4 &      3.7 &  \bf     0.6$^*$ &      2.1 &  0.38$ - $ 0.63\\
43 &       28.8 &     18.2 &  \bf    14.1$^*$ &     15.4 &  0.25$ - $ 0.82\\
44 &    \bf	1.2$^*$ &      1.2 &	  1.5 &      2.1 &  0.36$ - $ 1.00\\
45 &    \bf    11.5$^*$ &     13.3 &	 16.0 &     22.7 &  0.45$ - $ 1.00\\

46 &    \bf	6.5$^*$ &      8.1 &	 23.6 &     50.3 &  0.72$ - $ 1.00\\ 

47 &       12.4 &      7.4 &	  4.7 &  \bf	 2.6$^*$ &  0.25$ - $ 0.38\\
48 &        6.8 &      3.6 &  \bf     3.5$^*$ &      5.7 &  0.31$ - $ 0.88\\
49 &    \bf	0.9$^*$ &      1.0 &	  1.2 &      1.6 &  0.47$ - $ 1.00\\
50 &        5.4 &      3.4 &  \bf     1.2$^*$ &      1.4 &  0.25$ - $ 0.63\\
51 &    \bf	2.6$^*$ &      3.3 &	  7.3 &     14.8 &  0.72$ - $ 1.00\\
52 &       42.2 &     32.3 &	 23.2 &  \bf	14.0$^*$ &  0.25$ - $ 0.44\\

 \hline
 \end{tabular}
 \end{table}

   \begin{figure} \centering
   \includegraphics[angle=-90,width=0.5\textwidth]{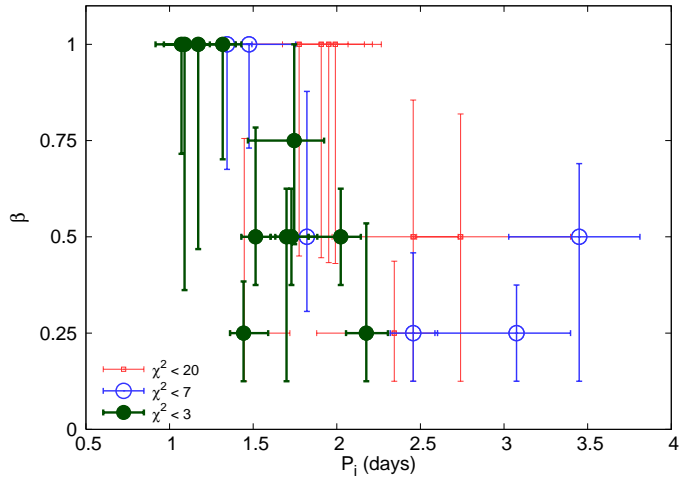}
   \caption{ The best fitting mass transfer efficiency $\beta$ is
   plotted against the best fitting initial orbital period $P_i$. The
   symbols indicate the magnitude of $\chi^2_{\min}$. The error bars
   indicate the $1\sigma$ range of the two fit parameters; if the
   best-fitting $\beta=0.25$ the error bar is extended downward by an
   arbitrary amount. No strong trend is found but there is a hint that
   systems with small initial orbital periods evolve more
   conservatively than wider systems.}
   \label{FigCorrBp} \end{figure}
%

   \begin{figure} \centering
   \includegraphics[angle=-90,width=0.5\textwidth]{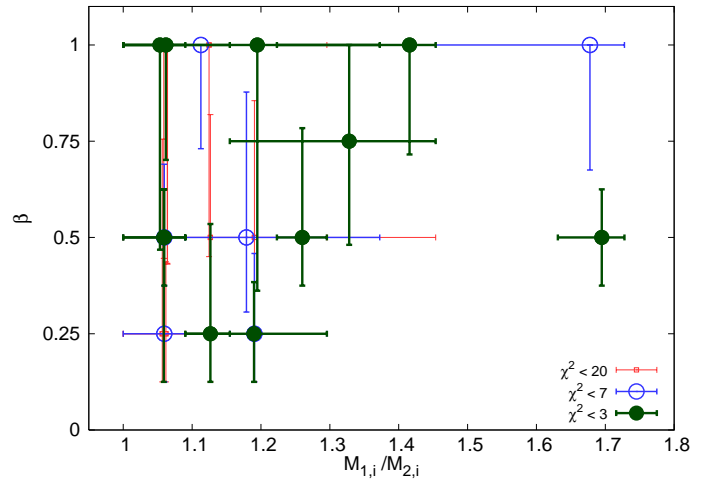}
   \caption{ As Fig.~\ref{FigCorrBp} $\beta$ against mass ratio $q_i$. }  
   \label{FigCorrBq} \end{figure}
%

   \begin{figure} \centering
   \includegraphics[angle=-90,width=0.5\textwidth]{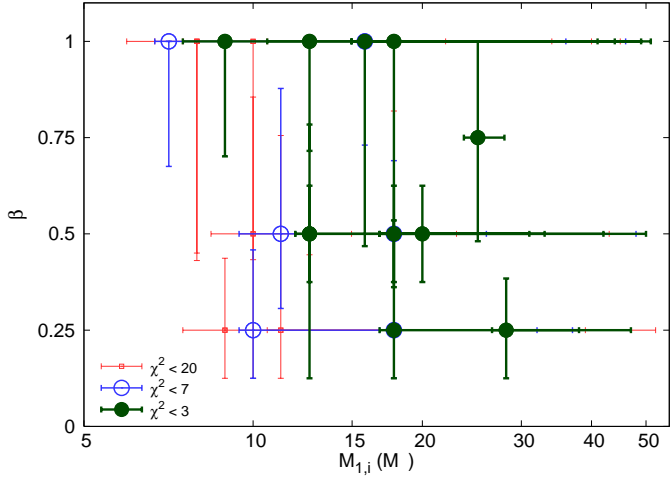}
   \caption{ As Fig.~\ref{FigCorrBp} for $\beta$ against primary mass $M_{1,i}$. }
   \label{FigCorrBm} \end{figure}
%

     Although a large fraction of the semi-detached systems cannot be
     fitted well with any assumption for the mass transfer efficiency,
     we will focus on those systems for which we do obtain reasonable
     fits to see if we can learn anything about the efficiency of mass
     transfer from the comparison of models and observations.  In
     Table~\ref{tab:chibeta} \chirmin\ is given as a function of the
     mass transfer efficiency. The last column gives an estimate for
     the range of $\beta$ for which a certain system can be fitted, as
     explained in Sect.~\ref{sec:method}.
     Some of the semi-detached systems fit significantly better
     against conservative models ($\beta \geq 0.7$), for example
     system 36, 41, 46 and 51, while other systems show a strong
     preference for the non-conservative models ($\beta < 0.5$), for
     example system 32, 37 and 47.  There is no single value of
     $\beta$ for which we can fit all systems.
     From this we conclude that the simplified assumption of
     conservative evolution is not valid. Neither is the often made
     assumption of $\beta = 0.5$. Binary systems with different
     initial parameters cannot be described with one fixed value of
     $\beta$. The spread in best-fitting $\beta$ values covers the
     full range of $\beta$.

     In order to understand the spread in $\beta$, we searched for
     correlations between the preferred $\beta$ and the initial
     parameters.  In Figure~\ref{FigCorrBp} the preferred $\beta$ is
     plotted against the initial orbital period. The correlation is
     not strong but there is a region in the upper right part of the
     diagram where we find no systems: the initially wider systems
     prefer less conservative models.  If real, this correlation might
     be explained by the interplay of two mechanisms: spin up of the
     accreting star and tidal interaction.  The mass-accreting star is
     spun up as it accretes angular momentum. Tidal interaction tends
     to synchronize the stars with the orbit and prevent the accreting
     star from reaching critical rotation. In wider systems, where
     tidal interaction is less efficient, the accreting star rotates
     faster and may lose mass along its equator.

     One can expect a strong correlation of the mass transfer
     efficiency with initial mass ratio, because in systems with
     extreme mass ratios the accretor will expand rapidly such that a
     contact system is formed. Mass and angular momentum can than be
     lost from the outer Lagrangian point. However, we found no
     correlation of the mass transfer efficiency with the mass ratio
     in the expected direction, see Figure~\ref{FigCorrBq}.
     This might be because the range in current mass ratios available
     in the sample is small, as the sample is strongly biased towards
     systems with equal masses. No significant correlation with the
     initial primary mass was found either, see
     Figure~\ref{FigCorrBm}.

\section{Discussion}
\label{sec:discussion}

    It is striking that we can fit the detached systems in the sample
    quite well, while a large fraction of the semi-detached systems
    cannot be explained with our models. We seem to understand the
    evolution of detached systems well, i.e. essentially the
    evolution of two single stars, while our models fail to account
    for a large number of systems in slow mass transfer, especially
    their temperatures. 

    In our models we assume a constant mass transfer efficiency
    throughout the whole evolution. However, mass loss from the system
    is more likely to occur mainly during the rapid phase of mass
    transfer. A more realistic description of the mass transfer phases
    is to be preferred. For instance, spin-up by accretion and
    associated rotation-induced mass loss can be taken into account,
    as was done by \citet{2001PhDT..........B} and
    \citet{2005A&A...435..247P}. However, this treatment introduces
    other uncertainties, e.g. for the strength of rotation-induced
    mass loss and the timescale for tidal interaction, which are both
    uncertain.

    For non-conservative evolution we assume that the lost mass
    carries the specific orbital angular momentum of the accreting
    star. While the underlying assumption (that the accreting star
    ejects the surplus of mass isotropically) is reasonable, it may be
    worthwhile to test the effect of different assumptions. For
    instance, somewhat stronger angular momentum loss than we have
    assumed here may improve some of the fits exhibiting the
    $R$-conflict.

    Our models with extreme initial mass ratios show that some systems
    undergo a temporary contact phase, after which the system restores
    thermal equilibrium 
    and becomes semi-detached again. A significant fraction of our
    models with extreme initial mass ratios fail to converge during
    rapid mass transfer and we stop our calculations if deep contact
    occurs, as our code contains only a crude model for these contact
    situations.  Improvement of the description of the contact phase
    might enable us to fit more systems properly.

    We assume the Schwarzschild criterion for stability against
    convection. As a result mass-accreting stars always rejuvenate
    (i.e. increase the size of their convective core). The question of
    whether stars rejuvenate or not depends on the still poorly
    understood efficiency of semi-convection. It is possible that the
    steep molecular weight gradient, built up before the onset of mass
    transfer, prevents the star from rejuvenating
    \citep{1995A&A...297..483B}. This shortens the remaining lifetime
    of the accreting star and affects its position in the H-R diagram
    as a function of time. Whether this effect would improve any of
    the fits remains to be investigated.

    In principle mass transfer can lead to extra mixing, for example
    because of spin up and resulting differential rotation, which is
    not included in our models. If the extra mixing is significant,
    then chemical abundance profiles in the star will become
    flatter. In general this results in more compact and therefore
    hotter stars. As this only affects the mass-accreting star, which
    is the hotter component, it will lead to more extreme temperature
    ratios. This might help to solve the discrepancy between the
    observed and modeled temperature ratios.

    To account for the discreteness of our model grid we estimated the
    spacing in our grid by the initial spacing. The actual grid
    spacing changes during the evolution and it might even depend on
    $\beta$.  Interpolating between the models would be a solution but
    this is not simple considering the many dimensions in the grid.
    We investigated the effect of different values for $\sigma_{\rm
    th}$ and we found no evidence for systematically under- or
    overestimating the actual grid spacing for evolved systems.

    Our fit method might also be improved by using more directly
    measured observables. We recall that the most commonly occurring
    conflicts between our models and the observations are related to
    the temperature ratio and the radius ratio. For this sample the
    temperature ratios were derived from the $I$-band flux ratios as
    determined from the light-curve solution. Rather than fitting to
    the temperature ratio, we can use the $I$-band flux ratio and its
    error directly as an observable.  The light-curve solution often
    yields the sum of the radii to a higher accuracy than the ratio of
    the radii, as a result of which the errors in the individual radii are
    correlated. It may be preferable to use the sum and the ratio of
    the radii as observables, rather than the individual radii.

\section{Summary and outlook}
\label{sec:summary}


    We computed an extensive grid of detailed binary evolution tracks
    at a metallicity $Z = 0.004$ for a variety of mass transfer
    efficiencies.  Our models are available to the astronomical
    community. We plan to extend the grid to a larger range of
    binary parameters and to other metallicities in the future.

    We have compared our models to observations of 50 double-lined
    eclipsing binaries, by fitting evolution tracks to each individual
    system. For the detached systems in the sample we find generally
    good agreement between observations and models. Our models can
    also explain a large fraction of the semi-detached systems
    although the overall agreement is less good. We identified three
    common conflicts between the models and the observations.

%
    (I) Several semi-detached systems in which the stars have fairly
    equal masses are observed to have a large temperature ratio, more
    extreme (in some cases much more extreme) than our models predict.
%
    (II) In some systems the ratio of donor radius to accretor radius,
    $\Rdon/\Racc$, is smaller than shown by our models, in which case
    the radius of the accreting star is often under-predicted.
%
    (III) A third problem is the combination of a large observed
    primary mass with a relatively low temperature which cannot be
    matched simultaneously by our models. The inverse (small mass
    together with high temperature) also occurs in a few systems.
%
%

    Conservative mass transfer cannot explain these
    case A mass transfer systems.  We find a large spread in the best
    fitting mass transfer efficiency. There is no single assumption
    for $\beta$ which can explain all semi-detached systems. We
    conclude that the often made assumption of a single constant
    $\beta$ for all case A binaries is not valid.
    We find a hint that initially wider systems fit better to
    non-conservative models. This might indicate the importance of
    spin up and tidal interaction in determining the efficiency of
    mass transfer.


    We have shown that, assuming non-conservative mass transfer, more
    systems will avoid contact during rapid mass transfer and that the
    duration of the slow mass transfer phase becomes longer. This can
    in principle give a strong test on the mass transfer efficiency by
    comparing the number of detached and semi-detached systems. Good
    understanding of the selection effects and the initial mass ratio
    distribution are needed for such a test.

      Although the observed sample we used is currently the largest
     single set of stellar parameters determined for massive binaries
     in any galaxy, it is limited in primary masses, mass ratios and
     orbital periods. It would be interesting to extend this study to
     all currently known semi-detached double-lined eclipsing binaries
     in our galaxy and the Magellanic Clouds. This would enable us to
     study a larger range of primary masses. Studying a larger range
     of orbital periods and mass ratios will however be difficult as
     all such samples are biased towards small periods and equal mass
     ratios.

     The largest discrepancies between models and observations are
     seen in the temperature ratios.  The $I$-band flux ratios from
     which these are determined are not very sensitive to temperature
     differences as they cover only the Raleigh-Jeans tail of the
     spectrum. Light curves in the $U$ and $B$-band would lead to more
     reliable determinations of the temperature ratio.  From high
     quality spectra accurate temperatures can be obtained by
     fitting model atmospheres.

     Our results indicate that surface abundance determinations of
     nitrogen and carbon can potentially constrain the mass transfer
     efficiency strongly. 
     If mass transfer is a non-conservative process, then in order to
     reach a given mass ratio the donor must lose a larger fraction of
     its mass than for conservative mass transfer. Therefore deeper
     layers of the star are revealed, in which CN-cycling has
     increased the N/C ratio. In Section~\ref{sec:ressemi} we have
     indicated several particularly interesting and promising targets
     for further study.
     In order to determine accurate temperatures and surface
     abundances, high-resolution spectroscopic follow-up observations
     are needed.

\begin{acknowledgements}

We would like to thank Frank Verbunt, Cees Bassa, Rob Izzard and Evert
Glebbeek for fruitful discussions and comments and the referee Peter
Eggleton for his useful suggestions.

\end{acknowledgements}

\bibliographystyle{aa}
\bibliography{art}


\Online

  \begin{figure*}
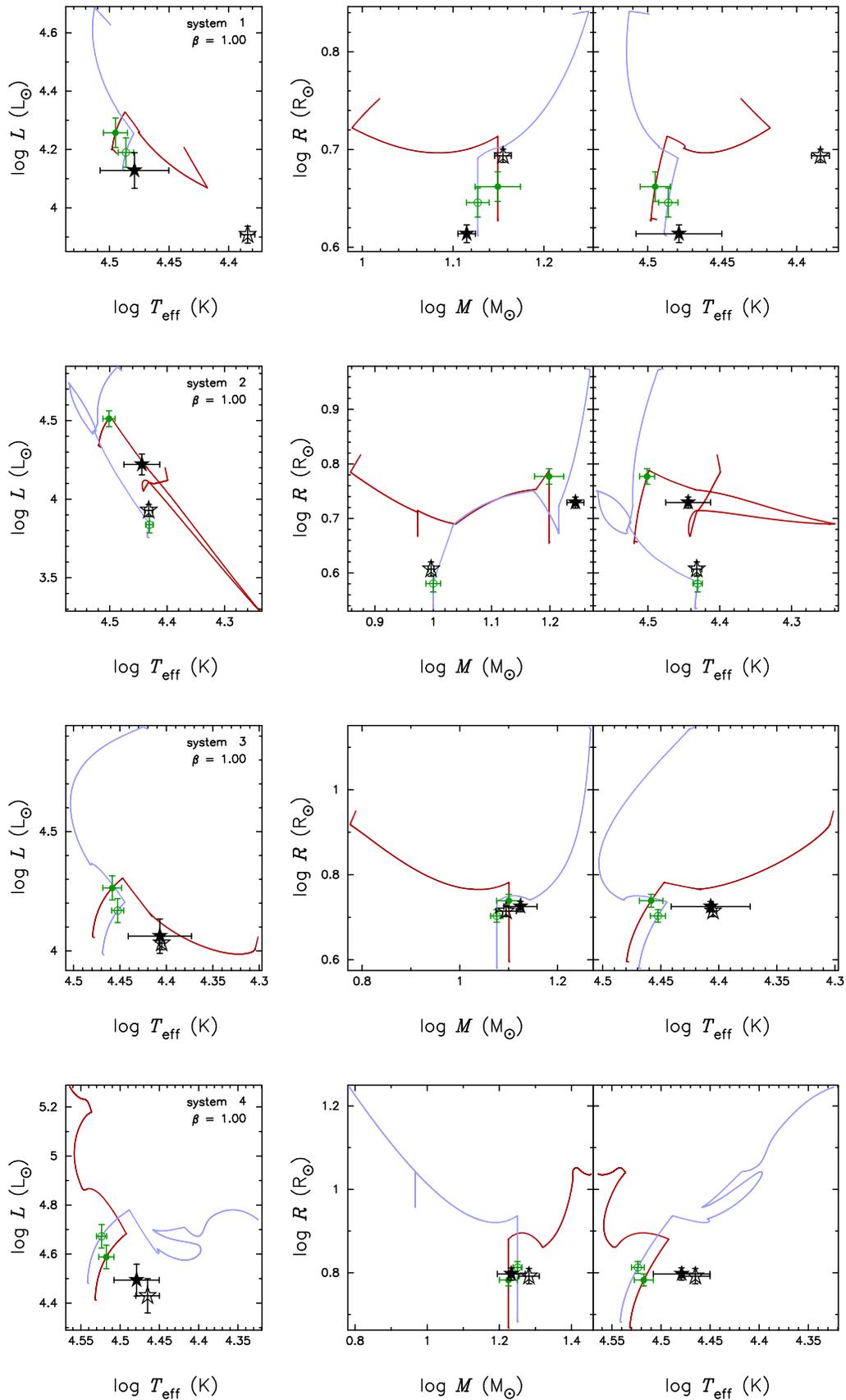
 \centering
\includegraphics[bb=20 30 520 240,clip,width=0.75\textwidth]{7007FA01}
\includegraphics[bb=20 30 520 240,clip,width=0.75\textwidth]{7007FA02}
\includegraphics[bb=20 30 520 240,clip,width=0.75\textwidth]{7007FA03}
\includegraphics[bb=20 30 520 240,clip,width=0.75\textwidth]{7007FA04}
 \caption{ See Fig~\ref{fig:detfitex} for description of the symbols
   and the units. The observed parameters are plotted together with
   the best fitting model, see table~\ref{tab:det1} and
   \ref{tab:sem1}. For the detached systems we plotted conservative tracks.
 }
 \end{figure*}
\addtocounter{figure}{-1}
  \begin{figure*} \centering
\includegraphics[bb=20 30 520 240,clip,width=0.75\textwidth]{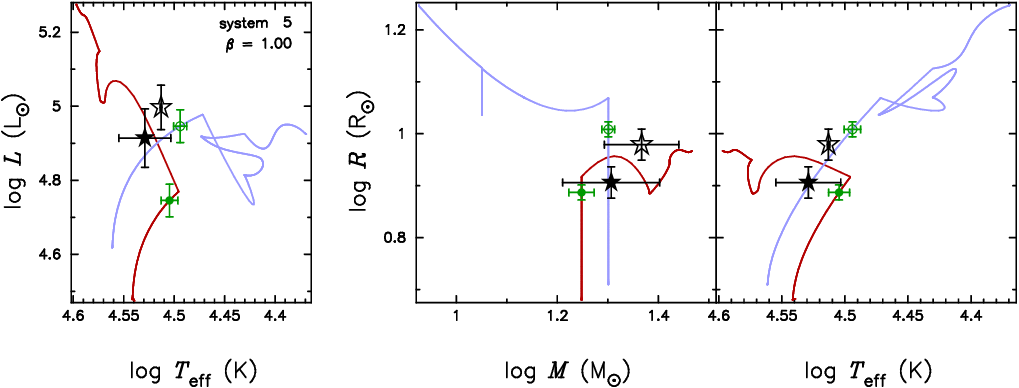}
\includegraphics[bb=20 30 520 240,clip,width=0.75\textwidth]{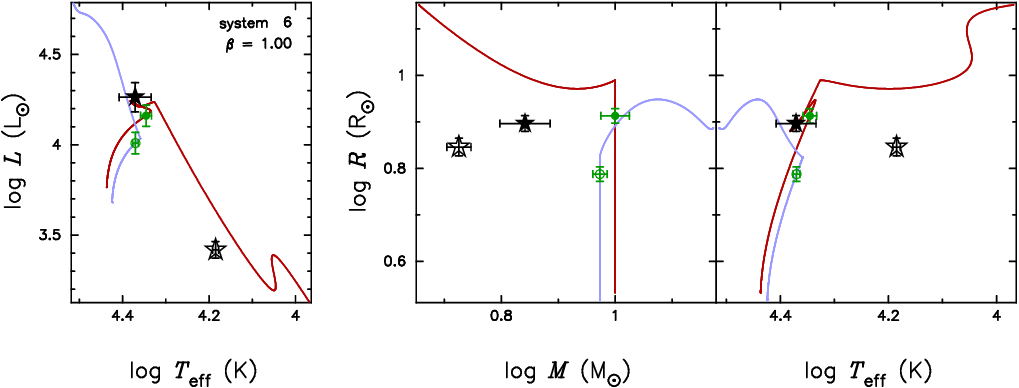}
\includegraphics[bb=20 30 520 240,clip,width=0.75\textwidth]{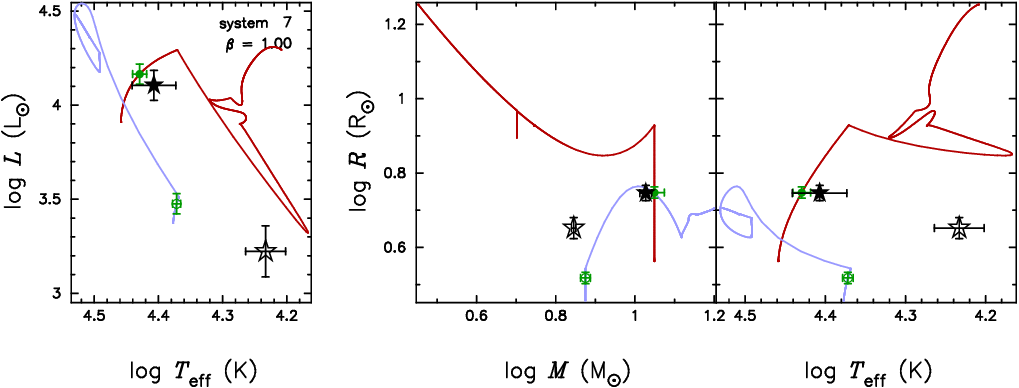}
\includegraphics[bb=20 30 520 240,clip,width=0.75\textwidth]{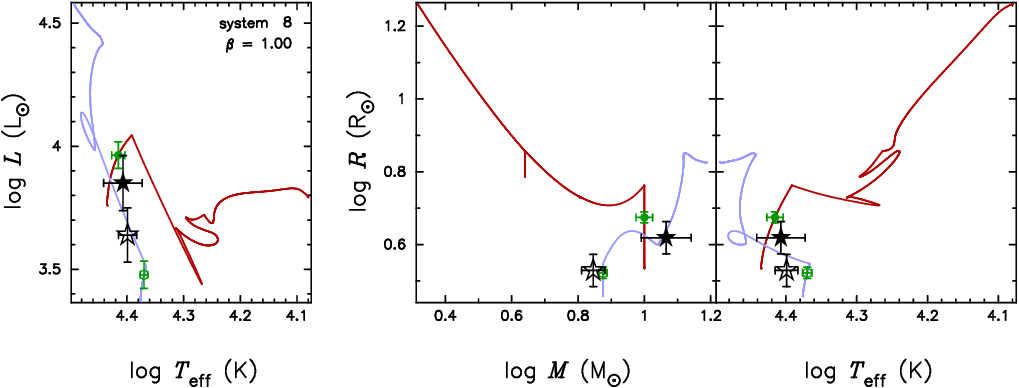}
 \caption{ 
  Continued.
 }
 \end{figure*}
\addtocounter{figure}{-1}
  \begin{figure*} \centering
\includegraphics[bb=20 30 520 240,clip,width=0.75\textwidth]{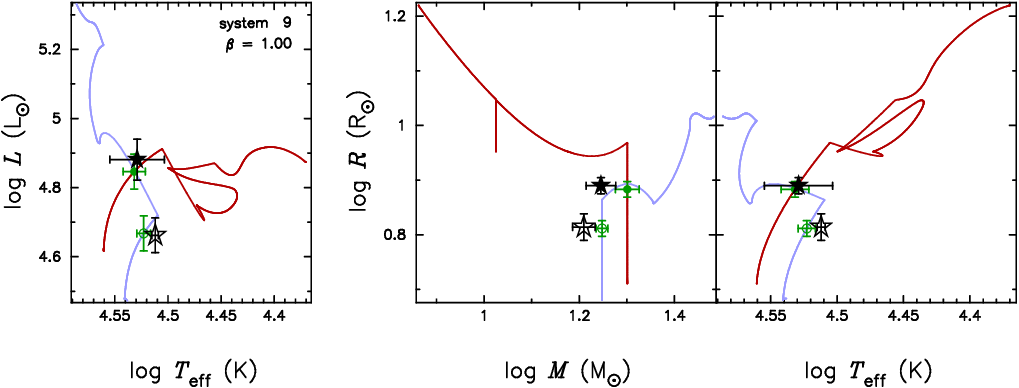}
\includegraphics[bb=20 30 520 240,clip,width=0.75\textwidth]{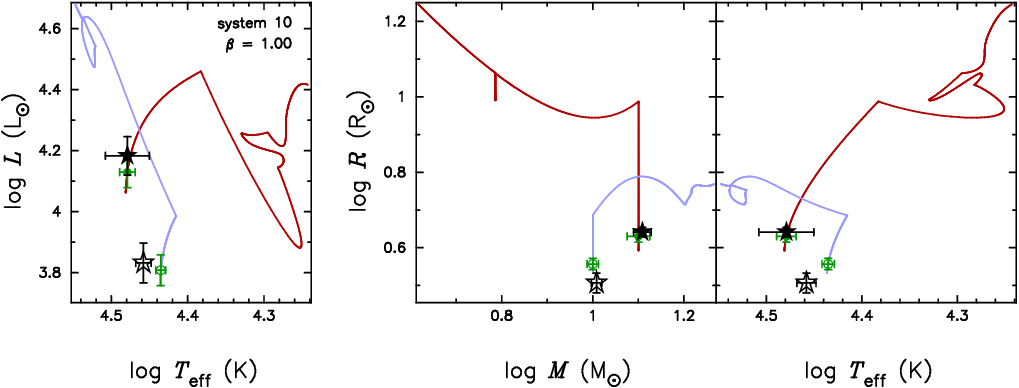}
\includegraphics[bb=20 30 520 240,clip,width=0.75\textwidth]{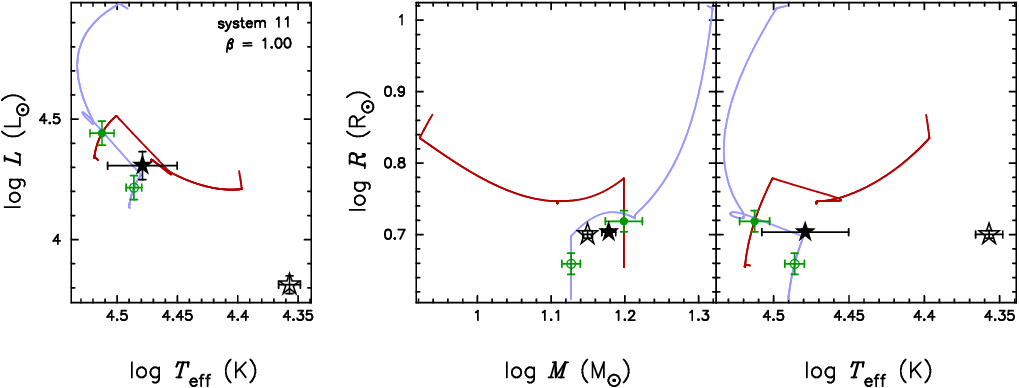}
\includegraphics[bb=20 30 520 240,clip,width=0.75\textwidth]{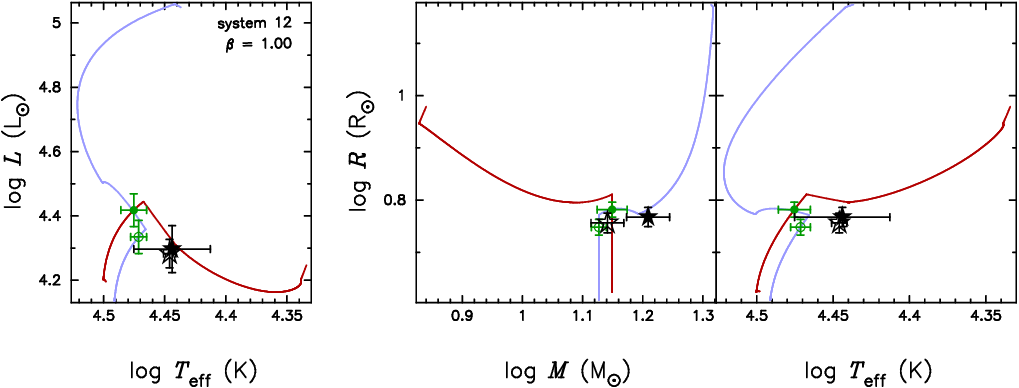}
 \caption{ 
  Continued.
 }
 \end{figure*}
\addtocounter{figure}{-1}
  \begin{figure*} \centering
\includegraphics[bb=20 30 520 240,clip,width=0.75\textwidth]{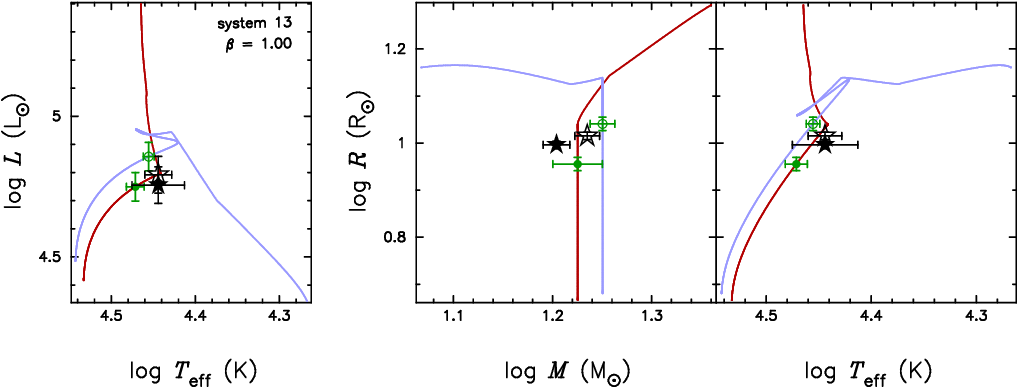}
\includegraphics[bb=20 30 520 240,clip,width=0.75\textwidth]{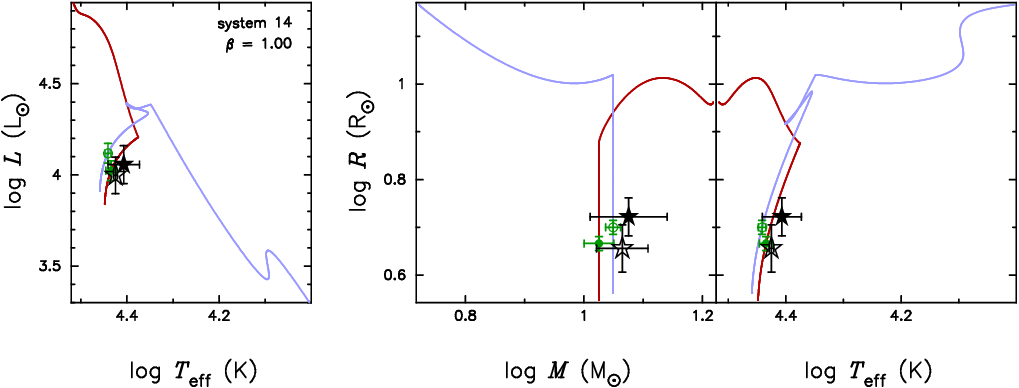}
\includegraphics[bb=20 30 520 240,clip,width=0.75\textwidth]{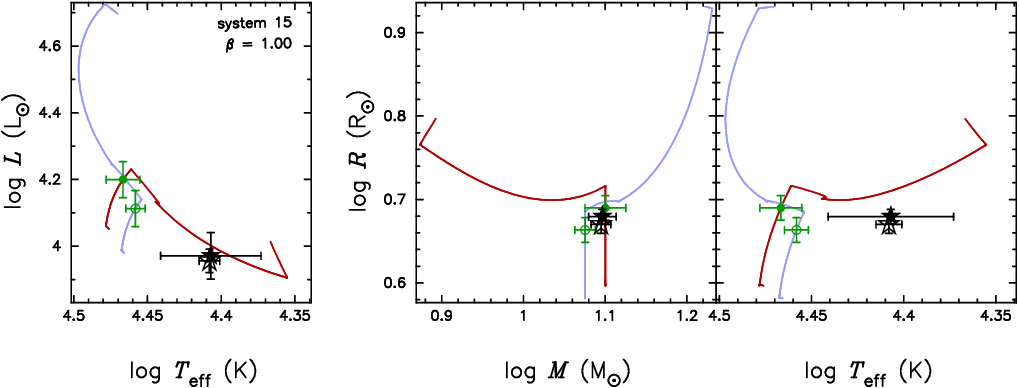}
\includegraphics[bb=20 30 520 240,clip,width=0.75\textwidth]{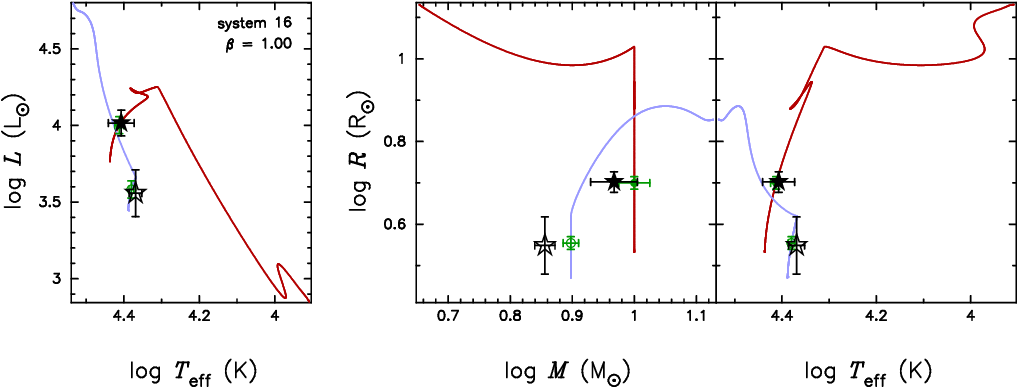}
 \caption{ 
  Continued.
 }
 \end{figure*}
\addtocounter{figure}{-1}
  \begin{figure*} \centering
\includegraphics[bb=20 30 520 240,clip,width=0.75\textwidth]{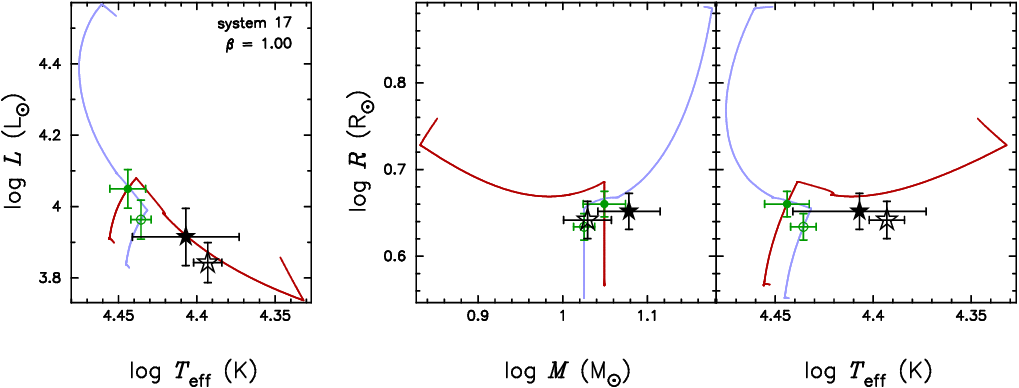}
\includegraphics[bb=20 30 520 240,clip,width=0.75\textwidth]{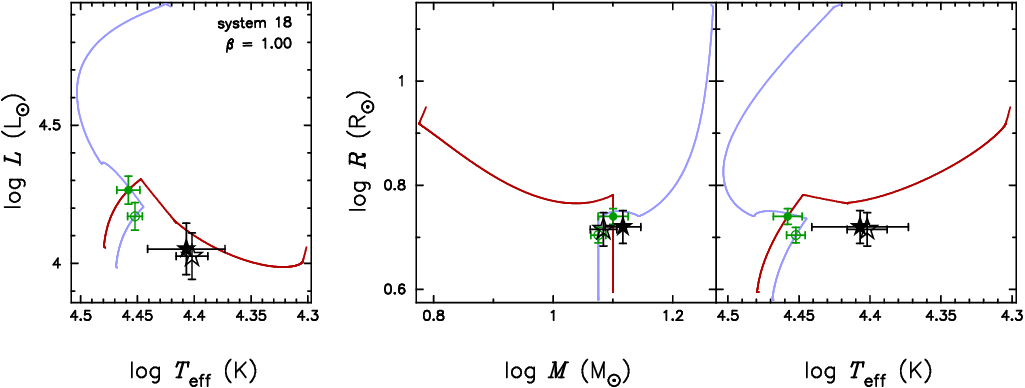}
\includegraphics[bb=20 30 520 240,clip,width=0.75\textwidth]{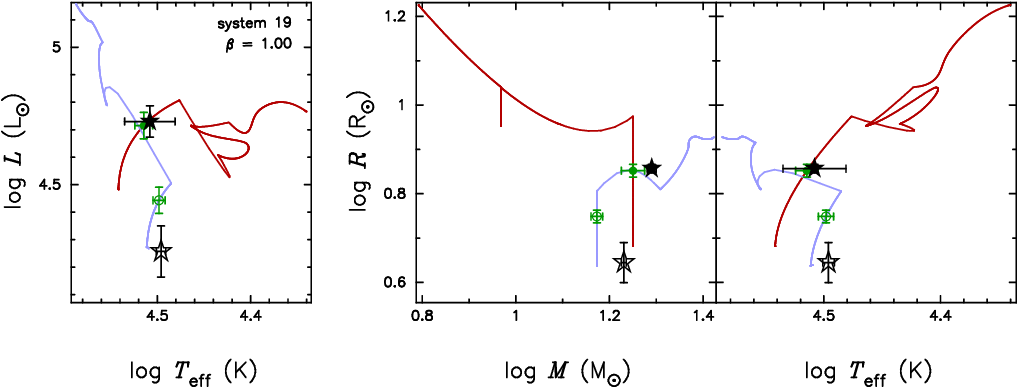}
\includegraphics[bb=20 30 520 240,clip,width=0.75\textwidth]{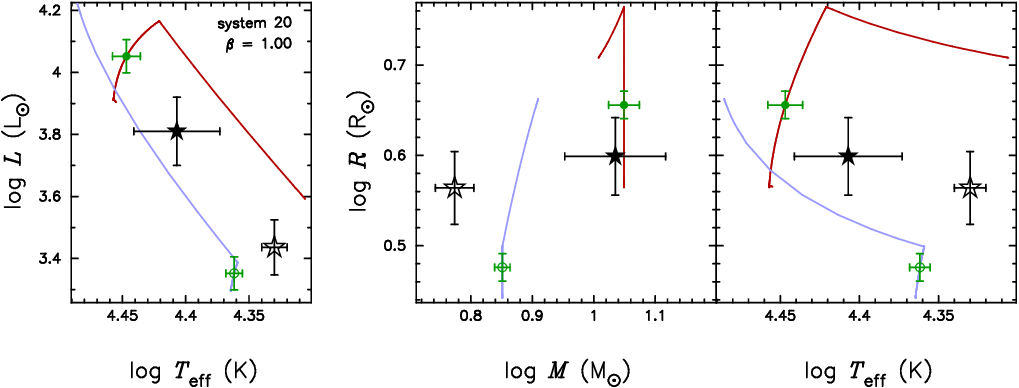}
 \caption{ 
  Continued.
 }
 \end{figure*}
\addtocounter{figure}{-1}
  \begin{figure*} \centering
\includegraphics[bb=20 30 520 240,clip,width=0.75\textwidth]{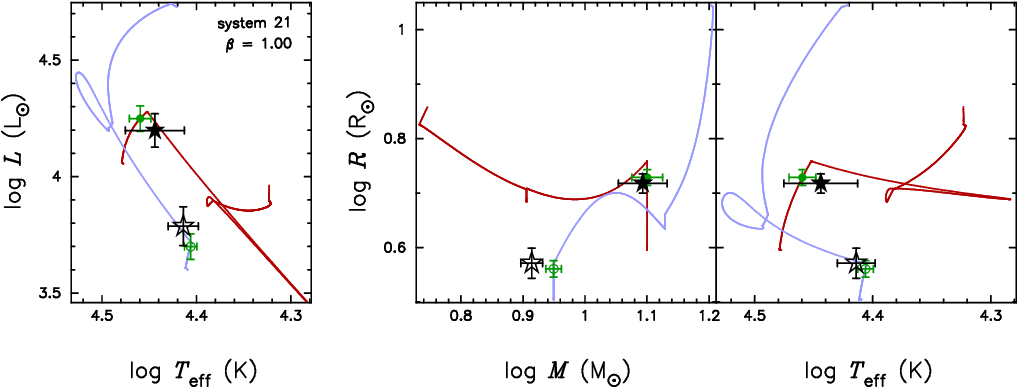}
\includegraphics[bb=20 30 520 240,clip,width=0.75\textwidth]{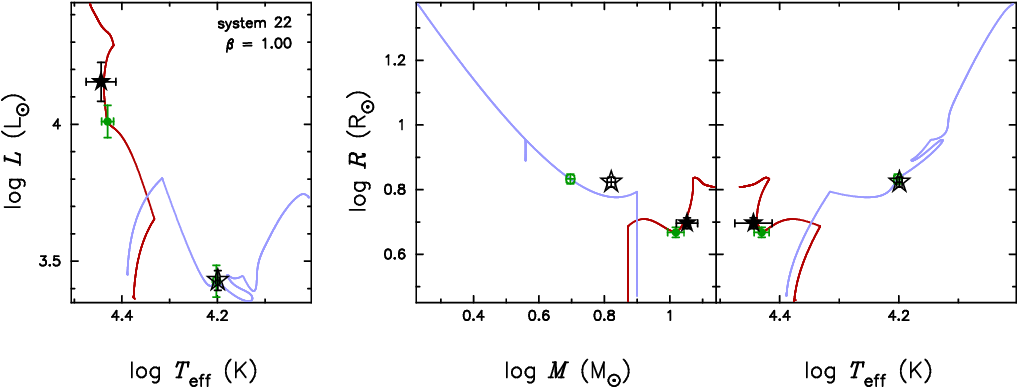}
\includegraphics[bb=20 30 520 240,clip,width=0.75\textwidth]{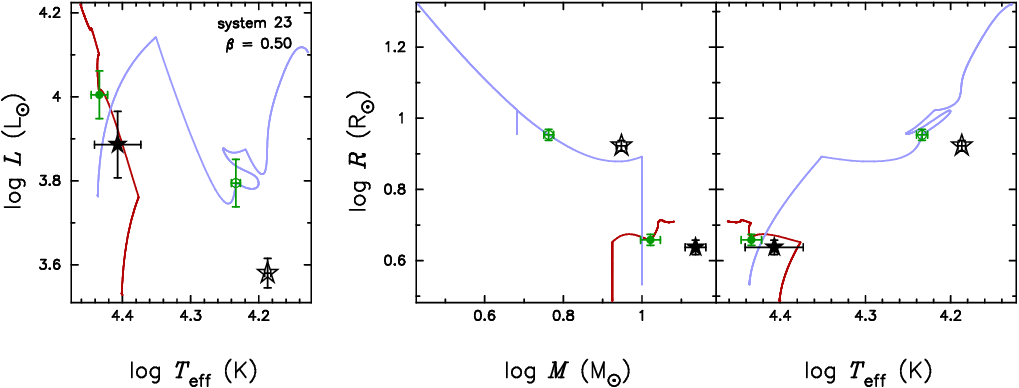}
\includegraphics[bb=20 30 520 240,clip,width=0.75\textwidth]{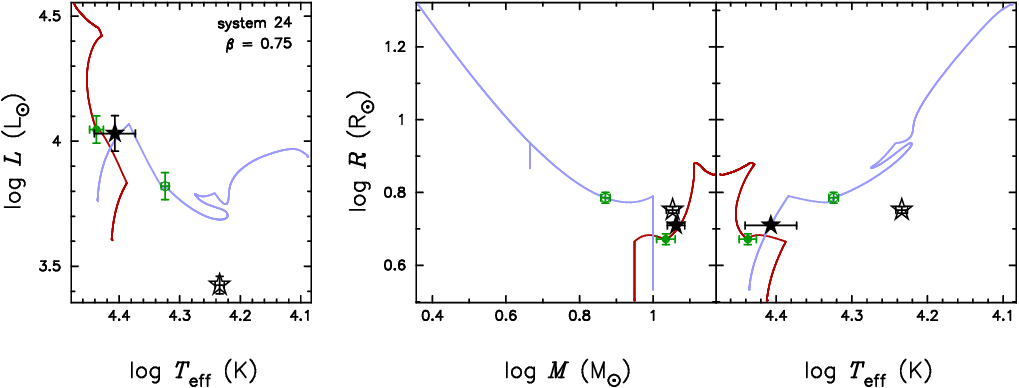}
 \caption{ 
  Continued.
 }
 \end{figure*}
\addtocounter{figure}{-1}
  \begin{figure*} \centering
\includegraphics[bb=20 30 520 240,clip,width=0.75\textwidth]{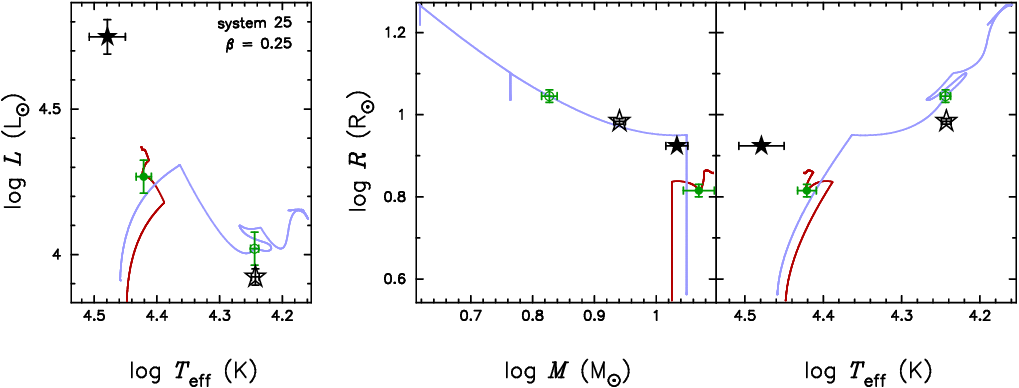}
\includegraphics[bb=20 30 520 240,clip,width=0.75\textwidth]{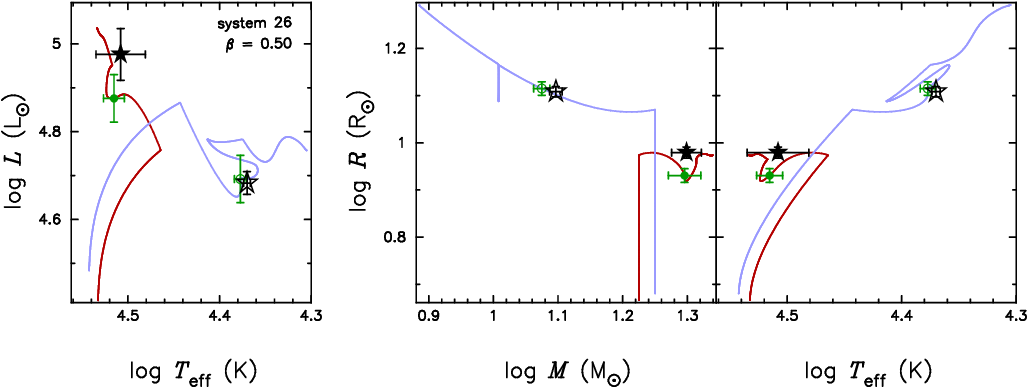}
\includegraphics[bb=20 30 520 240,clip,width=0.75\textwidth]{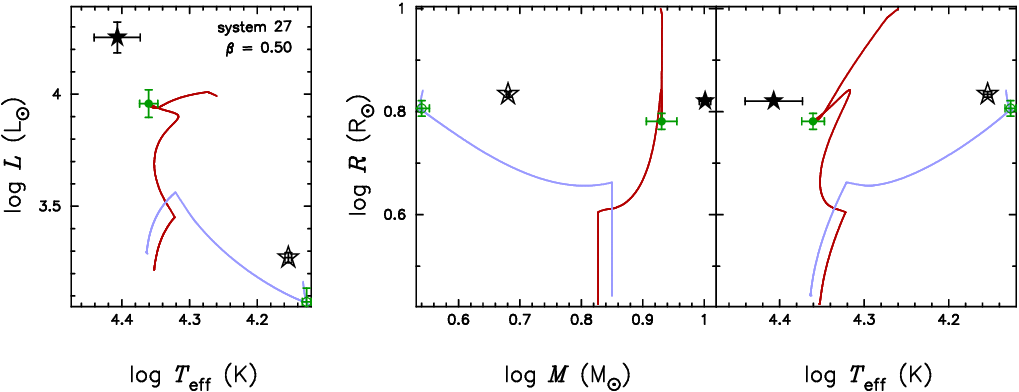}
\includegraphics[bb=20 30 520 240,clip,width=0.75\textwidth]{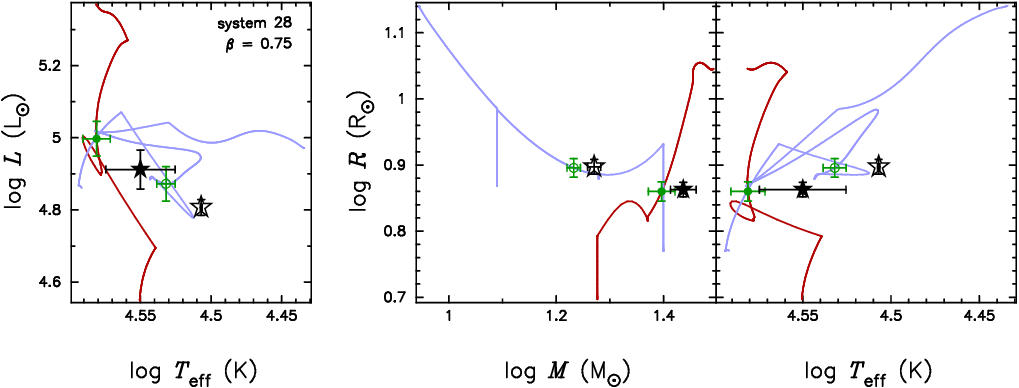}
 \caption{ 
  Continued.
 }
 \end{figure*}
\addtocounter{figure}{-1}
  \begin{figure*} \centering
\includegraphics[bb=20 30 520 240,clip,width=0.75\textwidth]{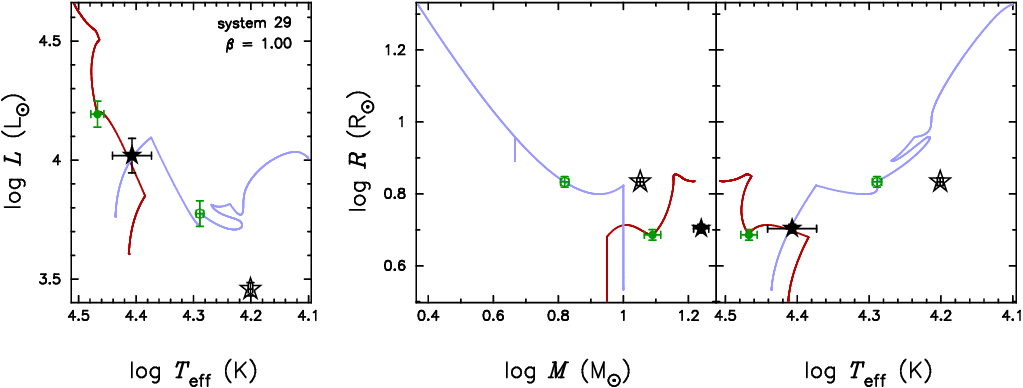}
\includegraphics[bb=20 30 520 240,clip,width=0.75\textwidth]{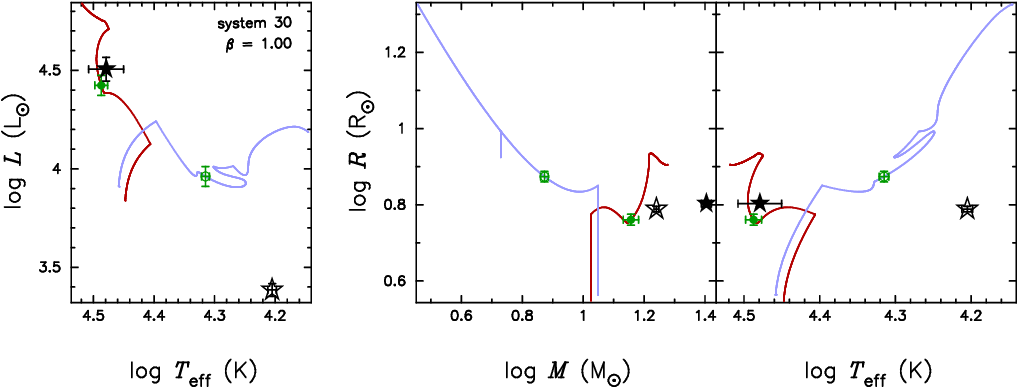}
\includegraphics[bb=20 30 520 240,clip,width=0.75\textwidth]{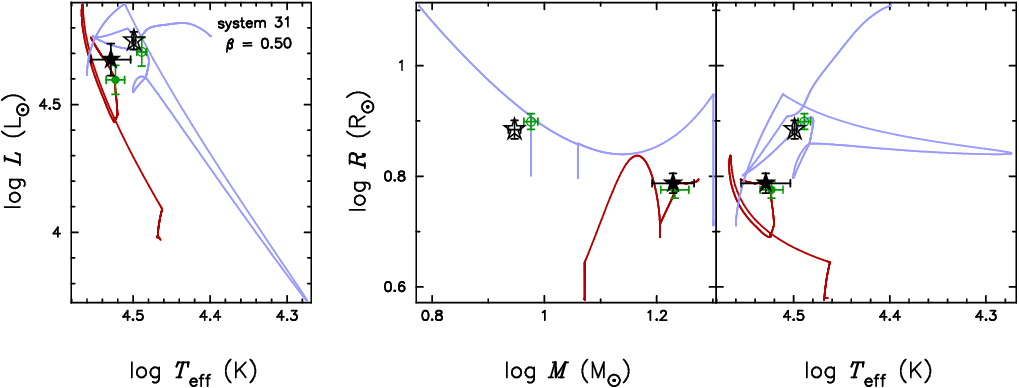}
\includegraphics[bb=20 30 520 240,clip,width=0.75\textwidth]{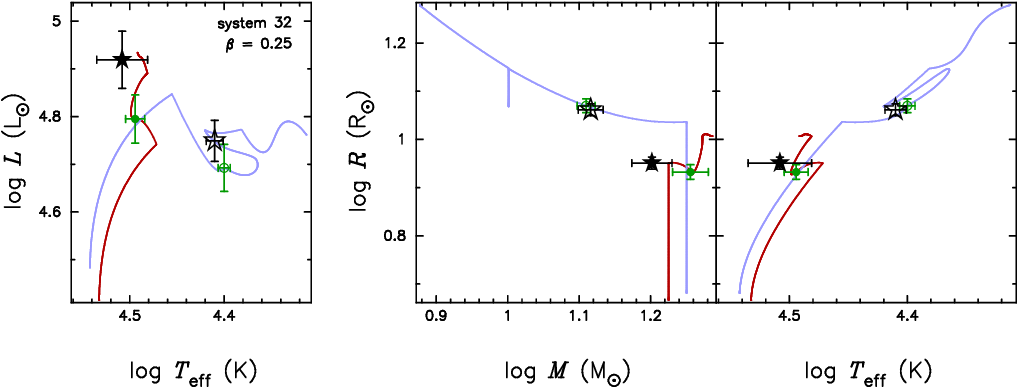}
 \caption{ 
  Continued.
 }
 \end{figure*}
\addtocounter{figure}{-1}
  \begin{figure*} \centering
\includegraphics[bb=20 30 520 240,clip,width=0.75\textwidth]{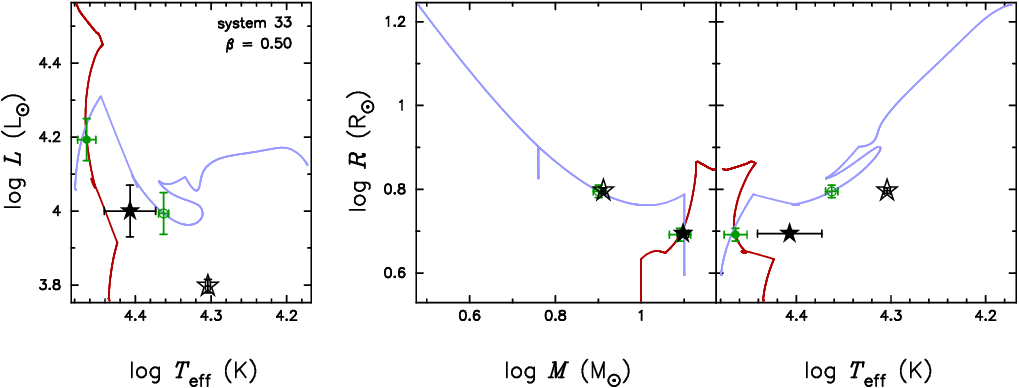}
\includegraphics[bb=20 30 520 240,clip,width=0.75\textwidth]{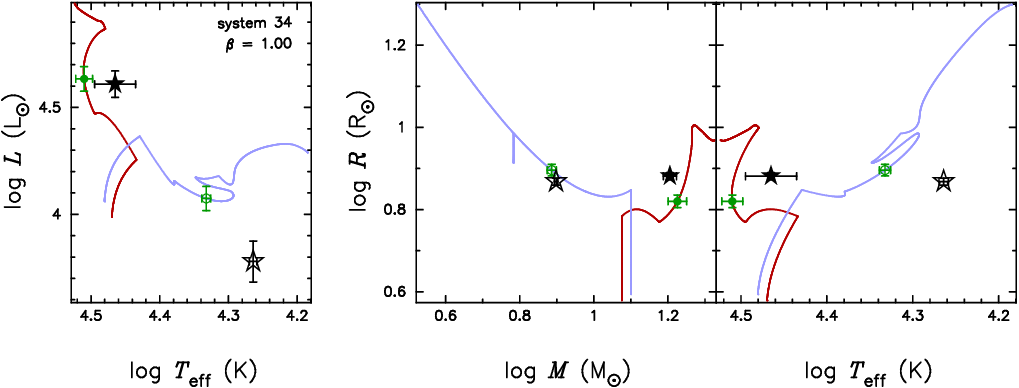}
\includegraphics[bb=20 30 520 240,clip,width=0.75\textwidth]{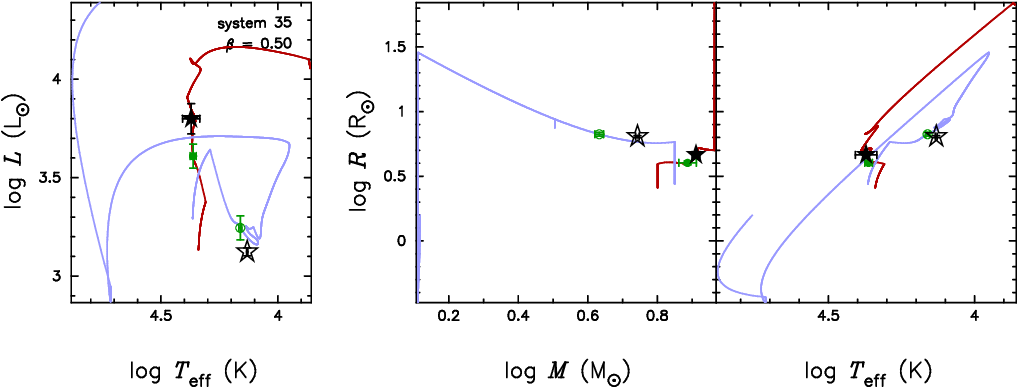}
\includegraphics[bb=20 30 520 240,clip,width=0.75\textwidth]{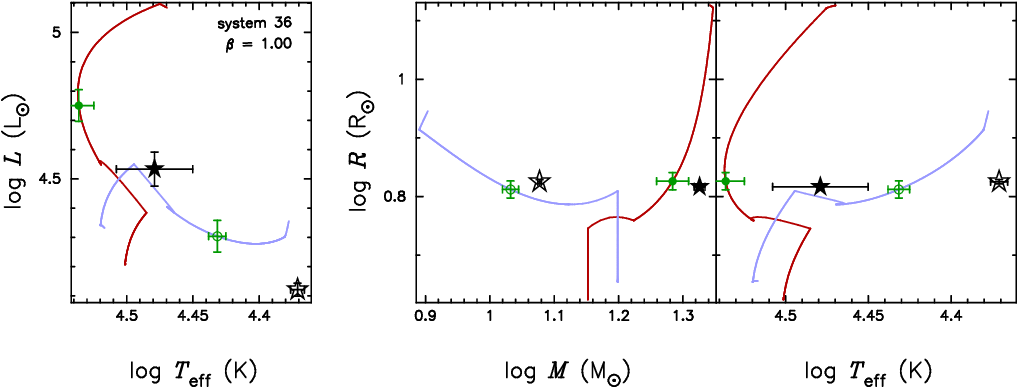}
 \caption{ 
  Continued.
 }
 \end{figure*}
\addtocounter{figure}{-1}
  \begin{figure*} \centering
\includegraphics[bb=20 30 520 240,clip,width=0.75\textwidth]{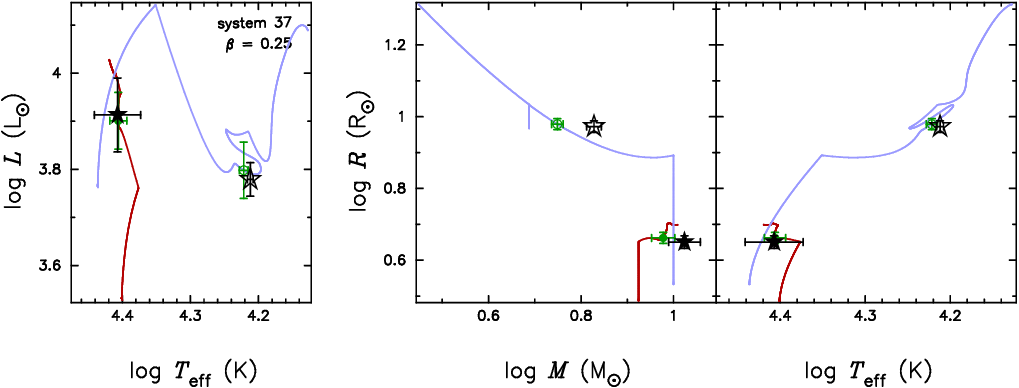}
\includegraphics[bb=20 30 520 240,clip,width=0.75\textwidth]{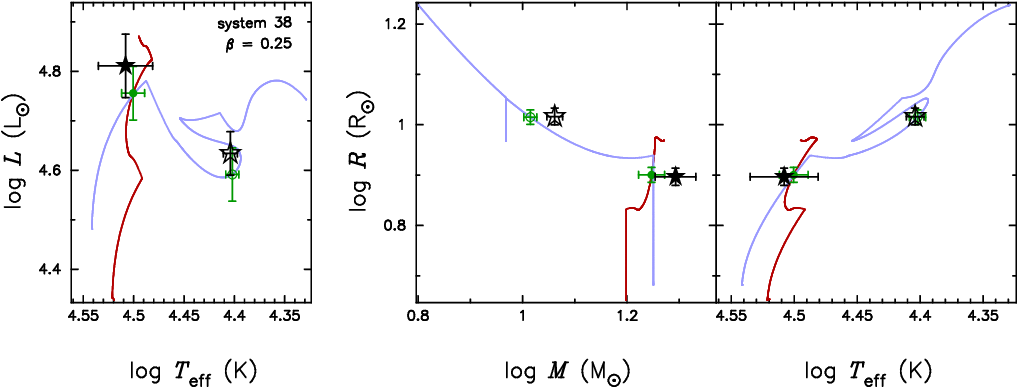}
\includegraphics[bb=20 30 520 240,clip,width=0.75\textwidth]{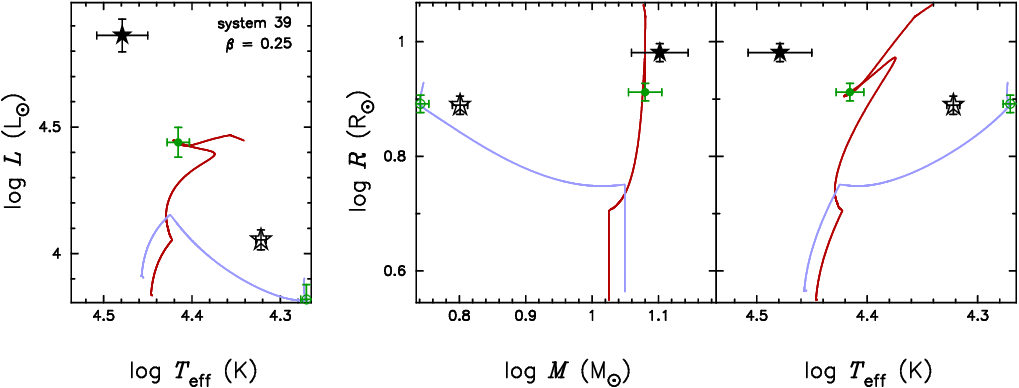}
\includegraphics[bb=20 30 520 240,clip,width=0.75\textwidth]{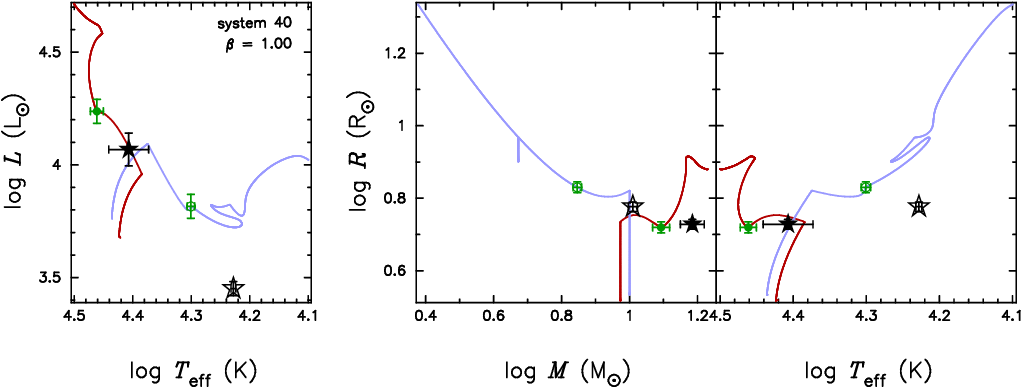}
 \caption{ 
  Continued.
 }
 \end{figure*}
\addtocounter{figure}{-1}
  \begin{figure*} \centering
\includegraphics[bb=20 30 520 240,clip,width=0.75\textwidth]{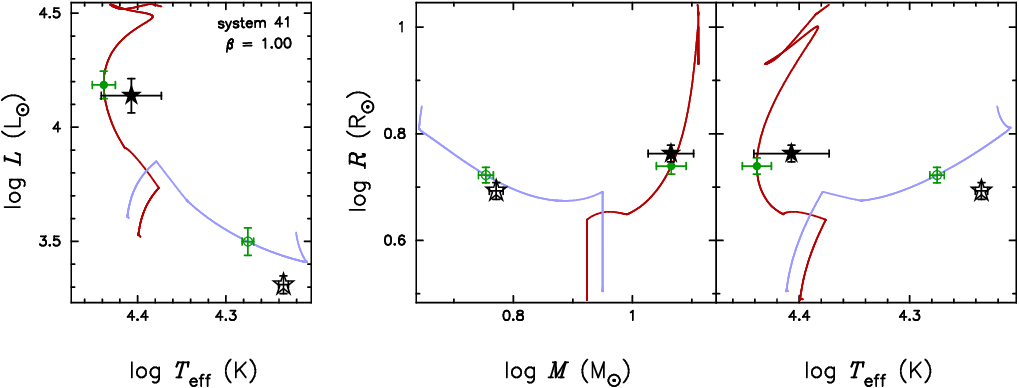}
\includegraphics[bb=20 30 520 240,clip,width=0.75\textwidth]{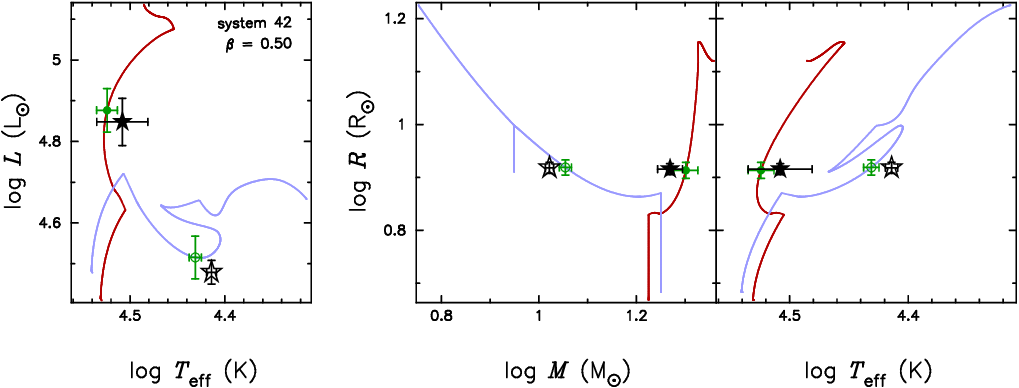}
\includegraphics[bb=20 30 520 240,clip,width=0.75\textwidth]{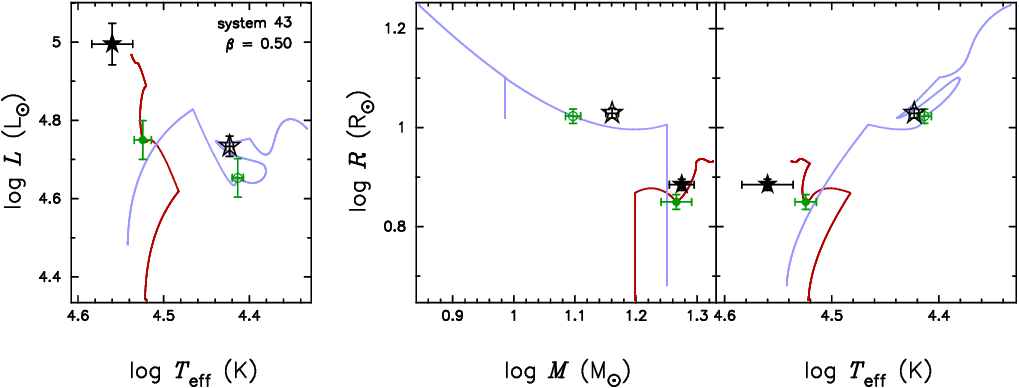}
\includegraphics[bb=20 30 520 240,clip,width=0.75\textwidth]{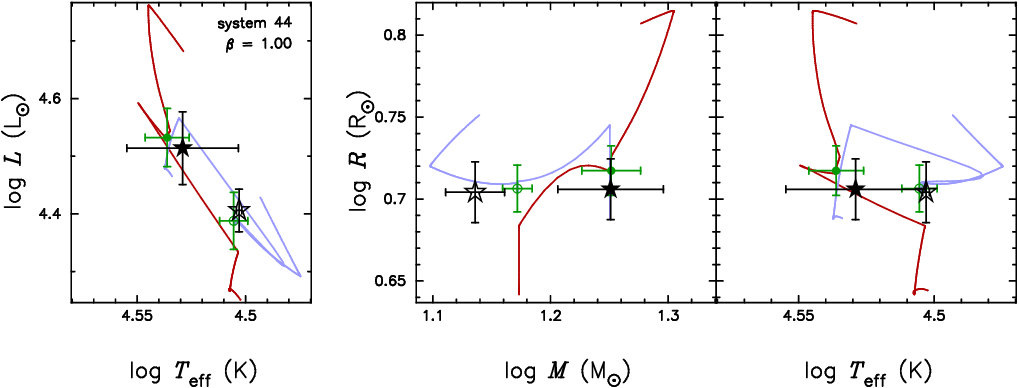}
 \caption{ 
  Continued.
 }
 \end{figure*}
\addtocounter{figure}{-1}
  \begin{figure*} \centering
\includegraphics[bb=20 30 520 240,clip,width=0.75\textwidth]{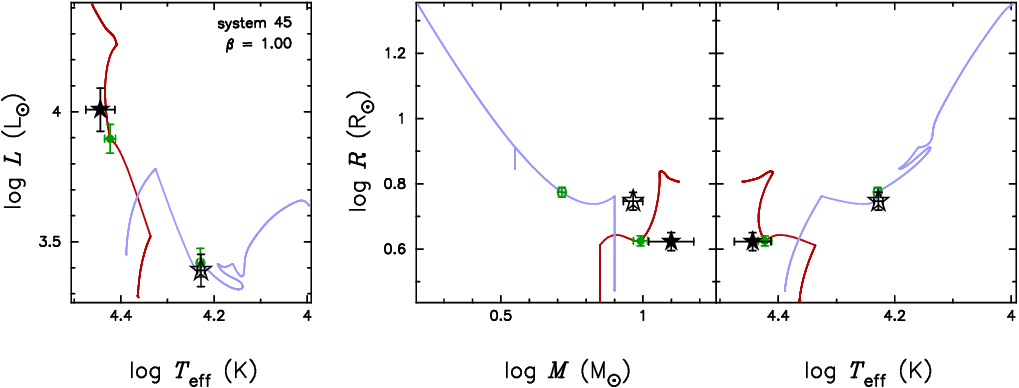}
\includegraphics[bb=20 30 520 240,clip,width=0.75\textwidth]{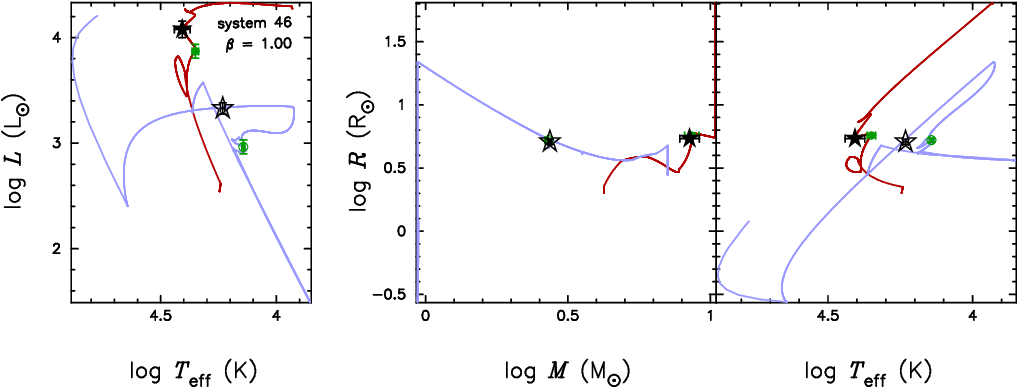}
\includegraphics[bb=20 30 520 240,clip,width=0.75\textwidth]{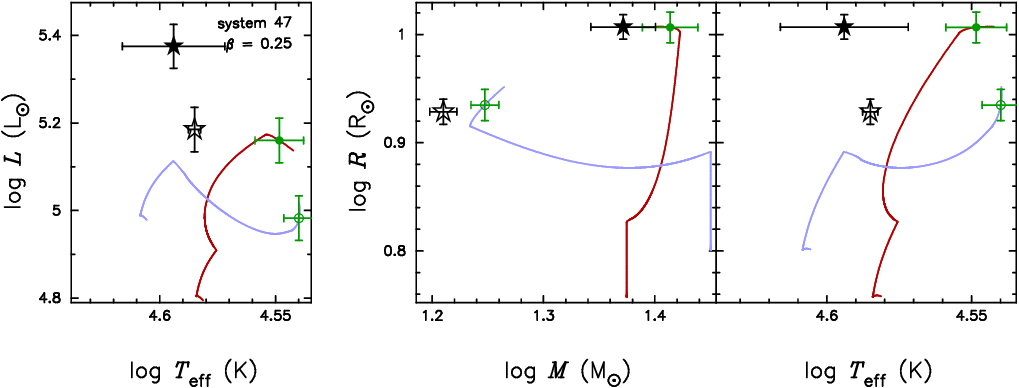}
\includegraphics[bb=20 30 520 240,clip,width=0.75\textwidth]{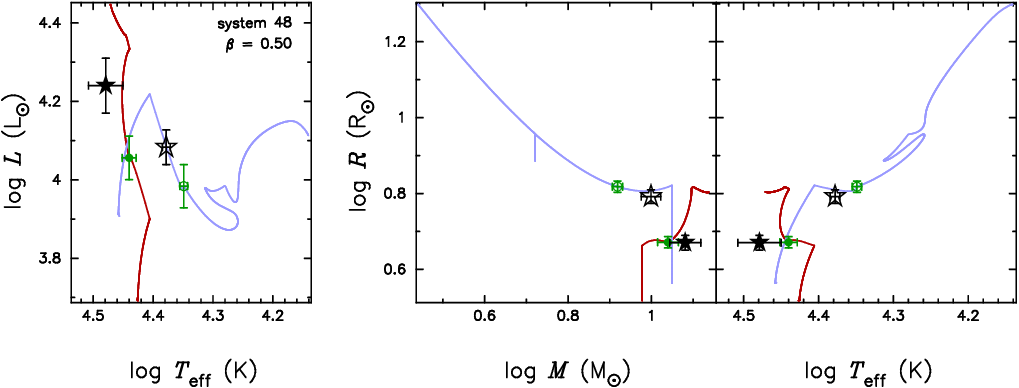}
 \caption{ 
  Continued.
 }
 \end{figure*}
\addtocounter{figure}{-1}
  \begin{figure*} \centering
\includegraphics[bb=20 30 520 240,clip,width=0.75\textwidth]{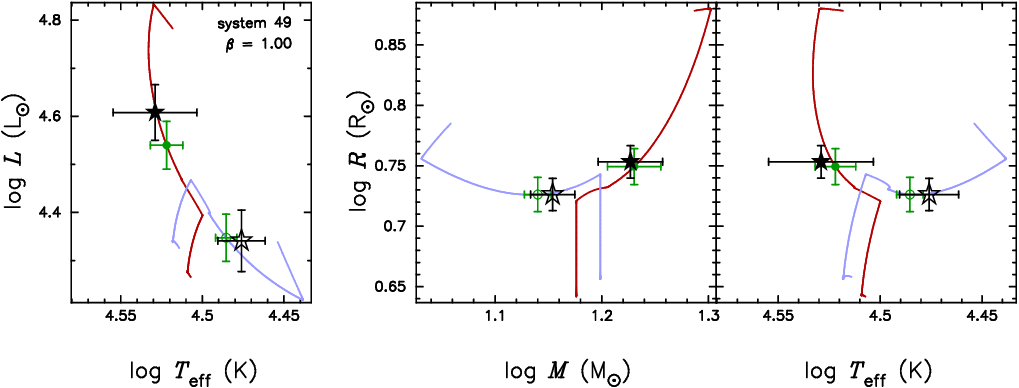}
\includegraphics[bb=20 30 520 240,clip,width=0.75\textwidth]{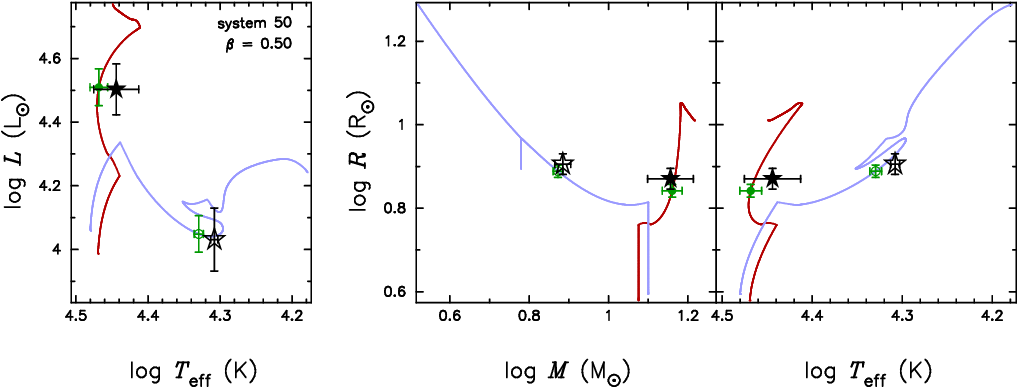}
\includegraphics[bb=20 30 520 240,clip,width=0.75\textwidth]{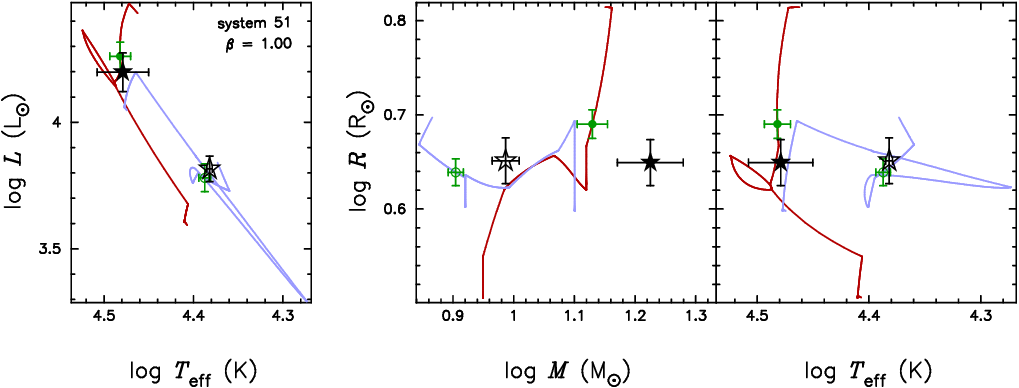}
\includegraphics[bb=20 30 520 240,clip,width=0.75\textwidth]{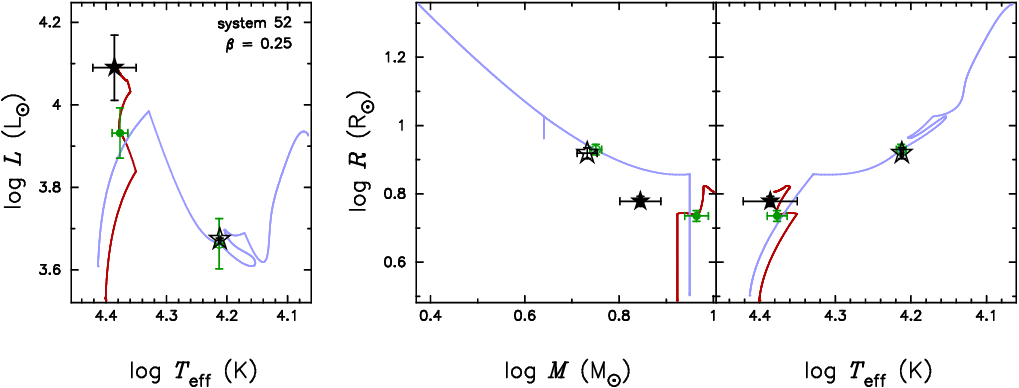}
 \caption{ 
  Continued.
 }
 \end{figure*}

%

\end{document}